\documentclass[12pt]{article}

\usepackage{cite}
\usepackage{amsmath}
\usepackage{amssymb}
\usepackage{bm}
\usepackage{color}
\usepackage{graphicx}
\numberwithin{equation}{section}

\allowdisplaybreaks

\DeclareMathOperator{\Det}{Det}

\DeclareMathOperator{\sign}{sign}

\newcommand{\llangle}{\langle\!\langle}
\newcommand{\rrangle}{\rangle\!\rangle}

\usepackage{collectbox}

\makeatletter
\newcommand{\sqbox}{%
    \collectbox{%
        \@tempdima=\dimexpr\width-\totalheight\relax
        \ifdim\@tempdima<\z@
            \fbox{\hbox{\hspace{-.5\@tempdima}\BOXCONTENT\hspace{-.5\@tempdima}}}%
        \else
            \ht\collectedbox=\dimexpr\ht\collectedbox+.5\@tempdima\relax
            \dp\collectedbox=\dimexpr\dp\collectedbox+.5\@tempdima\relax
            \fbox{\BOXCONTENT}%
        \fi
    }%
}
\makeatother

\newcounter{aff}

\begin{document}
\begin{titlepage}
\begin{flushright}
{\footnotesize NITEP 23, OCU-PHYS 504, YITP-19-58}
\end{flushright}
\begin{center}
{\LARGE\bf
Hanany-Witten Transition in Quantum Curves}\\
\bigskip\bigskip
{\large
Naotaka Kubo\,\footnote{\tt naotaka.kubo@yukawa.kyoto-u.ac.jp}
\quad
and
\quad
Sanefumi Moriyama\,\footnote{\tt moriyama@sci.osaka-cu.ac.jp}
}\\
\bigskip
${}^{*}$\,{\it Center for Gravitational Physics, 
Yukawa Institute for Theoretical Physics,}\\
{\it Kyoto University, Sakyo-ku, Kyoto 606-8502, Japan}\\[3pt]
${}^{\dagger}$\,{\it Department of Physics, Graduate School of Science,}\\
{\it Osaka City University, Sumiyoshi-ku, Osaka 558-8585, Japan}\\[3pt]
${}^{\dagger}$\,{\it Nambu Yoichiro Institute of Theoretical and Experimental Physics (NITEP),}\\
{\it Osaka City University, Sumiyoshi-ku, Osaka 558-8585, Japan}\\[3pt]
${}^\dagger$\,{\it Osaka City University Advanced Mathematical Institute (OCAMI),}\\
{\it Osaka City University, Sumiyoshi-ku, Osaka 558-8585, Japan}
\end{center}

\begin{abstract}
It was known that the U$(N)^4$ super Chern-Simons matrix model describing the worldvolume theory of D3-branes with two NS5-branes and two $(1,k)$5-branes in IIB brane configuration (dual to M2-branes after taking the T-duality and the M-theory lift) corresponds to the $D_5$ quantum curve.
For deformations of these two objects, on one hand the super Chern-Simons matrix model has three degrees of freedom (of relative rank deformations interpreted as fractional branes in brane configurations), while on the other hand the $D_5$ curve has five degrees of freedom (characterized by point configurations of asymptotic values).
To identify the three-dimensional parameter space of brane configurations in the five-dimensional space of point configurations, we propose the necessity to cut the compact T-duality circle (or the circular quiver diagram) open, which is similar to the idea of ``fixing a reference frame'' or ``fixing a local chart''.
Since the parameter space of curves enjoys the $D_5$ Weyl group beautifully, we are naturally led to conjecture that M2-branes are not only deformed by fractional branes but more obscure geometrical backgrounds.
\end{abstract}

\end{titlepage}

\tableofcontents

\section{Introduction}

In Newtonian mechanics the first step in studying motion of objects is to fix a reference frame.
Though transformation laws between frames are studied afterwards, without fixing a reference it is not even possible to describe the location of the objects by coordinates.
The importance of fixing a reference appears similarly in studying the super Chern-Simons matrix models.

The simplest super Chern-Simons matrix model describing M2-branes is the ABJM matrix model.
The ABJM theory \cite{ABJM,HLLLP2,ABJ} is the ${\cal N}=6$ superconformal Chern-Simons theory with gauge group $\text{U}(N_1)_k\times\text{U}(N_2)_{-k}$ and two pairs of bifundamental matters where the subscripts denote the Chern-Simons levels.
The ABJM matrix model is the partition function of the ABJM theory on $S^{3}$, which is originally defined by the infinite-dimensional path integral and reduces to a finite-dimensional matrix integration after applying the localization technique \cite{KWY}.
This theory describes the worldvolume theory of $\min(N_1,N_2)$ M2-branes with $|N_2-N_1|$ fractional M2-branes on the background geometry ${\mathbb{C}}^{4}/{\mathbb{Z}}_{k}$.
The description is understood from the brane configuration in type IIB string theory.
Hinted by the number of unbroken supercharges, it was known that the theory is realized in the brane configuration of D3-branes on a circle $S^{1}$ with a perpendicular NS5-brane and a $(1,k)$5-brane relatively tilted by an angle parametrized by $k$ where the numbers of D3-branes are $N_1$ and $N_2$ in each interval. 
After performing T-duality and lifting to M-theory, we obtain the background geometry of M2-branes.

The relation among matrix models, spectral theories and topological strings is revealed through the study of instanton expansion in the ABJM matrix model.
Though the relation was eventually established for general rank deformations, the analysis starts from the simplest case with equal ranks $N_2=N_1=N$.
On one hand, in studying the expression of the instanton corrections, a crucial proposal of the Fermi gas formalism was made in \cite{MP}.
Namely, it was found that the grand canonical partition function without rank deformations is expressed by the Fredholm determinant 
\begin{align}
\Xi_{k}(z)=\Det(1+z\widehat{H}^{-1}).\label{eq:GPF_SD0}
\end{align}
The spectral operator $\widehat H$ takes the form 
\begin{align}
\widehat{H}=\widehat{{\cal Q}}\widehat{{\cal P}},
\end{align}
with 
\begin{align}
\widehat{{\cal Q}}=2\cosh\frac{\widehat{q}}{2},\quad\widehat{{\cal P}}=2\cosh\frac{\widehat{p}}{2},
\end{align}
where $\widehat{q}$ and $\widehat{p}$ are the canonical coordinate and momentum operators satisfying the canonical commutation relation $[\widehat{q},\widehat{p}]=i\hbar$ with the identification $\hbar=2\pi k$.
This is reminiscent of the ${\mathbb{P}}^1\times{\mathbb{P}}^1$ geometry \cite{MPtop,MP} if we introduce 
\begin{align}
\widehat{Q}=e^{\widehat{q}},\quad\widehat{P}=e^{\widehat{p}},
\end{align}
and express $\widehat H$ by these canonical operators $\widehat H=(\widehat{Q}^{\frac{1}{2}}+\widehat{Q}^{-\frac{1}{2}})(\widehat{P}^{\frac{1}{2}}+\widehat{P}^{-\frac{1}{2}})$ where the Newton polygon of the resulting curve is nothing but that of $\mathbb{P}^1\times\mathbb{P}^1$ after a change of variables.
On the other hand, the large $N$ behavior $N^{\frac{3}{2}}$ of the degrees of freedom of $N$ M2-branes known from the gravity side \cite{KT} was reproduced by computing the free energy of the ABJM matrix model \cite{DMP1,HKPT}.
Subsequently various corrections were studied including the sum of all perturbative corrections \cite{FHM,MP}, worldsheet instantons \cite{DMP1,DMP2}, membrane instantons \cite{DMP2,MP,HMO2,CM} and their bound states \cite{HMO3}.
Interestingly, it was found that, although both the coefficients of the worldsheet instantons and those of the membrane instantons are divergent, the divergences are all canceled and the sum is free of divergences \cite{HMO2}.
Finally, from all the expansions and the cancellation mechanism, it was found that the final expression of the instanton corrections is given by the sum of the free energy of topological strings and the derivative of its refinement \cite{HMMO} on local $\mathbb{P}^1\times\mathbb{P}^1$ geometry.
After removing the connection to the matrix models, these observations further led \cite{GHM1} to conjecture that the Fredholm determinant of a general spectral operator is equal to the free energy of topological strings on a background read off from the spectral operator.

There are several generalizations of this theory.
One interesting direction is to increase the numbers of NS5-branes and $(1,k)5$-branes.
Then, the brane configuration is labeled by a digit sequence $\{s_a\}_{a=1}^R$ with $s_a=\pm 1$ where $s_{a}=+1$ and $s_a=-1$ correspond to an NS5-brane and a $(1,k)5$-brane respectively.
The worldvolume theory of $\{s_a\}_{a=1}^R$ is a quiver $\text{U}(N)^{R}$ ${\cal N}=4$ superconformal Chern-Simons theory of circular type with Chern-Simons levels given by \cite{IK4}
\begin{align}
k_{a}=\frac{k}{2}(s_{a}-s_{a-1}).
\end{align}
We often refer this theory and the corresponding matrix model obtained from the localization technique as the $(p_1,q_1,p_2,q_2,\cdots)$ theory and the $(p_1,q_1,p_2,q_2,\cdots)$  model when the digit sequence is
\begin{align}
\{s_a\}_{a=1}^R
=\{\underbrace{+1,\cdots,+1}_{p_1},\underbrace{-1,\cdots,-1}_{q_1},
\underbrace{+1,\cdots,+1}_{p_2},\underbrace{-1,\cdots,-1}_{q_2},\cdots\}.
\end{align}
As in the case of the ABJM theory, the relation among super Chern-Simons matrix models, spectral theories and topological strings holds again.
In \cite{MN1}, it was found that, for the $(p_1,q_1,p_2,q_2,\cdots)$ model, the grand canonical partition function without rank deformations is expressed by the Fredholm determinant \eqref{eq:GPF_SD0} of the spectral operator
\begin{align}
\widehat{H}=\cdots\widehat{{\cal Q}}^{q_2}\widehat{{\cal P}}^{p_2}\widehat{{\cal Q}}^{q_1}\widehat{{\cal P}}^{p_1},
\end{align}
in the inverse order of $(p_1,q_1,p_2,q_2,\cdots)$.
This generalization of the Fermi gas formalism was used in \cite{MN3} to study the $(2,2)$ model extensively.
After all the studies of the instanton effects it was found that the geometrical background of topological strings is local del Pezzo $D_5$.
The appearance of local del Pezzo $D_5$ is again natural from the viewpoint of \cite{GHM1} since the spectral operator $\widehat{H}=\widehat{{\cal Q}}^{2}\widehat{{\cal P}}^{2}$ gives exactly the Newton polygon of local del Pezzo $D_5$.

To understand the relation in more details, rank deformations of the $(2,2)$ model were studied in \cite{MNN}.
Combined with the results obtained from rank deformations of the $(1,1,1,1)$ model \cite{HM} through the Hanany-Witten transition \cite{HW}, it was found that the parameter spaces of both models are connected smoothly.
Among others it was pointed out that, though in rank deformations we have several ranks appearing, for the relation to topological strings to work correctly, we have to fix the power of the fugacity so that it matches to one of the ranks in defining the grand canonical partition function \cite{MNN}.
Then, the grand canonical partition function with rank deformations is described by the free energy of topological strings if we assume that the BPS indices are split suitably.
The split of the BPS indices which form representations of $D_5$ is further explained \cite{MNY} by assuming an unbroken subgroup of $D_5$ and studying the decomposition of the representations into the subgroup.
Especially, it was found that the unbroken subgroups of the $(2,2)$ model and the $(1,1,1,1)$ model without rank deformations are $D_4$ and $A_2\times(A_1)^2$ respectively.

It is curious to ask whether we can explain the unbroken subgroup directly from the matrix model.
In \cite{KMN} the idea of quantum curves was introduced by identifying the spectral operator with those obtained from similarity transformations.
Then, as the classical curves enjoying the $D_5$ Weyl symmetry, it can be shown that its quantum cousin also satisfies the same Weyl symmetry with a slight modification of the parameters.
After identifying the location of our matrix model in the parameter space, we can ask what subgroup of the $D_5$ Weyl symmetry the matrix model preserves.
After our full analysis in \cite{KMN}, it turns out that the unbroken subgroups match completely with the results from topological strings.
Thus, the symmetry breaking patterns of the matrix models without rank deformations were explained clearly from the study of quantum curves.
In \cite{KMN} some rank deformations were also identified, though the full studies of rank deformations were postponed.

In this paper, we head for the identification of the full rank deformations in the parameter space of quantum curves.
For this purpose, we need to reconsider the identification even without rank deformations.
In \cite{KMN} it was explained that, depending on the unbroken symmetry, we can consider cosets transforming among the parameters of quantum curves, which generate several points in the parameter space and invalidate the one-to-one correspondence of the ranks and the parameters.
The main idea to avoid the difficulty comes back to the idea appearing at the beginning of this paper, the introduction of \textit{reference frames}.
With the idea of fixing a reference frame we are able to get rid of the ambiguity.

More concretely, by looking back to each of the main characters in the correspondence, the brane configuration, the spectral theory and the topological string theory, we find without difficulty that the idea of the reference frame is omnipresent.
Let us explain each of them separately.
Firstly, in discussing the brane configuration, we often exchange branes with the Hanany-Witten transition.
As already pointed out in \cite{HW}, the rule of the Hanany-Witten transition can be derived from the NS/R charge conservation.
Namely, by requiring that the NS/R charges computed from the numbers of branes on the left and on the right are conserved, we can derive the rules of the Hanany-Witten transition.
However, in discussing the branes on the left or on the right in a compact circle $S^1$ with two sides identified, we need to specify asymptotic D3-branes, which break the circle into a segment.
This idea of cutting the circle open serves the role of fixing a reference frame in the brane configuration.
Secondly, from the viewpoint of the spectral theory, although the spectrum of a quantum operator is generally invariant under similarity transformations and we defined quantum curves with the identification of the similarity transformations in \cite{KMN}, the expression of the spectral operator $\widehat H$ itself and the parameters of the quantum curve are always given after fixing the order of the operators.
Also, in one of the Fermi gas formalisms for the matrix model with rank deformations (called the closed string formalism), the spectral operator or the quantum curve is obtained by integrating out all of the fractional brane backgrounds.
In this sense, we need to fix the closed string background to be that with minimal number of D3-branes to consider the spectral operator.
Thirdly, in the topological sting theory, as we mentioned previously, for the correspondence between matrix models and topological strings, we need to match the power of the fugacity with one of the ranks of the partition function, which serves as fixing a reference frame.

After clarifying the idea of fixing a reference frame, we can identify the three-dimensional space of rank deformations in brane configurations in the five-dimensional space of quantum curves.
By comparing the two parameter spaces and their symmetries, we find a novel symmetry for brane configurations which is not obtained from the Hanany-Witten transition or other well-known discrete symmetries.
We also find that the identification of rank deformations in quantum curves is consistent with the description in spectral theories and topological strings.

The organization of this paper is as follows.
In section \ref{secMatrixModel}, we elaborate the idea of fixing a reference frame in reviewing various aspects of the correspondence, such as brane configurations, super Chern-Simons matrix models and quantum curves.
After that, using the idea of fixing a reference we identify the rank deformations in the parameter space of the quantum curves in section \ref{sec:MM_QC}.
Then we present some non-trivial checks from the relation to spectral theories in section \ref{sec:MM_ST} and from the relation to topological strings in section \ref{sec:QCandTS}.
Finally we conclude with some further directions.
Appendix \ref{fgf} is devoted to clarification of the closed string formalism which is helpful for us to study the relation to spectral theories in section \ref{sec:MM_ST}, while appendix \ref{tsappendix} is a collection of non-perturbative effects and characters for the study in section \ref{sec:QCandTS}.

\section{Reference frame\label{secMatrixModel}}

In this section we review brane configurations in type IIB string theory, super Chern-Simons matrix models obtained from the brane configurations by the localization technique and quantum curves obtained in the analysis of the matrix models.
In reviewing each topic we emphasize that we have often unconsciously taken the idea of fixing a reference frame for granted.
We believe that the importance of fixing a frame in discussing the correspondence was not pointed out explicitly previously and we try to explain our idea carefully through the reviews of various aspects.

\subsection{Brane configurations\label{subsec:BC_duality}}

\begin{figure}[t!]
\centering\includegraphics[scale=0.6,angle=-90]{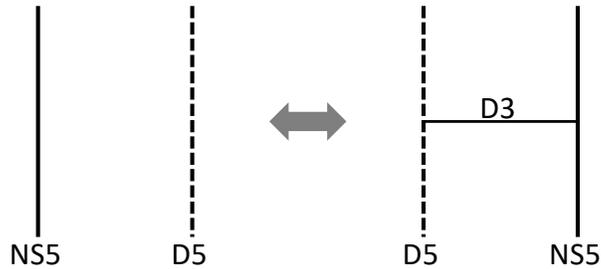}
\caption{The simplest example of the Hanany-Witten transition.
After the exchange of two 5-branes, a D3-brane is generated.}
\label{fig:HW}
\end{figure}

\begin{figure}[t!]
\centering\includegraphics[scale=0.6,angle=-90]{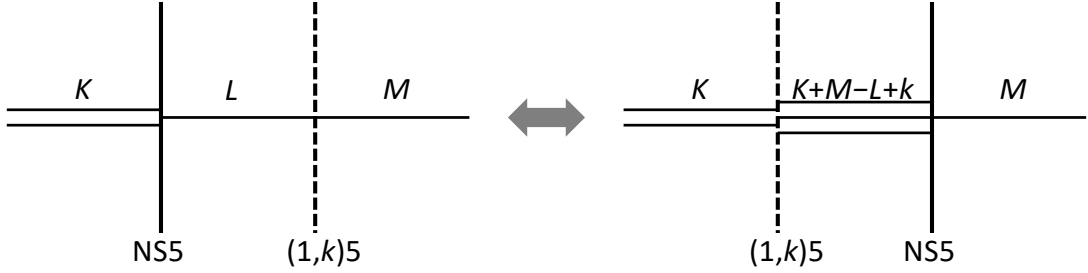}
\caption{The Hanany-Witten transition of our main interest in this paper.
In the following we often denote the brane configuration by $\cdots K\bullet L\circ M\cdots\simeq\cdots K\circ(K+M-L+k)\bullet M\cdots$ for simplicity, where $\bullet$ is an NS5-brane and $\circ$ is a $(1,k)$5-brane.}
\label{fig:HWCS}
\end{figure}

In this subsection, we review the brane configurations of our interest and explain the idea of fixing a reference frame in it.
Before it, we start with recapitulating the Hanany-Witten transition.
In \cite{HW} a supersymmetric brane configuration in type IIB string theory was considered which consists of a NS5-brane (in the $012456$ plane) and a D5-brane (in the $012789$ plane) placed at different positions on a line (along the $3$ direction) (see figure \ref{fig:HW}).
It was proposed that, when the two 5-branes move across, a D3-brane (in the $0123$ plane) stretching between the two 5-branes is generated.
The physics for these two brane configurations are considered to be equivalent and the equivalence in changing the configurations is called the Hanany-Witten transition. 
The Hanany-Witten transition is further generalized to a supersymmetric system with two general types of 5-branes and general numbers of D3-branes on each interval, where to preserve the supersymmetry the $(p,q)$5-brane is placed in the $012[4,7]_\theta[5,8]_\theta[6,9]_\theta$ plane with $[a,b]_\theta$ being the direction of $\vec e_a\cos\theta+\vec e_b\sin\theta$ and $\tan\theta=q/p$.
For our purpose, we consider a configuration with the two types of 5-branes being an NS5-brane and a $(1,k)$5-brane with $k>0$ and the numbers of D3-branes in each interval being\footnote{The rank deformations are restricted by supersymmetries \cite{HW} though we only consider the deformations formally without referring to the restriction.} $K$, $L$ and $M$ (see figure \ref{fig:HWCS}).
Then, the Hanany-Witten transition claims that, when the two 5-branes are exchanged, the number of D3-branes between two 5-branes becomes $K+M-L+k$.
Namely, if we denote the NS5-brane by $\bullet$, the $(1,k)$5-brane by $\circ$ and the D3-branes by their numbers, the Hanany-Witten transition claims the equivalences
\begin{align}
\cdots K\bullet L\circ M\cdots\simeq\cdots K\circ(K+M-L+k)\bullet M\cdots,\nonumber\\
\cdots K\circ L\bullet M\cdots\simeq\cdots K\bullet(K+M-L+k)\circ M\cdots,
\label{LMN1}
\end{align}
where we express the Hanany-Witten transition by the equivalence $\simeq$.
We also apply the transition to trivial exchanges of the same type and obtain
\begin{align}
\cdots K\bullet L\bullet M\cdots\simeq\cdots K\bullet(K+M-L)\bullet M\cdots,\nonumber\\
\cdots K\circ L\circ M\cdots\simeq\cdots K\circ(K+M-L)\circ M\cdots.
\label{LMN2}
\end{align}
Note that an overall addition of the numbers of D3-branes, $K\to K+N, L\to L+N, M\to M+N$, does not affect the relative numbers of D3-branes in the Hanany-Witten transition.

As explained in \cite{HW} the transition can be understood from charge conservation.
Namely, if we focus on the NS5-brane $\bullet$, the Hanany-Witten transition can be derived by requiring that the charge
\begin{align}
q_{\text{RR}}&=-\frac{(\#\text{D5})|_\text{L}-(\#\text{D5})|_\text{R}}{2}
+(\#\text{D3})|_\text{L}-(\#\text{D3})|_\text{R},
\label{charge}
\end{align}
is preserved under the exchange of 5-branes.
Here $(\#\text{D5})|_\text{L/R}$ denotes the number of D5-branes located to the left/right of the original NS5-brane $\bullet$ while $(\#\text{D3})|_\text{L/R}$ means the number of D3-branes ending on the NS5-brane $\bullet$ from the left/right.
Already at this point we easily find that if we consider the $3$ direction to be a compact circle $S^1$ instead of a line (as in the brane configuration of the ABJM theory), the concept of left or right is ambiguous unless we specify an interval between two 5-branes as a reference frame and do not consider the exchange of 5-branes across this interval.
In other words, we cut the compact circle $S^1$ open into a segment and bring the two ends to the infinity.

\begin{figure}[t!]
\centering\includegraphics[scale=0.6,angle=-90]{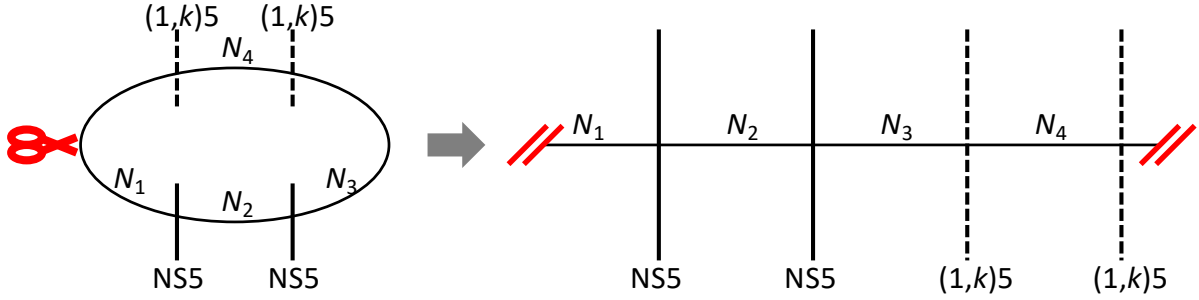}
\caption{The brane configuration corresponding to \eqref{eq:BraneConfig0}.}
\label{fig:BraneConfig0}
\end{figure}

Now let us turn to the supersymmetric brane configuration of our main interest\footnote{Our arguments apply to general brane configurations as well such as that for the original ABJM theory.
We mainly focus on this model because its abundance actually simplifies our arguments.}
with two NS5-branes and two $(1,k)$5-branes on a compact circle $S^1$ (see figure \ref{fig:BraneConfig0}).
We denote the brane configuration by a bracket
\begin{align}
\langle N_1\bullet N_2\bullet N_3\circ N_4\,\circ\rangle,
\label{eq:BraneConfig0}
\end{align}
where we place the reference interval which 5-branes do not move across at the ends and denote the number of D3-branes in each interval between two 5-branes along this line as $N_1,N_2,N_3,N_4$ respectively.
We omit displaying the number of D3-branes after the last $(1,k)$5-brane $\circ$, which is of course $N_1$ from the original periodicity of $S^1$.
Namely in the present case we fix the interval with $N_1$ D3-branes as the reference and do not consider the exchange of 5-branes across this interval.
From the Hanany-Witten transition explained above, we obtain many non-trivial relations of physically equivalent brane configurations including
\begin{align}
\langle N_1\bullet N_2\bullet N_3\circ N_4\,\circ\rangle
\simeq\langle N_1\bullet N_2\circ(N_2+N_4-N_3+k)\bullet N_4\,\circ\rangle,
\label{HWex}
\end{align}
which played an important role in computing the partition function of the super Chern-Simons matrix model in \cite{MNN}.

Note that, besides the Hanany-Witten effect, it is natural to assume that the brane configuration also enjoys a few rather trivial symmetries, similar to the charge conjugation or the parity in usual field theories.
If we reverse the $789$ directions we effectively change the signs of $k$ and find
\begin{align}
\langle N_1\bullet N_2\bullet N_3\circ N_4\,\circ\rangle
\simeq\langle N_1\circ N_2\circ N_3\bullet N_4\,\bullet\rangle,
\label{eq:ExchangeDual}
\end{align}
while if we reverse all of the spacetime directions, we obtain the relation
\begin{align}
\langle N_1\bullet N_2\bullet N_3\circ N_4\,\circ\rangle
\simeq\langle N_1\circ N_4\circ N_3\bullet N_2\,\bullet\rangle.
\label{eq:InverseDual}
\end{align}
The symmetries discussed in this subsection generate a large number of symmetries.
For example, by combining \eqref{eq:ExchangeDual} and \eqref{eq:InverseDual}, we immediately find
\begin{align}
\langle N_1\bullet N_2\bullet N_3\circ N_4\,\circ\rangle
\simeq\langle N_1\bullet N_4\bullet N_3\circ N_2\,\circ\rangle.
\label{1432}
\end{align}
Also, by exchanging two $(1,k)$5-branes with two NS5-branes in \eqref{eq:InverseDual} using the Hanany-Witten transition \eqref{LMN1} so that the order of the 5-branes is preserved, we find
\begin{align}
\langle N_1\bullet N_2\bullet N_3\circ N_4\,\circ\rangle
\simeq\langle N_1\bullet N_1+N_2-N_3+2k\bullet 2N_1-N_3+4k\circ N_1-N_3+N_4+2k\,\circ\rangle,
\label{eq:24Duality}
\end{align}
while by the trivial exchange of the Hanany-Witten transition \eqref{LMN2} we obtain
\begin{align}
\langle N_1\bullet N_2\bullet N_3\circ N_4\,\circ\rangle
\simeq\langle N_1\bullet N_1+N_3-N_2\bullet N_3\circ N_4\,\circ\rangle,\nonumber\\
\langle N_1\bullet N_2\bullet N_3\circ N_4\,\circ\rangle
\simeq\langle N_1\bullet N_2\bullet N_3\circ N_1+N_3-N_4\,\circ\rangle.
\label{13-4}
\end{align}

\subsection{Super Chern-Simons matrix models\label{subsec:MM}}

In this subsection we recapitulate matrix models associated to brane configurations discussed in the previous subsection and explain again how the concept of fixing a reference frame appears in the matrix models.
It was known that the worldvolume theory of the D3-branes for those supersymmetric brane configurations on a circle (which are dual to the M2-branes on supersymmetric backgrounds after taking T-duality and the M-theory lift) is described by the supersymmetric Chern-Simons theory of the $\widehat A$ quiver \cite{GW,ABJM,HLLLP2,ABJ} and the partition function on $S^3$ reduces to a matrix model using the localization technique \cite{P,KWY}.
Concretely, the partition function of the worldvolume theory of the D3-branes with $R$ perpendicular 5-branes reduces to a matrix model
\begin{align}
Z^{\{s_a\}_{a=1}^R}_{k}(\{N_a\}_{a=1}^R)=\prod_{a=1}^Re^{i\Theta_a}
\int\prod_{a=1}^R\frac{D^{N_a}\lambda_a}{N_a!(2\pi)^{N_a}}
\prod_{a=1}^R
Z(N_a,N_{a+1};\lambda_a,\lambda_{a+1}),
\label{eq:PFdef}
\end{align}
where each component is\footnote{The phase factor is a natural generalization from those of the ABJM matrix model.
The sign function is defined by $\sign(k_a)=(+1,0,-1)$ for $k_a=(+k,0,-k)$ respectively.}
\begin{align}
&e^{i\Theta_a}=i^{-\frac{1}{2}\sign(k_a)N_a^2},\quad
D^{N_a}\lambda_a=\prod_{l_a=1}^{N_a}D\lambda_{a,l_a},\quad
D\lambda_{a,l_a}=d\lambda_{a,l_a}\exp\biggl(\frac{ik_a}{4\pi}\lambda_{a,l_a}^2\biggr),\nonumber\\
&Z(N_a,N_{a+1};\lambda_a,\lambda_{a+1})
=\frac{\prod_{l_a<l'_{a}}^{N_a}2\sinh\frac{\lambda_{a,l_a}-\lambda_{a,l'_a}}{2}
\prod_{l_{a+1}<l'_{a+1}}^{N_{a+1}}2\sinh\frac{\lambda_{a+1,l_{a+1}}-\lambda_{a+1,l'_{a+1}}}{2}}
{\prod_{l_{a+1}=1}^{N_{a+1}}\prod_{l_a=1}^{N_a}2\cosh\frac{\lambda_{a+1,l_{a+1}}-\lambda_{a,l_a}}{2}},
\label{component}
\end{align}
The Chern-Simons level $k_a$ is determined by $(k>0)$
\begin{align}
k_a=\frac{k}{2}(s_a-s_{a-1}),
\end{align}
where the sign $s_a=\pm 1$ represents the type of the $a$-th 5-brane, with $s_a=+1$ and $s_a=-1$ being the NS5-brane and the $(1,k)$5-brane respectively and the two ends identified by $s_0=s_R$.
Thus, the sequence of the two types of 5-branes on a circle $S^1$ in the previous subsection is translated into the digit sequence of $\{s_a\}_{a=1}^{R}$ in the matrix model.
The argument $N_a$ of the partition function \eqref{eq:PFdef} originating from the number of D3-branes in each interval denotes the rank of the gauge group and we continue to call them ranks even in the matrix model.
After fixing a matrix model with a digit sequence $\{s_a\}_{a=1}^{R}$, we often omit displaying $\{s_a\}_{a=1}^{R}$ explicitly.
Note that at this stage the concept of the reference frame has not appeared.

In connecting the super Chern-Simons matrix models to spectral theories or topological string theories, it is important to move to the grand canonical ensemble, where we regard a rank of the group as the particle number and introduce a fugacity $z$ dual to it.
Although there are multiple ranks, as we have mentioned around \eqref{LMN1} and \eqref{LMN2} for the corresponding brane configuration, the overall number of D3-branes decouples from the other relative numbers in the Hanany-Witten transition and we naturally identify this overall rank as the particle number to be dualized.
Also, as noted in \cite{MNN}, for the correspondence to the topological string theory, we need to fix the power of the fugacity to be one of the ranks, which we identify as the reference frame.
Namely, we define the grand canonical partition function of the super Chern-Simons matrix model with the $n$-th rank being the reference as\footnote{The overall phase was not investigated in \cite{MNN}.
Hence, strictly speaking, to discuss the correspondence to topological strings, we need to take the absolute value for the partition function here and later for example in \eqref{Xi1} and \eqref{Xi2}.}
\begin{align}
\Xi_{k,{\bm M}}^{(n)}(z)
\simeq\sum_{N}^{\infty}z^{N+N'_n}Z_{k}(N+N'_1,N+N'_2,N+N'_3,\cdots).
\label{eq:GPFdef}
\end{align}
Here the summation is taken over the overall rank $N$ with relative ranks $(N'_1,N'_2,N'_3,\cdots)$ fixed.
We allow ambiguities in \eqref{eq:GPFdef} where $\simeq$ stands for a possible correction by an overall normalization factor independent of the fugacity $z$ and we do not specify explicitly the lower bound of summation.
From the discussions on the correspondence to the Fredholm determinant $\Det(1+z\widehat H^{-1})$ of a spectral operator $\widehat H^{-1}$ \cite{MP,GHM1,MST}, we vaguely believe that by adjusting these ambiguities we can define the grand canonical partition function\footnote{As noted in \cite{MNN} the overall normalization can be divergent and require a regularization.} so that it has the expansion $\Xi_{k,{\bm M}}^{(n)}(z)=1+{\cal O}(z)$.
Since we have moved to the grand canonical ensemble by dualizing the overall rank $N$, the grand canonical partition function is labeled only by the Chern-Simons level $k$ and the relative ranks which we have collectively denoted as $\bm M$ in \eqref{eq:GPFdef}.

When there are no rank deformations with all the relative ranks vanishing ${\bm M}={\bm 0}$, the reference is irrelevant
\begin{align}
\Xi_{k,{\bm 0}}(z)=\sum_{N=0}^\infty z^NZ_k(N,N,\cdots,N),
\end{align}
and it is especially simple to see that the grand canonical partition function reduces to the Fredholm determinant of a spectral operator.
Namely in \cite{MP,MN1} it was shown that when the digit sequence $\{s_a\}_{a=1}^R$ is given by
\begin{align}
\{s_{a}\}_{a=1}^R
=\{\underbrace{+1,\cdots,+1}_{p_1},\underbrace{-1,\cdots,-1}_{q_1},\underbrace{+1,\cdots,+1}_{p_2},\underbrace{-1,\cdots,-1}_{q_2},\cdots\}.
\end{align}
the grand canonical partition function is given by the Fredholm determinant
\begin{align}
\Xi_{k,{\bm 0}}(z)=\Det\bigl(1+z\widehat{H}^{-1}\bigr),
\label{eq:GPF_SD}
\end{align}
of a spectral operator
\begin{align}
\widehat{H}=\cdots\widehat{{\cal Q}}^{q_2}\widehat{{\cal P}}^{p_2}\widehat{{\cal Q}}^{q_1}\widehat{{\cal P}}^{p_1},\quad
\widehat{{\cal Q}}=2\cosh\frac{\widehat{q}}{2},\quad
\widehat{{\cal P}}=2\cosh\frac{\widehat{p}}{2}.
\label{eq:NDQCGen}
\end{align}
Here $\widehat q$ and $\widehat p$ are the canonical coordinate and momentum operators satisfying the commutation relation
\begin{align}
[\widehat q,\widehat p]=i\hbar,
\end{align}
with the identification $\hbar=2\pi k$.
The derivation of \eqref{eq:NDQCGen} was given in \cite{MN1} by a direct change of integration variables following previous computations in \cite{MP}.
Here we sketch the derivation slightly differently in the operator formalism in appendix \ref{derivation}.
Note that in \eqref{eq:NDQCGen} the sequence of the canonical operators $\widehat{\cal Q}$ and $\widehat{\cal P}$ appears in the reverse order from the sequence of 5-branes $\{s_a\}_{a=1}^R$.

For certain rank deformations, the spectral theory was generalized \cite{MM} by correcting the Fredholm determinant \eqref{eq:GPF_SD} with expectation values of the spectral operator while keeping the spectral operator \eqref{eq:NDQCGen} fixed.
This was named the open string formalism in \cite{PTEP} since the spectral operator seems to reflect the closed string background after expanding the Fredholm determinant with traces and the correction by expectation values is reminiscent of the idea of taking care of the deformations by adding open string fluctuations to a fixed closed string background.
Another generalization by correcting the spectral operator \eqref{eq:NDQCGen} while keeping the expression of the Fredholm determinant \eqref{eq:GPF_SD} fixed was also proposed in \cite{AHS,H,HO,MS2,MN5,KM}.
This was named the closed string formalism since now we try to take care of the deformations by changing the spectral operator for the closed string background.
The expression keeping the Fredholm determinant \eqref{eq:GPF_SD} seems more elegant which leads \cite{GHM1} to remove the role of the matrix models and propose a conjecture between spectral theories and topological strings.
We stress however that, from the viewpoint of matrix models, the open string formalism is more efficient and allows us to compute various rank deformations (and reveal some integrable structures \cite{HHMO,MMg,FM1,KuMo,FM2}).

For the brane configurations with two NS5-branes and two $(1,k)$5-branes without rank deformations, the spectral operators for the cases with $(p_1,q_1)=(2,2)$ and $(p_1,q_1,p_2,q_2)=(1,1,1,1)$ are respectively
\begin{align}
\widehat H_{(2,2)}=\widehat{\cal Q}^2\widehat{\cal P}^2,\quad
\widehat H_{(1,1,1,1)}=\widehat{\cal Q}\widehat{\cal P}\widehat{\cal Q}\widehat{\cal P}.
\label{Hnorank}
\end{align}
Note that, from the invariance of determinants \eqref{eq:GPF_SD} under similarity transformations, the expression of the spectral operator $\widehat H$ is subject to ambiguities.
Namely, by similarity transformations, we can alternatively present the operator for $(p_1,q_1)=(2,2)$ as
$\widehat{{\cal Q}}\widehat{{\cal P}}^2\widehat{{\cal Q}}$,
$\widehat{{\cal P}}^2\widehat{{\cal Q}}^2$ or
$\widehat{{\cal P}}\widehat{{\cal Q}}^2\widehat{{\cal P}}$
and the operator for $(p_1,q_1,p_2,q_2)=(1,1,1,1)$ as
$\widehat{{\cal P}}\widehat{{\cal Q}}\widehat{{\cal P}}\widehat{{\cal Q}}$.
Clearly, to fix the ambiguities in the expression of the spectral operator, we need to avoid the uncritical use of similarity transformations.
More importantly, since we have emphasized in the previous subsection that the Hanany-Witten transition for the sequence of 5-branes in the brane configuration on a circle is discussed unambiguously from the charge conservation \eqref{charge} only after we cut the circle open and that the sequence of 5-branes is translated into the sequence of the canonical operators $\widehat{\cal Q}$ and $\widehat{\cal P}$ reversely \eqref{eq:NDQCGen}, it is natural to expect that the reference frame should also be taken into account for the spectral operators.

By applying this formalism for the $(2,2)$ model and the $(1,1,1,1)$ model with the Fredholm determinant \eqref{eq:GPF_SD} and the spectral operator \eqref{Hnorank}, the exact values of the partition function were studied carefully in \cite{HM,MN3} and it was found that the result is given in terms of the free energy of the topological string theory as in the ABJM case \cite{HMMO}.
The background for topological strings was found to be local del Pezzo $D_5$ geometry.

The analysis was further generalized to rank deformations.
Let us parameterize the three relative ranks of U$(N_1)\times$U$(N_2)\times$U$(N_3)\times$U$(N_4)$ by $\bm{M}=(M_1,M_2,M_3)$ with the identification
\begin{align}
(N_1,N_2,N_3,N_4)=(N+M_2+M_3,N+M_1+2M_3,N+2M_1+M_2+M_3,N+M_1),
\label{eq:RankDefDef}
\end{align}
or inversely
\begin{align}
M_1=\frac{-N_1+N_3}{2},\quad
M_2=\frac{N_1-N_2+N_3-N_4}{2},\quad
M_3=\frac{N_2-N_4}{2}.
\label{MfromN'}
\end{align}
In other words, we consider the brane configuration
\begin{align}
\langle N+M_2+M_3\bullet
N+M_1+2M_3\bullet
N+2M_1+M_2+M_3\circ
N+M_1\,\circ\rangle,
\label{canonicaldeform}
\end{align}
and parametrize the brane configuration by the relative ranks as
\begin{align}
{\cal C}_\text{B}=\{(M_1,M_2,M_3)\}.
\label{CB}
\end{align}

For the special case of the $M_1$ and $M_2$ rank deformations, the correction was identified in \cite{MNN} following the open string formalism \cite{MM}.
Using this formalism and the same formalism applied to the brane configurations obtained from the Hanany-Witten transition \eqref{HWex}, it was possible to see that the description by the free energy of topological strings on local del Pezzo $D_5$ geometry is still valid in the rank deformations.
It was found \cite{MNN} that the total integral BPS indices \cite{HKP} are split to various combinations in different rank deformations.
In \cite{MNY} the integral BPS indices was further identified as representations of the $D_5$ algebra and the split was identified as the decomposition of the representations into a subgroup.

Before closing this subsection, we comment on the symmetries of the partition function \eqref{eq:PFdef} for $R=4$.
It is clear that the partition function is invariant under reversing the order of integrations in the partition function
\begin{align}
Z_k(N_1,N_2,N_3,N_4)=Z_{k}(N_1,N_4,N_3,N_2),
\end{align}
which corresponds to the symmetry in the brane configurations \eqref{1432} discussed at the end of the previous subsection if we identify the brane configuration $\langle N_1\bullet N_2\bullet N_3\circ N_4\,\circ\rangle$ with the partition function $Z_k(N_1,N_2,N_3,N_4)$.
Furthermore, in moving to the grand canonical ensemble in \eqref{eq:GPFdef}, we can translate the symmetries \eqref{1432}, \eqref{eq:24Duality} and \eqref{13-4} found for the brane configurations into
\begin{align}
\Xi_{k,(M_1,M_2,M_3)}(z)&=\Xi_{k,(M_1,-M_3,-M_2)}(z),\nonumber\\
\Xi_{k,(M_1,M_2,M_3)}(z)&=\Xi_{k,(M_1,M_2,-M_3)}(z),\nonumber\\
\Xi_{k,(M_1,M_2,M_3)}(z)&=\Xi_{k,(M_1,M_3,M_2)}(z),\nonumber\\
\Xi_{k,(M_1,M_2,M_3)}(z)&=\Xi_{k,(2k-M_1,M_2,M_3)}(z),
\label{symXi}
\end{align}
in terms of the relative ranks ${\bm M}$ defined in \eqref{eq:RankDefDef} when fixing the reference to be the first rank.

\subsection{Quantum curves}\label{qcurves}

In the previous two subsections, we have reviewed the brane configurations and the matrix models derived from the brane configurations.
In this subsection we recapitulate the analysis of the matrix models for our brane configuration with two NS5-branes and two $(1,k)$5-branes from a more general viewpoint of spectral theories.

In the previous subsection we have explained that the matrix model obtained from the brane configuration with two NS5-branes and two $(1,k)$5-branes is described by the free energy of topological strings on local del Pezzo $D_5$ geometry.
The appearance of local del Pezzo $D_5$ is natural from the fact that the spectral operators \eqref{Hnorank} falls into the family of the $D_5$ quantum curve consisting of nine terms,
\begin{align}
\widehat Q^\alpha\widehat P^\beta,\quad\alpha,\beta=-1,0,+1,
\label{nine}
\end{align}
if we introduce
\begin{align}
\widehat{\cal Q}=\widehat Q^{\frac{1}{2}}+\widehat Q^{-\frac{1}{2}},\quad
\widehat{\cal P}=\widehat P^{\frac{1}{2}}+\widehat P^{-\frac{1}{2}},\quad
\widehat Q=e^{\widehat q},\quad\widehat P=e^{\widehat p},
\label{QQPP}
\end{align}
and use the canonical commutation relation
\begin{align}
\widehat P^\beta\widehat Q^\alpha=e^{-i\hbar\alpha\beta}\widehat Q^\alpha\widehat P^\beta.
\label{QPrelation}
\end{align}

Since the matrix model, on one hand, corresponds to the Fredholm determinant of the $D_5$ spectral operator and, on the other hand, corresponds to the free energy of topological strings on local del Pezzo $D_5$ geometry, the correspondence was advertised as the ST/TS (Spectral-Theory/Topological-String) correspondence \cite{GHM1,MST} after removing the consideration of the matrix model.
After seeing that the spectral operators without rank deformations \eqref{Hnorank} fall into the $D_5$ curve and some rank deformations still correspond to the topological string theory on local del Pezzo $D_5$, it is natural to consider that the expression of the Fredholm determinant \eqref{eq:GPF_SD} is still valid when we introduce rank deformations for the matrix model.

\begin{figure}[t!]
\centering\includegraphics[scale=0.7,angle=-90]{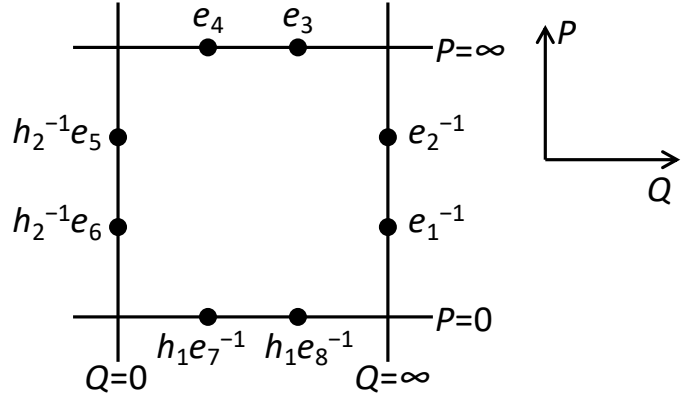}
\caption{Asymptotic values of the $D_5$ curve \eqref{eqClassicalCurve1}.
After applying the normal ordering, the asymptotic values are $\{e_1^{-1},e_2^{-1}\}$, $\{e_3,e_4\}$, $\{h_2^{-1}e_5,h_2^{-1}e_6\}$ and $\{h_1e_7^{-1},h_1e_8^{-1}\}$.
We omit the minus signs in displaying the asymptotic values for simplicity.}
\label{fig:asymptoticvalues}
\end{figure}

Following the progress \cite{NS,MiMo,ACDKV} and many others, in \cite{KMN} the framework to study the spectral operator was provided.
Namely, the $D_5$ quantum curve is defined as a linear combination of nine terms \eqref{nine}.
More explicitly, we parameterize the curve as
\begin{align}
&\widehat{H}/\alpha
=\widehat{Q}\widehat{P}+(e_3+e_4)\widehat{P}+e_3e_4\widehat{Q}^{-1}\widehat{P}\nonumber\\
&\quad+(e_1^{-1}+e_2^{-1})\widehat{Q}+E/\alpha+h_2^{-1}e_3e_4(e_5+e_6)\widehat{Q}^{-1}\nonumber\\
&\quad+(e_1e_2)^{-1}\widehat{Q}\widehat{P}^{-1}+h_1(e_1e_2)^{-1}(e_7^{-1}+e_8^{-1})\widehat{P}^{-1}
+h_1^2(e_1e_2e_7e_8)^{-1}\widehat{Q}^{-1}\widehat{P}^{-1},
\label{eqClassicalCurve1}
\end{align}
with the constraint
\begin{align}
(h_1h_2)^2=\prod_{i=1}^8e_i.
\label{eq:e7Fix}
\end{align}
Since the operators in different orders have to be distinguished, we adopt the normal ordering by taking $\widehat Q$ to the left and $\widehat P$ to the right.
The coefficients of the curve are parameterized so that the asymptotic values of its classical cousin in $Q\to\infty$, $P\to\infty$, $Q\to 0$ and $P\to 0$ can be expressed as $\{e_1^{-1},e_2^{-1}\}$, $\{e_3,e_4\}$, $\{h_2^{-1}e_5,h_2^{-1}e_6\}$ and $\{h_1e_7^{-1},h_1e_8^{-1}\}$ respectively (see figure \ref{fig:asymptoticvalues}).
We omit the minus signs in displaying the asymptotic values for simplicity.
These asymptotic values are called the point configuration in \cite{KNY} and determine the quantum curve (along with the two parameters $\alpha$ and $E$).
Since the spectral operator in the Fredholm determinant is invariant under similarity transformations, quantum curves were defined up to similarity transformations in \cite{KMN}.
Namely, an adjoint transformation of quantum curves by any operator $\widehat{G}$ is considered to be equivalent
\begin{align}
\widehat{G}\widehat{H}\widehat{G}^{-1}\sim\widehat{H}.
\label{eq:EqClass}
\end{align}

Totally the quantum curve $\widehat{H}$ is parametrized by twelve parameters $(h_1,h_2,e_1,\cdots,e_8,\alpha,E)$ with the constraint \eqref{eq:e7Fix}.
Since we have only nine terms in \eqref{eqClassicalCurve1}, two degrees of freedom are redundant.
Furthermore, if we choose $\widehat{G}=A^{\frac{i}{\hbar}\widehat{p}}$ or $\widehat{G}=B^{-\frac{i}{\hbar}\widehat{q}}$ in \eqref{eq:EqClass}, we find that quantum curves with $(\widehat{Q},\widehat{P})$ and $(A\widehat{Q},B\widehat{P})$ should be identified, which reduce two more parameters.
By using these four degrees of freedom, we can adopt the gauge fixing condition
\begin{align}
e_2=e_4=e_6=e_8=1,
\label{eq:GaugeFix}
\end{align}
with the constraint $(h_1h_2)^2=e_1e_3e_5e_7$ \eqref{eq:e7Fix} fixing the value of $e_7$.
Also, the parameter $E$ is irrelevant since similarity transformations \eqref{eq:EqClass} do not affect the value of it and we ignore $\alpha$ since this value does not affect the structure of symmetry \cite{KMN}.
After removing these parameters, quantum curves are characterized by five parameters forming a five-dimensional space of point configurations
\begin{align}
{\cal C}_\text{P}=\{(h_1,h_2,e_1,e_3,e_5)\}.
\label{CP}
\end{align}

\begin{figure}[t!]
\centering\includegraphics[scale=0.6,angle=-90]{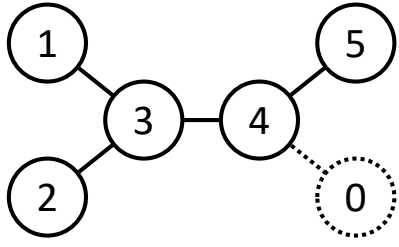}
\caption{Dynkin diagram of the $D_5$ algebra.
The number in circles corresponds to the subscript of the generators of the Weyl symmetry \eqref{eq:SimRootMap}.
Solid circles and lines denote the Dynkin diagram of the ordinary $D_5$ algebra, while the dashed one is for the lowest element $s_0$ which is generated by the other generators \eqref{s0}.}
\label{fig:D5Dynkin}
\end{figure}

The equivalence \eqref{eq:EqClass} further generates discrete symmetries in the point configuration \cite{KMN}.
As in the classical case \cite{KNY}, the discrete symmetries consist of
\begin{align}
s_1&:(\overline{h}_1,\overline{h}_2,e_1,e_3,e_5)
\mapsto\biggl(\frac{e_1e_3e_5}{\overline{h}_1\overline{h}_2^{2}},\overline{h}_2,
e_1,e_3,e_5\biggr),\nonumber\\
s_2&:(\overline{h}_1,\overline{h}_2,e_1,e_3,e_5)
\mapsto\biggl(\frac{\overline{h}_1}{e_3},\overline{h}_2,
e_1,\frac{1}{e_3},e_5\biggr),\nonumber\\
s_3&:(\overline{h}_1,\overline{h}_2,e_1,e_3,e_5)
\mapsto\biggl(\overline{h}_1,\frac{e_1e_5}{\overline{h}_1\overline{h}_2},
e_1,\frac{e_1e_3e_5}{\overline{h}_1\overline{h}_2^{2}},e_5\biggr),\nonumber\\
s_4&:(\overline{h}_1,\overline{h}_2,e_1,e_3,e_5)
\mapsto\biggl(\frac{\overline{h}_1\overline{h}_2}{e_1e_5},\overline{h}_2,
\frac{\overline{h}_2}{e_5},e_3,\frac{\overline{h}_2}{e_1}\biggr),\nonumber\\
s_5&:(\overline{h}_1,\overline{h}_2,e_1,e_3,e_5)
\mapsto\biggl(\overline{h}_1,\frac{\overline{h}_2}{e_1},
\frac{1}{e_1},e_3,e_5\biggr),
\label{eq:SimRootMap}
\end{align}
and generate the Weyl group of $D_5$, which is denoted as $W(D_5)$ (see figure \ref{fig:D5Dynkin} for labels of the simple roots).
Here we have introduced the shifted parameters
\begin{align}
(\overline{h}_1,\overline{h}_2,e_1,e_3,e_5)=(qh_1,q^{-1}h_2,e_1,e_3,e_5),
\label{eq:PointModuli}
\end{align}
with $q=e^{i\hbar}$.
It is also convenient to introduce the lowest element (corresponding to the affine element)
\begin{align}
s_0=s_4s_3s_2s_5s_4s_3s_1s_3s_4s_5s_2s_3s_4:
(\overline{h}_1,\overline{h}_2,e_1,e_3,e_5)
\mapsto\biggl(\overline{h}_1,\frac{\overline{h}_2}{e_5},e_1,e_3,\frac{1}{e_5}\biggr).
\label{s0}
\end{align}
The generators of the Weyl symmetry $s_1$, $s_2$, $s_5$, $s_0$ originate respectively from trivial exchanges of the asymptotic values $h_1e_7^{-1}\leftrightarrow h_1e_8^{-1}$, $e_3\leftrightarrow e_4$, $e_1^{-1}\leftrightarrow e_2^{-1}$, $h_2^{-1}e_5\leftrightarrow h_2^{-1}e_6$, while the generators $s_3$ and $s_4$ are more non-trivial.

In \cite{KMN}, it was further found that the spectral operators for the $(2,2)$ model and the $(1,1,1,1)$ model without rank deformations \eqref{Hnorank} are identified as
\begin{align}
(\overline{h}_1,\overline{h}_2,e_1,e_3,e_5)^{(2,2)}=(q,q^{-1},1,1,1),\quad
(\overline{h}_1,\overline{h}_2,e_1,e_3,e_5)^{(1,1,1,1)}
=(q,q^{-1},q^{-\frac{1}{2}},q^{\frac{1}{2}},q^{-\frac{1}{2}}),
\end{align}
respectively in the parameter space of point configurations ${\cal C}_\text{P}$ \eqref{CP} and respect the remaining symmetry $W(D_4)$ and $W(A_2\times(A_1)^2)$ (which is consistent with the split of the BPS indices in topological strings).
This immediately implies that cosets of broken symmetries map the parameters into those with the same unbroken symmetry where there are 10 point configurations for the $(2,2)$ model while 80 for the $(1,1,1,1)$ model.
It is bewildering that we have many equivalent point configurations if we try to identify the parameter space of brane configurations ${\cal C}_\text{B}$ \eqref{CB} in physics with that of point configurations ${\cal C}_\text{P}$ \eqref{CP} in geometry.
As we have stressed below \eqref{Hnorank}, the expression of spectral operators is obtained only after fixing a reference frame, while in quantum curves the $D_5$ Weyl symmetry is obtained from all of the similarity transformations \eqref{eq:EqClass} including those changing reference frames.
This suggests that, for the identification between brane configurations and point configurations to work, we are not supposed to apply similarity transformations uncritically and the concept of fixing a reference frame should also be taken into consideration in quantum curves as well.

\section{Brane configurations and quantum curves\label{sec:MM_QC}}

In the previous section, we have reviewed various aspects of the M2-brane physics, including brane configurations, matrix models and quantum curves.
In each aspect we find that we have often unconsciously taken the concept of fixing a reference frame for granted.
In this section, we explain that by fixing a reference in each aspect we can identify the three-dimensional parameter space of brane configurations ${\cal C}_\text{B}=\{(M_1,M_2,M_3)\}$ \eqref{CB} in the five-dimensional parameter space of point configurations ${\cal C}_\text{P}=\{(\overline{h}_1,\overline{h}_2,e_1,e_3,e_5)\}$ \eqref{CP} clearly.
After the identification we investigate the symmetry structure of the three-dimensional subspace, where we specify some as novel symmetries not known previously after identifying the known symmetries from brane configurations.

\subsection{From brane configurations to point configurations\label{subsec:BCPC}}

The main purpose of this section is to identify the three-dimensional parameter space of brane configurations ${\cal C}_\text{B}=\{(M_1,M_2,M_3)\}$ in the five-dimensional parameter space of point configurations ${\cal C}_\text{P}=\{(\overline{h}_1,\overline{h}_2,e_1,e_3,e_5)\}$.
The concept of fixing a reference frame plays an important role in the identification.

In section \ref{subsec:BC_duality} it has been emphasized that the Hanany-Witten transition for the brane configuration with a sequence of 5-branes is discussed unambiguously from the charge conservation \eqref{charge} only after we fix a reference interval where 5-branes do not move across.
In section \ref{subsec:MM}, it was found that the sequence of the canonical operators $\widehat{\cal Q}$, $\widehat{\cal P}$ in the spectral operator is determined from the sequence of 5-branes reversely \eqref{eq:NDQCGen}.
From these observations we are naturally led to introducing a reference frame for spectral operators as well.
After it, we can distinguish the $(2,2)$ model and the $(1,1,1,1)$ model without rank deformations from those obtained by transformations of the cosets in the parameter space of point configurations ${\cal C}_\text{P}$.

Since the sequence of 5-branes in brane configurations is directly translated to the sequence of the canonical operators in spectral operators, besides fixing a reference frame, we also distinguish all of the 5-branes in brane configurations and all of the canonical operators in spectral operators.
Concretely, on the side of brane configurations, we consider the sequence of two NS5-branes and two $(1,k)$5-branes from left to right as the standard order and label them by $2,1,3,4$ respectively as\footnote{The order of $2,1,3,4$ is related to the gauge fixing condition \eqref{eq:GaugeFix}, as will be clear later.}
\begin{align}
&\bigl\langle N_1\stackrel{2}{\bullet}N_2\stackrel{1}{\bullet}
N_3\stackrel{3}{\circ}N_4\stackrel{4}{\circ}\bigr\rangle\nonumber\\
&=\bigl\langle N+M_2+M_3\stackrel{2}{\bullet}
N+M_1+2M_3\stackrel{1}{\bullet}
N+2M_1+M_2+M_3\stackrel{3}{\circ}
N+M_1\stackrel{4}{\circ}\bigr\rangle,
\label{eq:CanonicalConfig}
\end{align}
by adding the information of labels to our notation of the brane configuration \eqref{eq:BraneConfig0}.

Correspondingly, on the side of spectral operators, we also label the canonical operators.
Namely, we do not only distinguish the spectral operators with different references as discussed below \eqref{Hnorank}, but also label the canonical operators by $4,3,1,2$ (reversely from brane configurations \eqref{eq:CanonicalConfig}) and consider
\begin{align}
\widehat{H}=\widehat{{\cal Q}}_4\widehat{{\cal Q}}_3\widehat{{\cal P}}_1\widehat{{\cal P}}_2,
\label{eq:CanonicalQC}
\end{align}
as the standard order.
For example, we consider two spectral operators $\widehat{\cal P}_2\widehat{\cal Q}_4\widehat{\cal Q}_3\widehat{\cal P}_1$ and $\widehat{\cal P}_1\widehat{\cal Q}_4\widehat{\cal Q}_3\widehat{\cal P}_2$ for the $(2,2)$ model to be different.

In this setup we can already find an interesting correspondence from the spectral operators without rank deformations as follows.
We pick up a spectral operator of the $(2,2)$ model or the $(1,1,1,1)$ model without rank deformations with the reference and the labels fixed.
If we find out the corresponding point configuration, we can plot the spectral operator in the five-dimensional parameter space ${\cal C}_\text{P}$.
On one hand, for point configurations of spectral operators, we can perform the normal ordering by moving $\widehat{\cal Q}$ to the left while moving $\widehat{\cal P}$ to the right using the commutation relations in \eqref{QPrelation} and read off the parameters in ${\cal C}_\text{P}$.
On the other hand, for the brane configurations without relative rank deformations, we can change the order of 5-branes into the standard one $2,1,3,4$ \eqref{eq:CanonicalConfig} using the Hanany-Witten transition and find out the relative rank deformations $(M_1,M_2,M_3)$ in the brane configuration ${\cal C}_\text{B}$.
By comparing these two computations, we can identify rank deformations $(M_1,M_2,M_3)$ in ${\cal C}_\text{P}$.
For this purpose, let us first regard each spectral operator with the reference fixed as the quantum curve \eqref{eqClassicalCurve1} and compute four pairs of the asymptotic values.
The result is given in table \ref{pointconfig}.
For example, asymptotic values of the spectral operator $\widehat{\cal P}\widehat{\cal Q}\widehat{\cal Q}\widehat{\cal P}$ are read off from the expansion
\begin{align}
\widehat{\cal P}\widehat{\cal Q}\widehat{\cal Q}\widehat{\cal P}/q^{-\frac{1}{2}}
=\bigl[\widehat Q+2q^{\frac{1}{2}}+q\widehat Q^{-1}\bigr]\widehat P+(q+1)\widehat Q+4q^{\frac{1}{2}}+(q+1)\widehat Q^{-1}+\bigl[q\widehat Q+2q^{\frac{1}{2}}+\widehat Q^{-1}\bigr]\widehat P^{-1}.
\end{align}

\begin{table}[t!]
\begin{centering}
\begin{tabular}{c|c|c|c|c}
type&$\{e_1^{-1},e_2^{-1}\}$&$\{e_3,e_4\}$&$\{h_2^{-1}e_5,h_2^{-1}e_6\}$&$\{h_1e_7^{-1},h_1e_8^{-1}\}$\\
\hline
\hline
$\widehat{{\cal Q}}\widehat{{\cal Q}}\widehat{{\cal P}}\widehat{{\cal P}}$
& $\{1,1\}$ & $\{1,1\}$ & $\{1,1\}$ & $\{1,1\}$\\
\hline 
$\widehat{{\cal Q}}\widehat{{\cal P}}\widehat{{\cal Q}}\widehat{{\cal P}}$
& $\{q^{\frac{1}{2}},1\}$ & $\{q^{\frac{1}{2}},1\}$ & $\{q^{-\frac{1}{2}},1\}$ & $\{q^{-\frac{1}{2}},1\}$\\
\hline 
$\widehat{{\cal Q}}\widehat{{\cal P}}\widehat{{\cal P}}\widehat{{\cal Q}}$
& $\{q^{\frac{1}{2}},q^{\frac{1}{2}}\}$ & $\{q,1\}$ & $\{q^{-\frac{1}{2}},q^{-\frac{1}{2}}\}$ & $\{q^{-1},1\}$\\
\hline 
$\widehat{{\cal P}}\widehat{{\cal Q}}\widehat{{\cal Q}}\widehat{{\cal P}}$
& $\{q,1\}$ & $\{q^{\frac{1}{2}},q^{\frac{1}{2}}\}$ & $\{q^{-1},1\}$ & $\{q^{-\frac{1}{2}},q^{-\frac{1}{2}}\}$\\
\hline 
$\widehat{{\cal P}}\widehat{{\cal Q}}\widehat{{\cal P}}\widehat{{\cal Q}}$
& $\{q,q^{\frac{1}{2}}\}$ & $\{q,q^{\frac{1}{2}}\}$ & $\{q^{-1},q^{-\frac{1}{2}}\}$ & $\{q^{-1},q^{-\frac{1}{2}}\}$\\
\hline 
$\widehat{{\cal P}}\widehat{{\cal P}}\widehat{{\cal Q}}\widehat{{\cal Q}}$
& $\{q,q\}$ & $\{q,q\}$ & $\{q^{-1},q^{-1}\}$ & $\{q^{-1},q^{-1}\}$\\
\end{tabular}
\par
\end{centering}
\caption{Asymptotic values of the spectral operators after taking the normal ordering.}
\label{pointconfig}
\end{table}

To identify the parameter space of brane configurations ${\cal C}_\text{B}=\{(M_1,M_2,M_3)\}$ in that of point configurations ${\cal C}_\text{P}=\{(\overline{h}_1,\overline{h}_2,e_1,e_3,e_5)\}$ from this setup, however, we need to clarify a few points.
Although in fixing a reference frame in spectral operators we avoid uncritical similarity transformations, to adopt the gauge-fixing condition \eqref{eq:GaugeFix} we still need to identify the quantum curve with $(\widehat Q,\widehat P)$ and that with $(A\widehat Q,B\widehat P)$.
Hence we distinguish the exponential linear operators $\widehat{G}=A^{\frac{i}{\hbar}\widehat{p}}$ and $\widehat{G}=B^{-\frac{i}{\hbar}\widehat{q}}$ as {\it small} similarity transformations from general similarity transformations with general $\widehat G$ and only allow the small similarity transformations.

Besides, although in table \ref{pointconfig} we have identified four pairs of asymptotic values, it is unclear how to distinguish between the pairs of $\{e_1^{-1},e_2^{-1}\}$, $\{e_3,e_4\}$, $\{h_2^{-1}e_5,h_2^{-1}e_6\}$ and $\{h_1e_7^{-1},h_1e_8^{-1}\}$.
Putting it more directly, although in the previous paragraph we allow the small similarity transformations for the gauge fixing \eqref{eq:GaugeFix}, when we fix the gauge $e_2=e_4=1$ by rescaling $(\widehat{Q},\widehat{P})\to(A\widehat{Q},B\widehat{P})$, a priori we do not know which in the pair of $\{e_1^{-1},e_2^{-1}\}$ or $\{e_3,e_4\}$ should be set to $1$.

Before directly answering this question, we first reduce the question by relating the asymptotic values.
In the normal ordering for the spectral operator, the asymptotic values $\{e_1^{-1},e_2^{-1}\}$, $\{e_3,e_4\}$, $\{h_2^{-1}e_5,h_2^{-1}e_6\}$ and $\{h_1e_7^{-1},h_1e_8^{-1}\}$ comes respectively by commuting $\widehat{{\cal P}}$ with all of $\widehat{Q}^{\frac{1}{2}}$ in the right, by commuting $\widehat{{\cal Q}}$ with all of $\widehat{P}^{\frac{1}{2}}$ in the left, by commuting $\widehat{{\cal P}}$ with all of $\widehat{Q}^{-\frac{1}{2}}$ in the right and by commuting $\widehat{{\cal Q}}$ with all of $\widehat{P}^{-\frac{1}{2}}$ in the left.
Since all of the operators $\widehat{Q}^{\pm\frac{1}{2}}$ and $\widehat{P}^{\pm\frac{1}{2}}$ come from $\widehat{\cal Q}$ and $\widehat{\cal P}$ \eqref{QQPP}, due to the commutation relation 
\begin{align}
\widehat{{\cal P}}\widehat{Q}^{\frac{n}{2}}
&=\widehat{Q}^{\frac{n}{2}}(q^{-\frac{n}{4}}\widehat{P}^{\frac{1}{2}}
+q^{\frac{n}{4}}\widehat{P}^{-\frac{1}{2}}),
&
\widehat{P}^{\frac{n}{2}}\widehat{{\cal Q}}
&=(q^{-\frac{n}{4}}\widehat{Q}^{\frac{1}{2}}
+q^{\frac{n}{4}}\widehat{Q}^{-\frac{1}{2}})\widehat{P}^{\frac{n}{2}},
\nonumber\\
\widehat{{\cal P}}\widehat{Q}^{-\frac{n}{2}}
&=\widehat{Q}^{-\frac{n}{2}}(q^{\frac{n}{4}}\widehat{P}^{\frac{1}{2}}
+q^{-\frac{n}{4}}\widehat{P}^{-\frac{1}{2}}),
&
\widehat{P}^{-\frac{n}{2}}\widehat{{\cal Q}}
&=(q^{\frac{n}{4}}\widehat{Q}^{\frac{1}{2}}
+q^{-\frac{n}{4}}\widehat{Q}^{-\frac{1}{2}})\widehat{P}^{-\frac{n}{2}},
\end{align}
it is clear that when one of the asymptotic values in $\{e_1^{-1},e_2^{-1}\}$ is $q^{\frac{n}{2}}$ one of the asymptotic values in $\{h_2^{-1}e_5,h_2^{-1}e_6\}$ has to be $q^{-\frac{n}{2}}$ and when one of the asymptotic values in $\{e_3,e_4\}$ is $q^{\frac{n}{2}}$, one of the asymptotic values in $\{h_1e_7^{-1},h_1e_8^{-1}\}$ has to be $q^{-\frac{n}{2}}$.
In this sense the asymptotic values of $\{e_1^{-1},e_2^{-1}\}$ and $\{h_2^{-1}e_5,h_2^{-1}e_6\}$ are correlated and the asymptotic values of $\{e_3,e_4\}$ and $\{h_1e_7^{-1},h_1e_8^{-1}\}$ are correlated as well.

Then, the above argument of correlating the asymptotic values indicates that we do not have $2^4=16$ choices in identifying each choice in $\{e_1^{-1},e_2^{-1}\}$, $\{e_3,e_4\}$, $\{h_2^{-1}e_5,h_2^{-1}e_6\}$, $\{h_1e_7^{-1},h_1e_8^{-1}\}$ separately.
Instead, we can combine the pairs of the reciprocal numbers $q^{\pm\frac{n}{2}}$ as $(e_4,h_1e_8^{-1})$, $(e_3,h_1e_7^{-1})$, $(e_1^{-1},h_2^{-1}e_5)$, $(e_2^{-1},h_2^{-1}e_6)$ so that there are only $2^2=4$ choices and two operators $\widehat{\cal Q}_{(4,8)}$, $\widehat{\cal Q}_{(3,7)}$ are responsible for the asymptotic values $(e_4,h_1e_8^{-1})$, $(e_3,h_1e_7^{-1})$ while two operators $\widehat{\cal P}_{(1,5)}$, $\widehat{\cal P}_{(2,6)}$ are responsible for $(e_1^{-1},h_2^{-1}e_5)$, $(e_2^{-1},h_2^{-1}e_6)$.
Since our gauge fixing condition \eqref{eq:GaugeFix} indicates that $e_2=e_4=1$, it is convenient to identify
\begin{align}
\widehat{{\cal Q}}_{(4,8)}=\widehat{{\cal Q}}_4,\quad
\widehat{{\cal Q}}_{(3,7)}=\widehat{{\cal Q}}_3,\quad
\widehat{{\cal P}}_{(1,5)}=\widehat{{\cal P}}_1,\quad
\widehat{{\cal P}}_{(2,6)}=\widehat{{\cal P}}_2.
\label{48371526}
\end{align}
since the standard ordering $4,3,1,2$, where $\widehat{{\cal Q}}_4$ is already located to the left of $\widehat{{\cal Q}}_3$ and $\widehat{{\cal P}}_2$ is to the right of $\widehat{{\cal P}}_1$, matches the gauge fixing condition which simplifies the values of $e_4$ and $e_2$.
This is why we have adopted $4,3,1,2$ as the standard order for spectral operators.

\begin{table}[t!]
\begin{centering}
\begin{tabular}{c|c|c|c}
Type&Quantum curve&$(\overline{h}_1,\overline{h}_2,e_1,e_3,e_5)$&$(h,e,f)$\\
\hline
\hline
$\widehat{{\cal Q}}\widehat{{\cal Q}}\widehat{{\cal P}}\widehat{{\cal P}}$
&$\widehat{{\cal Q}}_4\widehat{{\cal Q}}_3\widehat{{\cal P}}_1\widehat{{\cal P}}_2$
&$(q,q^{-1},1,1,1)$&$(q^{-1},1,1)$\\
\hline
$\widehat{{\cal Q}}\widehat{{\cal P}}\widehat{{\cal Q}}\widehat{{\cal P}}$
&$\widehat{{\cal Q}}_4\widehat{{\cal P}}_1\widehat{{\cal Q}}_3\widehat{{\cal P}}_2$
&$(q,q^{-1},q^{-\frac{1}{2}},q^{\frac{1}{2}},q^{-\frac{1}{2}})$&$(q^{-\frac{1}{2}},q^{\frac{1}{2}},1)$\\
\cline{2-4}
&$\widehat{{\cal Q}}_3\widehat{{\cal P}}_1\widehat{{\cal Q}}_4\widehat{{\cal P}}_2$
&$(1,q^{-1},q^{-\frac{1}{2}},q^{-\frac{1}{2}},q^{-\frac{1}{2}})$&$(q^{-\frac{1}{2}},1,q^{-\frac{1}{2}})$\\
\cline{2-4}
&$\widehat{{\cal Q}}_4\widehat{{\cal P}}_2\widehat{{\cal Q}}_3\widehat{{\cal P}}_1$
&$(q,1,q^{\frac{1}{2}},q^{\frac{1}{2}},q^{\frac{1}{2}})$&$(q^{-\frac{1}{2}},1,q^{\frac{1}{2}})$\\
\cline{2-4}
&$\widehat{{\cal Q}}_3\widehat{{\cal P}}_2\widehat{{\cal Q}}_4\widehat{{\cal P}}_1$
&$(1,1,q^{\frac{1}{2}},q^{-\frac{1}{2}},q^{\frac{1}{2}})$&$(q^{-\frac{1}{2}},q^{-\frac{1}{2}},1)$\\
\hline
$\widehat{{\cal Q}}\widehat{{\cal P}}\widehat{{\cal P}}\widehat{{\cal Q}}$
&$\widehat{{\cal Q}}_4\widehat{{\cal P}}_1\widehat{{\cal P}}_2\widehat{{\cal Q}}_3$
&$(q,1,1,q,1)$&$(1,q^{\frac{1}{2}},q^{\frac{1}{2}})$\\
\cline{2-4}
&$\widehat{{\cal Q}}_3\widehat{{\cal P}}_1\widehat{{\cal P}}_2\widehat{{\cal Q}}_4$
&$(q^{-1},1,1,q^{-1},1)$&$(1,q^{-\frac{1}{2}},q^{-\frac{1}{2}})$\\
\hline
$\widehat{{\cal P}}\widehat{{\cal Q}}\widehat{{\cal Q}}\widehat{{\cal P}}$
&$\widehat{{\cal P}}_1\widehat{{\cal Q}}_4\widehat{{\cal Q}}_3\widehat{{\cal P}}_2$
&$(1,q^{-1},q^{-1},1,q^{-1})$& $(1,q^{\frac{1}{2}},q^{-\frac{1}{2}})$\\
\cline{2-4}
&$\widehat{{\cal P}}_2\widehat{{\cal Q}}_4\widehat{{\cal Q}}_3\widehat{{\cal P}}_1$
&$(1,q,q,1,q)$&$(1,q^{-\frac{1}{2}},q^{\frac{1}{2}})$\\
\hline
$\widehat{{\cal P}}\widehat{{\cal Q}}\widehat{{\cal P}}\widehat{{\cal Q}}$
&$\widehat{{\cal P}}_1\widehat{{\cal Q}}_4\widehat{{\cal P}}_2\widehat{{\cal Q}}_3$
&$(1,1,q^{-\frac{1}{2}},q^{\frac{1}{2}},q^{-\frac{1}{2}})$&$(q^{\frac{1}{2}},q^{\frac{1}{2}},1)$\\
\cline{2-4}
&$\widehat{{\cal P}}_1\widehat{{\cal Q}}_3\widehat{{\cal P}}_2\widehat{{\cal Q}}_4$
&$(q^{-1},1,q^{-\frac{1}{2}},q^{-\frac{1}{2}},q^{-\frac{1}{2}})$&$(q^{\frac{1}{2}},1,q^{-\frac{1}{2}})$\\
\cline{2-4}
&$\widehat{{\cal P}}_2\widehat{{\cal Q}}_4\widehat{{\cal P}}_1\widehat{{\cal Q}}_3$
&$(1,q,q^{\frac{1}{2}},q^{\frac{1}{2}},q^{\frac{1}{2}})$&$(q^{\frac{1}{2}},1,q^{\frac{1}{2}})$\\
\cline{2-4}
&$\widehat{{\cal P}}_2\widehat{{\cal Q}}_3\widehat{{\cal P}}_1\widehat{{\cal Q}}_4$
&$(q^{-1},q,q^{\frac{1}{2}},q^{-\frac{1}{2}},q^{\frac{1}{2}})$&$(q^{\frac{1}{2}},q^{-\frac{1}{2}},1)$\\
\hline 
$\widehat{{\cal P}}\widehat{{\cal P}}\widehat{{\cal Q}}\widehat{{\cal Q}}$
&$\widehat{{\cal P}}_1\widehat{{\cal P}}_2\widehat{{\cal Q}}_4\widehat{{\cal Q}}_3$
&$(q^{-1},q,1,1,1)$&$(q,1,1)$\\
\end{tabular}
\par\end{centering}
\caption{Quantum curves with a reference frame and labels specified and the corresponding point configurations.}
\label{normalorder}
\end{table}

By now it is not difficult to identify the spectral operators with labels in the parameter space of point configurations ${\cal C}_\text{P}$.
For various orders, we can bring them to the standard order and identify the point configuration.
We list quantum curves and the corresponding point configurations in table \ref{normalorder}.
For example, for the quantum curve $\widehat{\cal P}_2\widehat{\cal Q}_4\widehat{\cal Q}_3\widehat{\cal P}_1$, from table \ref{pointconfig} the identification of the asymptotic values is unambiguous for $e_3=e_4=q^{\frac{1}{2}}$ and $h_1e_7^{-1}=h_1e_8^{-1}=q^{-\frac{1}{2}}$, while the rest should be identified as $e_1^{-1}=h_2^{-1}e_5=1$, $e_2^{-1}=q$ and $h_2^{-1}e_6=q^{-1}$ since we need to bring the leftmost $\widehat{{\cal P}}_2=\widehat{{\cal P}}_{(2,6)}$ responsible for $e_2^{-1}$ and $h_2^{-1}e_6$ to the rightmost.
After applying the small similarity transformation, we find $(h_1,h_2,e_1,e_3,e_5)=(q^{-1},q^2,q,1,q)$.

\begin{figure}[t!]
\centering\includegraphics[scale=0.6,angle=-90]{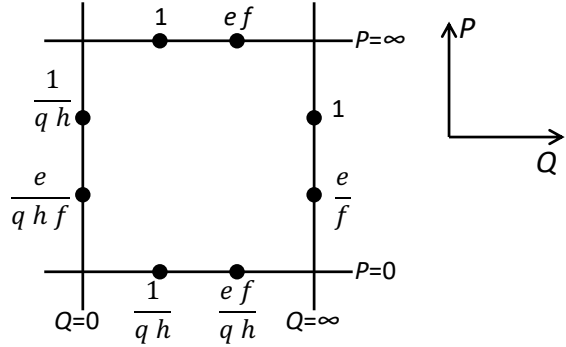}
\caption{Asymptotic values for parameters $(h,e,f)$ in the three-dimensional space of brane configurations ${\cal C}_\text{B}$.}
\label{asymptoticvaluesfixed}
\end{figure}

The first thing to note is that all of these $14$ points live in a three-dimensional subspace of the original five-dimensional parameter space ${\cal C}_\text{P}$.
This is an important sign indicating that we are performing the correct analysis by fixing a reference frame and labeling the canonical operators.
Let us parameterize the three-dimensional subspace by 
\begin{align}
(\overline{h}_1,\overline{h}_2,e_1,e_3,e_5)
=\biggl(\frac{ef}{h},\frac{hf}{e},\frac{f}{e},ef,\frac{f}{e}\biggr),
\label{hef}
\end{align}
and identify the $14$ points also by the parameters $(h,e,f)$ (see table \ref{normalorder}).
To summarize for now, we have successfully identified the three-dimensional parameter space of brane configurations ${\cal C}_\text{B}$ in the five-dimensional space of point configurations ${\cal C}_\text{P}$ by correctly taking the idea of fixing a reference frame and labels into consideration.
For later convenience, we depict again the asymptotic values for parameters in the three-dimensional subspace ${\cal C}_\text{B}$ in figure \ref{asymptoticvaluesfixed}.

Our remaining task is to identify each brane configuration corresponding to the spectral operator without relative rank deformations as a brane configuration in the standard order with rank deformations \eqref{eq:CanonicalQC} by applying the Hanany-Witten transition explicitly.
For example, by use of the Hanany-Witten transition, we can bring the brane configuration $1342$ into that in the standard order $2134$.
Since we are only interested in the relative rank difference, let us express the brane configuration $\bigl\langle N\stackrel{1}{\bullet}N\stackrel{3}{\circ}N\stackrel{4}{\circ}N\stackrel{2}{\bullet}\bigr\rangle$ simply as $\bigl\langle 0\stackrel{1}{\bullet}0\stackrel{3}{\circ}0\stackrel{4}{\circ}0\stackrel{2}{\bullet}\bigr\rangle$.
Then, by using \eqref{LMN1} and \eqref{LMN2} iteratively we find
\begin{align}
\bigl\langle 0\stackrel{1}{\bullet}0\stackrel{3}{\circ}0\stackrel{4}{\circ}0\stackrel{2}{\bullet}\bigr\rangle
\sim\bigl\langle 0\stackrel{1}{\bullet}0\stackrel{3}{\circ}0\stackrel{2}{\bullet}k\stackrel{4}{\circ}\bigr\rangle
\sim\bigl\langle 0\stackrel{1}{\bullet}0\stackrel{2}{\bullet}2k\stackrel{3}{\circ}k\stackrel{4}{\circ}\bigr\rangle
\sim\bigl\langle 0\stackrel{2}{\bullet}2k\stackrel{1}{\bullet}2k\stackrel{3}{\circ}k\stackrel{4}{\circ}\bigr\rangle.
\label{Nchange}
\end{align}
By comparing with \eqref{MfromN'}, it is clear that the corresponding relative rank deformation is $(M_1,M_2,M_3)=(k,-\frac{k}{2},\frac{k}{2})$.
Since the rank deformations are considered relatively from our standard order, correspondingly on the spectral operator side, we also define the relative parameter $\Delta(h,e,f)$ compared with that for the standard order $(h,e,f)|_{(4312)}=(q^{-1},1,1)$,
\begin{align}
\Delta(h,e,f)=\biggl(\frac{h}{h_{(4312)}},\frac{e}{e_{(4312)}},\frac{f}{f_{(4312)}}\biggr)=(qh,e,f).
\end{align}
In table \ref{relative22} we list the relative parameters of spectral operators $\Delta(h,e,f)$ and the parameters of rank deformations $(M_1,M_2,M_3)$ for all of the spectral operators considered in table \ref{normalorder}.
Then, it is not difficult to observe a clear identification 
\begin{align}
\Delta(h,e,f)=(e^{2\pi iM_1},e^{2\pi iM_2},e^{2\pi iM_3}),
\label{M123}
\end{align}
which shows that the three-dimensional subspace is nothing but that of the three relative rank deformations in the $(2,2)$ model.

\begin{table}[t!]
\begin{centering}
\begin{tabular}{c|c||c|c}
Spectral operator&$\Delta(h,e,f)$&Brane configuration&$\left(M_1,M_2,M_3\right)$\\
\hline\hline
$\widehat{{\cal Q}}_4\widehat{{\cal Q}}_3\widehat{{\cal P}}_1\widehat{{\cal P}}_2$&$(1,1,1)$&
$\langle 0\stackrel{2}{\bullet}0\stackrel{1}{\bullet}0\stackrel{3}{\circ}0\stackrel{4}{\circ}\rangle$&$(0,0,0)$\\
\hline
$\widehat{{\cal Q}}_4\widehat{{\cal P}}_1\widehat{{\cal Q}}_3\widehat{{\cal P}}_2$&$(q^{\frac{1}{2}},q^{\frac{1}{2}},1)$&
$\langle 0\stackrel{2}{\bullet}0\stackrel{3}{\circ}0\stackrel{1}{\bullet}0\stackrel{4}{\circ}\rangle$&$(\frac{k}{2},\frac{k}{2},0)$\\ 
\hline
$\widehat{{\cal Q}}_3\widehat{{\cal P}}_1\widehat{{\cal Q}}_4\widehat{{\cal P}}_2$&$(q^{\frac{1}{2}},1,q^{-\frac{1}{2}})$&
$\langle 0\stackrel{2}{\bullet}0\stackrel{4}{\circ}0\stackrel{1}{\bullet}0\stackrel{3}{\circ}\rangle$&$(\frac{k}{2},0,-\frac{k}{2})$\\
\hline
$\widehat{{\cal Q}}_4\widehat{{\cal P}}_2\widehat{{\cal Q}}_3\widehat{{\cal P}}_1$&$(q^{\frac{1}{2}},1,q^{\frac{1}{2}})$&
$\langle 0\stackrel{1}{\bullet}0\stackrel{3}{\circ}0\stackrel{2}{\bullet}0\stackrel{4}{\circ}\rangle$&$(\frac{k}{2},0,\frac{k}{2})$\\
\hline
$\widehat{{\cal Q}}_3\widehat{{\cal P}}_2\widehat{{\cal Q}}_4\widehat{{\cal P}}_1$&$(q^{\frac{1}{2}},q^{-\frac{1}{2}},1)$&
$\langle 0\stackrel{1}{\bullet}0\stackrel{4}{\circ}0\stackrel{2}{\bullet}0\stackrel{3}{\circ}\rangle$&$(\frac{k}{2},-\frac{k}{2},0)$\\
\hline
$\widehat{{\cal Q}}_4\widehat{{\cal P}}_1\widehat{{\cal P}}_2\widehat{{\cal Q}}_3$&$(q,q^{\frac{1}{2}},q^{\frac{1}{2}})$&
$\langle 0\stackrel{3}{\circ}0\stackrel{2}{\bullet}0\stackrel{1}{\bullet}0\stackrel{4}{\circ}\rangle$&$(k,\frac{k}{2},\frac{k}{2})$\\
\hline
$\widehat{{\cal Q}}_3\widehat{{\cal P}}_1\widehat{{\cal P}}_2\widehat{{\cal Q}}_4$&$(q,q^{-\frac{1}{2}},q^{-\frac{1}{2}})$&
$\langle 0\stackrel{4}{\circ}0\stackrel{2}{\bullet}0\stackrel{1}{\bullet}0\stackrel{3}{\circ}\rangle$&$(k,-\frac{k}{2},-\frac{k}{2})$\\
\hline
$\widehat{{\cal P}}_1\widehat{{\cal Q}}_4\widehat{{\cal Q}}_3\widehat{{\cal P}}_2$&$(q,q^{\frac{1}{2}},q^{-\frac{1}{2}})$&
$\langle 0\stackrel{2}{\bullet}0\stackrel{3}{\circ}0\stackrel{4}{\circ}0\stackrel{1}{\bullet}\rangle$&$(k,\frac{k}{2},-\frac{k}{2})$\\
\hline
$\widehat{{\cal P}}_2\widehat{{\cal Q}}_4\widehat{{\cal Q}}_3\widehat{{\cal P}}_1$&$(q,q^{-\frac{1}{2}},q^{\frac{1}{2}})$&
$\langle 0\stackrel{1}{\bullet}0\stackrel{3}{\circ}0\stackrel{4}{\circ}0\stackrel{2}{\bullet}\rangle$&$(k,-\frac{k}{2},\frac{k}{2})$\\
\hline
$\widehat{{\cal P}}_1\widehat{{\cal Q}}_4\widehat{{\cal P}}_2\widehat{{\cal Q}}_3$&$(q^{\frac{3}{2}},q^{\frac{1}{2}},1)$&
$\langle 0\stackrel{3}{\circ}0\stackrel{2}{\bullet}0\stackrel{4}{\circ}0\stackrel{1}{\bullet}\rangle$&$(\frac{3k}{2},\frac{k}{2},0)$\\
\hline
$\widehat{{\cal P}}_1\widehat{{\cal Q}}_3\widehat{{\cal P}}_2\widehat{{\cal Q}}_4$&$(q^{\frac{3}{2}},1,q^{-\frac{1}{2}})$&
$\langle 0\stackrel{4}{\circ}0\stackrel{2}{\bullet}0\stackrel{3}{\circ}0\stackrel{1}{\bullet}\rangle$&$(\frac{3k}{2},0,-\frac{k}{2})$\\
\hline
$\widehat{{\cal P}}_2\widehat{{\cal Q}}_4\widehat{{\cal P}}_1\widehat{{\cal Q}}_3$&$(q^{\frac{3}{2}},1,q^{\frac{1}{2}})$&
$\langle 0\stackrel{3}{\circ}0\stackrel{1}{\bullet}0\stackrel{4}{\circ}0\stackrel{2}{\bullet}\rangle$&$(\frac{3k}{2},0,\frac{k}{2})$\\
\hline
$\widehat{{\cal P}}_2\widehat{{\cal Q}}_3\widehat{{\cal P}}_1\widehat{{\cal Q}}_4$&$(q^{\frac{3}{2}},q^{-\frac{1}{2}},1)$&
$\langle 0\stackrel{4}{\circ}0\stackrel{1}{\bullet}0\stackrel{3}{\circ}0\stackrel{2}{\bullet}\rangle$&$(\frac{3k}{2},-\frac{k}{2},0)$\\
\hline
$\widehat{{\cal P}}_1\widehat{{\cal P}}_2\widehat{{\cal Q}}_4\widehat{{\cal Q}}_3$&$(q^{2},1,1)$&
$\langle 0\stackrel{3}{\circ}0\stackrel{4}{\circ}0\stackrel{2}{\bullet}0\stackrel{1}{\bullet}\rangle$&$(2k,0,0)$
\end{tabular}
\par\end{centering}
\caption{List of the quantum curves with a reference frame and labels specified in a general order and the corresponding brane configurations.
After changing into the standard order by commutation relation \eqref{QPrelation} for the quantum curves and by the Hanany-Witten transition \eqref{LMN1} and \eqref{LMN2}, we find a clear correspondence \eqref{M123} between $(M_1,M_2,M_3)$ and $\Delta(h,e,f)$.}
\label{relative22} 
\end{table}

To summarize, in this subsection, using the idea of fixing a reference frame and labeling the 5-branes, we have identified the three-dimensional subspace of the three relative rank deformations of brane configurations ${\cal C}_\text{B}$ in the five-dimensional parameter space of point configurations ${\cal C}_\text{P}$.
The parameter of the $D_5$ quantum curve $(h,e,f)$ is given by
\begin{align}
(h,e,f)=(e^{2\pi i(M_1-k)},e^{2\pi iM_2},e^{2\pi iM_3}),
\label{hef2134}
\end{align}
in terms of the relative rank deformation ${\bm M}=(M_1,M_2,M_3)$ for the standard order of 5-branes $2134$ \eqref{eq:CanonicalConfig}.
However, so far we have not discussed the matrix model itself.
In the following sections, by explicitly analyzing the matrix models, we have more checks and more discussions on the relation \eqref{hef2134} from various viewpoints such as the correspondence between matrix models and spectral theories or between matrix models and topological strings.
Before going there, in the remaining part of this section, we discuss the effects of changing frames and the symmetry structure.

\subsection{Change of frames\label{change}}

\begin{figure}[t!]
\centering\includegraphics[scale=0.6,angle=-90]{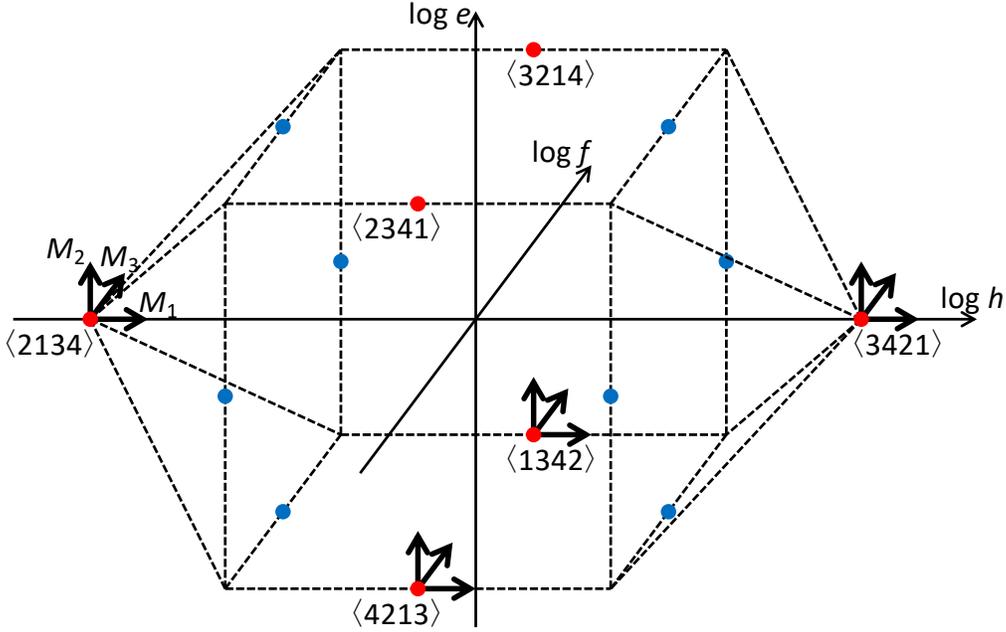}
\caption{The three-dimensional subspace of brane configurations ${\cal C}_\text{B}$ in the five-dimensional space of point configurations ${\cal C}_\text{P}$.
Red dots and blue dots denote the $(2,2)$ model and the $(1,1,1,1)$ model without rank deformations respectively.
Although the origins of the $(2,2)$ model without rank deformations are shifted in changing the frames, the directions of rank deformations $(M_1,M_2,M_3)$ are unchanged.}
\label{fig:3dStructure}
\end{figure}

So far we have stressed the importance of fixing a reference frame.
It is interesting to investigate the situation when frames are changed.
For this purpose we consider our labels of rank deformations
$\bigl\langle N+M_2+M_3\stackrel{2}{\bullet}
N+M_1+2M_3\stackrel{1}{\bullet}
N+2M_1+M_2+M_3\stackrel{3}{\circ}
N+M_1\stackrel{4}{\circ}\bigr\rangle$ in \eqref{eq:CanonicalConfig} and change the reference frame to the second one.
Namely we cyclically move to the brane configuration into
$\bigl\langle N+M_1+2M_3\stackrel{1}{\bullet}
N+2M_1+M_2+M_3\stackrel{3}{\circ}
N+M_1\stackrel{4}{\circ}
N+M_2+M_3\stackrel{2}{\bullet}\bigr\rangle$
and rearrange the 5-branes into our standard order $2134$ with the reference frame fixed.
As in \eqref{Nchange} since we are only interested in the relative rank difference we can simply consider
$\bigl\langle M_1+2M_3\stackrel{1}{\bullet}
2M_1+M_2+M_3\stackrel{3}{\circ}
M_1\stackrel{4}{\circ}
M_2+M_3\stackrel{2}{\bullet}\bigr\rangle$
and change the 5-brane with number 2 into the first one by using the Hanany-Witten transition as in \eqref{Nchange}.
Then, we find
\begin{align}
&\bigl\langle M_1+2M_3\stackrel{1}{\bullet}
2M_1+M_2+M_3\stackrel{3}{\circ}
M_1\stackrel{4}{\circ}
M_2+M_3\stackrel{2}{\bullet}\bigr\rangle\nonumber\\
&\quad\sim\bigl\langle M_1+2M_3\stackrel{2}{\bullet}
2M_1-M_2+3M_3+2k\stackrel{1}{\bullet}
3M_1+2M_3+2k\stackrel{3}{\circ}
2M_1-M_2+M_3+k\stackrel{4}{\circ}
\bigr\rangle.
\end{align}
With the parameterization we find that the relative rank difference given by \eqref{MfromN'} changes into
\begin{align}
(M'_1,M'_2,M'_3)=\biggl(M_1+k,M_2-\frac{k}{2},M_3+\frac{k}{2}\biggr).
\label{M123change}
\end{align}
Our result is consistent with the case without deformations \eqref{Nchange} by setting $M_1=M_2=M_3=0$.
Surprisingly, the shift by $(k,-\frac{k}{2},\frac{k}{2})$ is the only change for $(M_1,M_2,M_3)$ and the direction of $(M_1,M_2,M_3)$ is exactly the same as the original one.
In other words, after fixing a reference and labeling the 5-branes in brane configurations as fixing a reference frame in mechanics or taking a local coordinate in geometry, we are also able to change frames or change local charts.
In our case the transition map is rather trivial and we have only to shift the origins.
See figure \ref{fig:3dStructure} for locating various brane configurations and identifying directions of rank deformations.

Note however that, although the direction is the same, since the origin is different, the unbroken symmetry is in general different.
This is why it is important to fix a reference frame in our analysis.
In the next subsection we study the symmetry in this space carefully.

It is interesting to point out that the parameter space of point configurations for the Painlev\'e system enjoys the affine Weyl group which contains a shift generator \cite{KNY}.
Although it was observed that the matrix models are related to the $q$-Painlev\'e system \cite{BGT}, quantum curves defined by identifying those obtained by similarity transformations only enjoys the Weyl group without the affine element.
We unexpectedly encounter a shift generator in the change of frames.
It would be interesting to clarify the relation to the affine Weyl group.

\subsection{Weyl symmetries\label{unbrokensym}}

In the previous subsection we have identified the three-dimensional parameter space of brane configurations ${\cal C}_\text{B}$ in the five-dimensional parameter space of point configurations ${\cal C}_\text{P}$.
After the identification, let us proceed to the study of the symmetry of the subspace $(h,e,f)$ \eqref{hef}.

\begin{figure}[t!]
\centering\includegraphics[scale=0.5,angle=-90]{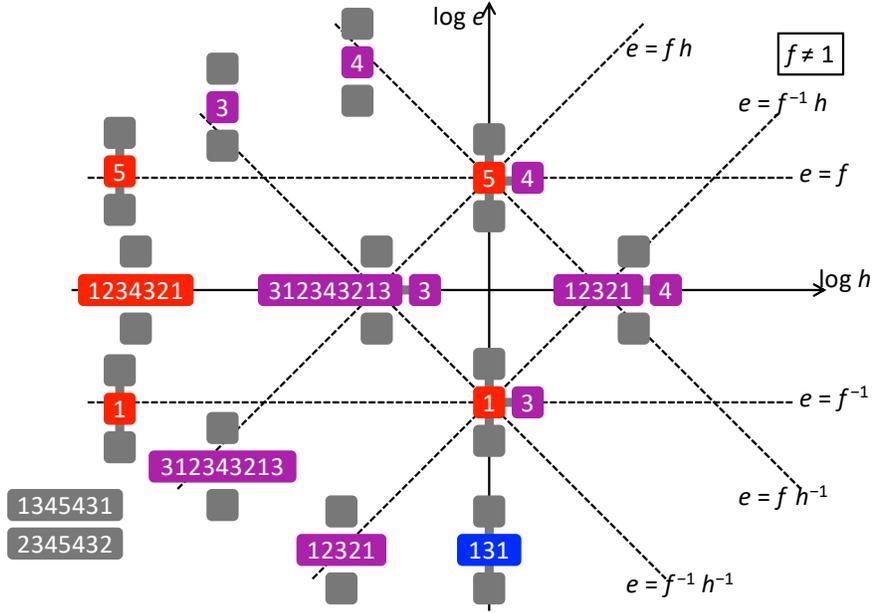}
\caption{Unbroken symmetries at each point in the three-dimensional space of brane configurations ${\cal C}_\text{B}$.
Every point enjoys the Weyl symmetry of $W((A_1)^2)$ generated by $s_1s_3s_4s_5s_4s_3s_1$ and $s_2s_3s_4s_5s_4s_3s_2$.
Along the axes or the dotted lines, an additional generator appears where the symmetry can be enhanced trivially to $W((A_1)^3)$ or drastically to $W(A_3)$.
At the points where only two dotted lines cross, the symmetry is enhanced trivially to $W(A_3\times A_1)$, which we do not depict explicitly, though the symmetry is enhanced drastically to $W(D_4)$ or $W(A_2\times(A_1)^2)$ if more than two lines cross.
}
\label{fig:f1}
\end{figure}

First, we can ask what subgroup of symmetries in $W(D_5)$ generated by \eqref{eq:SimRootMap} leaves each point invariant.
It is not difficult to find that a general point in this subspace satisfies the symmetry of $W((A_1)^2)$ generated by $s_1s_3s_4s_5s_4s_3s_1$ and $s_2s_3s_4s_5s_4s_3s_2$, while the symmetry is enhanced for special points.
In \cite{KMN} the symmetry in the two-dimensional subspace without the $M_3$ deformation or the $f$ deformation was already studied.
Here we present the study for $f\ne 1$ in figure \ref{fig:f1}.
It is interesting to find that, although the same $(2,2)$ models or the same $(1,1,1,1)$ models without rank deformations enjoy the same symmetries in different points in the three-dimensional subspace, after deforming with relative ranks, the symmetry is not the same any more.
This indicates the importance in fixing a reference frame.

\begin{figure}[t!]
\centering\includegraphics[scale=0.7,angle=-90]{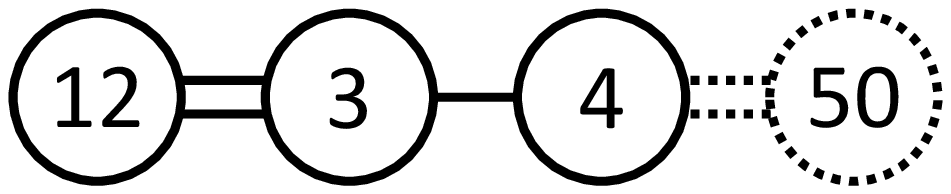}
\caption{Dynkin diagram of the $B_3=\text{so}(7)$ algebra.
The number in circles corresponds to the generator of the Weyl symmetry or its product.
If we restrict ourselves to ${\cal C}_\text{B}$, $s_5s_0$ can be generated by $s_1s_2$, $s_3$ and $s_4$ as in \eqref{eq:3dWeylHW}.}
\label{fig:B3Dynkin}
\end{figure}

Second, alternatively we can ask which subgroup in $W(D_5)$ preserves the three-dimensional subspace $(h,e,f)$ as a whole.
We find that the subgroup is $W(B_3)$, which is generated by $s_1s_2$, $s_3$ and $s_4$ and has 48 elements.
See figure \ref{fig:B3Dynkin} for the Dynkin diagram of this group.
The reason why this group preserve the subspace is by now quite apparent from the correlations between $(e_1^{-1},h_2^{-1}e_5)$, $(e_2^{-1},h_2^{-1}e_6)$, $(e_3,h_1e_7^{-1})$ and $(e_4,h_1e_8^{-1})$ as above \eqref{48371526}.
Namely the switch between $h_1e_7^{-1}$ and $h_1e_8^{-1}$ generated by $s_1$ should be accompanied by the switch between $e_3$ and $e_4$ generated by $s_2$, while the switch between $e_1^{-1}$ and $e_2^{-1}$ generated by $s_5$ should be accompanied by the switch between $h_2^{-1}e_5$ and $h_2^{-1}e_6$ generated by $s_0$.
The concrete form of these maps on the three-dimensional subspace is given by
\begin{align}
s_1s_2&:(h,e,f)\mapsto\biggl(h,\frac{1}{f},\frac{1}{e}\biggr),\nonumber\\
s_3&:(h,e,f)\mapsto\biggl(\frac{1}{ef},\sqrt{\frac{e}{hf}},\sqrt{\frac{f}{he}}\biggr),\nonumber\\
s_4&:(h,e,f)\mapsto\biggl(\frac{f}{e},\sqrt{\frac{ef}{h}},\sqrt{hef}\biggr),\nonumber\\
s_5s_0&:(h,e,f)\mapsto(h,f,e).
\label{eq:3dimMap}
\end{align}

\subsection{Hanany-Witten transition\label{subsec:HW_Weyl}}

After identifying the three-dimensional space of brane configurations ${\cal C}_\text{B}$ in the five-dimensional space of point configurations ${\cal C}_\text{P}$ as in \eqref{hef} and \eqref{M123} and studying its symmetry structure, we can compare the $B_3$ Weyl group \eqref{eq:3dimMap} with the symmetries of the $(2,2)$ model \eqref{symXi} generated by the Hanany-Witten transition and a few discrete symmetries.

\begin{figure}[t!]
\centering\includegraphics[scale=0.7,angle=-90]{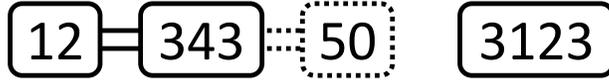}
\caption{Group-theoretical structure of the symmetries generated by the Hanany-Witten transition and a few trivial symmetries such as the charge conjugation and the parity.}
\label{fig:WeylandHW}
\end{figure}

We find that, corresponding to all of the dualities from \eqref{symXi}
\begin{align}
(M_1,M_2,M_3)&\sim(M_1,-M_3,-M_2),\nonumber\\
(M_1,M_2,M_3)&\sim(M_1,M_2,-M_3),\nonumber\\
(M_1,M_2,M_3)&\sim(M_1,M_3,M_2),\nonumber\\
(M_1,M_2,M_3)&\sim(2k-M_1,M_2,M_3),
\end{align}
there are elements of $W(B_3)$
\begin{align}
s_1s_2&:\Delta(h,e,f)\mapsto\Delta\biggl(h,\frac{1}{f},\frac{1}{e}\biggr),\nonumber\\
s_3s_4s_3&:\Delta(h,e,f)\mapsto\Delta\biggl(h,e,\frac{1}{f}\biggr),\nonumber\\
s_5s_0=s_3s_4s_3s_1s_2s_3s_4s_3&:\Delta(h,e,f)\mapsto\Delta(h,f,e),\nonumber\\
s_3s_1s_2s_3=s_4s_5s_0s_4&:\Delta(h,e,f)\mapsto\Delta\biggl(\frac{q^{2}}{h},e,f\biggr).
\label{eq:3dWeylHW}
\end{align}
Note that the equalities $s_5s_0=s_3s_4s_3s_1s_2s_3s_4s_3$ and $s_3s_1s_2s_3=s_4s_5s_0s_4$ hold only in this three-dimensional subspace ${\cal C}_\text{B}$.
We find that these elements generate a group isomorphic to $W(B_2\times A_1)$ of order 16 whose Dynkin diagram is depicted in figure \ref{fig:WeylandHW}.

So far we have identified the well-known symmetries of the $(2,2)$ model, such as the Hanany-Witten transition and the discrete symmetries, as $W(B_2\times A_1)$ in $W(B_3)$.
This indicates that there are novel symmetries or dualities for the brane configurations unknown from them.
A representative of these elements is $s_3$ or $s_4$.
To make contact with future studies from brane physics, let us express them in terms of brane configurations as
\begin{align}
s_3&:\langle N_1\stackrel{2}{\bullet}N_2\stackrel{1}{\bullet}N_3\stackrel{3}{\circ}N_4\stackrel{4}{\circ}\rangle\mapsto
\langle N_1\stackrel{2}{\bullet}N_2-N_3+N_4+k\stackrel{1}{\bullet}-N_3+2N_4+2k\stackrel{3}{\circ}N_4\stackrel{4}{\circ}\rangle,\nonumber\\
s_4&:\langle N_1\stackrel{2}{\bullet}N_2\stackrel{1}{\bullet}N_3\stackrel{3}{\circ}N_4\stackrel{4}{\circ}\rangle\mapsto
\langle N_1\stackrel{2}{\bullet}N_2\stackrel{1}{\bullet}2N_2-N_3+2k\stackrel{3}{\circ}N_2-N_3+N_4+k\stackrel{4}{\circ}\rangle.
\end{align}
Here we have rewritten the transformations $s_3$ and $s_4$ in \eqref{eq:3dimMap} in terms of the relative rank difference $(M_1,M_2,M_3)$ using the identification \eqref{hef2134} and expressed the results by fixing the reference rank $N_1$.
We can further rewrite them into a significant form by exchanging the NS5-brane $\stackrel{1}{\bullet}$ and the $(1,k)$5-brane $\stackrel{3}{\circ}$ with the Hanany-Witten transition.
Namely, by introducing $N'_3=N_2-N_3+N_4+k$, the transformations are given by
\begin{align}
s_3&:\langle N_1\stackrel{2}{\bullet}N_2\stackrel{3}{\circ}N'_3\stackrel{1}{\bullet}N_4\stackrel{4}{\circ}\rangle\mapsto
\langle N_1\stackrel{2}{\bullet}N'_3\stackrel{3}{\circ}N_2\stackrel{1}{\bullet}N_4\stackrel{4}{\circ}\rangle,\nonumber\\
s_4&:\langle N_1\stackrel{2}{\bullet}N_2\stackrel{3}{\circ}N'_3\stackrel{1}{\bullet}N_4\stackrel{4}{\circ}\rangle\mapsto
\langle N_1\stackrel{2}{\bullet}N_2\stackrel{3}{\circ}N_4\stackrel{1}{\bullet}N'_3\stackrel{4}{\circ}\rangle.
\label{s3s4N}
\end{align}
We have not been aware of any simple explanations for these transformations.

\section{Matrix models and spectral theories\label{sec:MM_ST}}

In the previous section, with the idea of fixing a reference frame and labeling the 5-branes, we have proposed to identify the three-dimensional space of brane configurations ${\cal C}_\text{B}$ in the five-dimensional space of point configurations ${\cal C}_\text{P}$.
However, so far we have only observed the similarity in the algebraic structure between the Hanany-Witten transition in brane configurations \eqref{LMN1} and the canonical commutation relation in point configurations \eqref{QPrelation} for several cases without rank deformations.
It is desirable to present a more qualitative comparison of matrix models to spectral theories or topological strings.
In this section and the next section, we establish the relation to spectral theories and topological strings respectively and present some non-trivial checks for our proposals.

\subsection{Correspondence}

As explained in section \ref{subsec:MM}, for the case without relative rank deformations, the relation between matrix models and spectral theories was given in \eqref{eq:GPF_SD} and \eqref{eq:NDQCGen} where the 5-branes and the spectral operators are aligned in the reverse order.
Generalizations in the open string formalism enabled us to compute the $(M_1,M_2)$ rank deformations in the $(2,2)$ model efficiently \cite{MNN} and find that the results are described by the free energy of topological strings following the discovery in the ABJM matrix model \cite{HMMO}.
However, the corresponding expression of spectral operators is not clear in this analysis.

Prior to the analysis, by removing the role of matrix models, in \cite{GHM1} the relation between spectral theories and topological strings (the ST/TS correspondence) was proposed.
Without referring to matrix models, the proposal states that the Fredholm determinant of spectral operators is described by the free energy of topological strings on a background associated to the spectral operators.
Since the spectral operator varies in the ST/TS correspondence while keeping the expression of the Fredholm determinant fixed, the idea of the ST/TS correspondence is related more directly to the closed string formalism instead of the open string formalism.
In this section we shall relate the matrix models to the spectral theories via the closed string formalism.
Again we find that the idea of fixing a reference frame plays an important role.

As we have explained in section \ref{subsec:MM}, the effect of fractional branes is regarded as the change of the closed string background in the closed string formalism.
Due to this reason it is natural to consider the lowest rank as the reference in the spectral theories by integrating out the effect of fractional branes.
Namely on the matrix model side we consider the grand canonical partition function $\Xi_{k,\bm{M}}^{(n)}(z)$ where the reference $n$-th rank is the lowest.
For example, the condition that the first rank is the lowest is given by $0\le M_1$ and $M_2-M_1\le |M_3|$.

Then, from the analysis in the previous section, our interpretation of the ST/TS correspondence is
\begin{align}
\Xi_{k,\bm{M}}^{(n)}(z)=\Det\bigl(1+z\widehat{H}_{(h,e,f)}^{-1}\bigr).
\label{eq:GPF_SD22}
\end{align}
The expression itself may seem familiar to most of the readers, though we stress that the identification of the parameters $(h,e,f)$ on the right-hand side was not clear in previous works before we introduce the idea of reference frames.
Namely, the subscript $(h,e,f)$ of the spectral operator is the parameter for the three-dimensional subspace of point configurations ${\cal C}_\text{P}$ given as follows.
After fixing the $n$-th rank to be the reference, we bring the order of 5-branes into the standard one $\langle\bullet\bullet\circ\,\circ\,\rangle$ with two NS5-branes and two $(1,k)$5-branes by the Hanany-Witten transition with the relative ranks labeled by 
$\langle N+M_2+M_3\bullet
N+M_1+2M_3\bullet
N+2M_1+M_2+M_3\circ
N+M_1\,\circ\rangle$
\eqref{canonicaldeform} and apply the identification \eqref{hef2134}
\begin{align}
(h,e,f)=(q^{-1}e^{2\pi iM_1},e^{2\pi iM_2},e^{2\pi iM_3}).
\label{hefq}
\end{align}
Our presentation indicates that we can describe the matrix models completely in terms of the group-theoretical language on the spectral theory side by specifying the relative rank difference $(M_1,M_2,M_3)$ in the parameter space of point configurations ${\cal C}_\text{P}$.

We note that, although fixing a reference frame is important in the identification \eqref{eq:GPF_SD22}, labeling the 5-branes as $2134$ or the canonical operators as ${\cal Q}_4{\cal Q}_3{\cal P}_1{\cal P}_2$ is not relevant.
In fact, on the matrix model side, the labels do not appear in the definition of the partition function \eqref{eq:PFdef} or the grand canonical partition function \eqref{eq:GPFdef}.
Also, on the spectral theory side, although different labels lead to different parameters $(h,e,f)$ of the spectral operator, the change \eqref{13-4} is generated by similarity transformations $s_1s_2$ and $s_5s_0$ as explained in section \ref{subsec:HW_Weyl}, which does not affect the value of the Fredholm determinant \eqref{eq:GPF_SD22}.

Nevertheless, as we find in a few examples in the next subsection, if we keep track of the labels of 5-branes carefully both on the matrix model side and the spectral theory side, we can still identify the asymptotic values of the spectral operator clearly without ambiguities of similarity transformations.

\subsection{Rank deformed spectral operators\label{subsec:M2Def}}

In this subsection we present a non-trivial check of our proposal in \eqref{eq:GPF_SD22} combined with our identification of brane configurations ${\cal C}_\text{B}$ in point configurations ${\cal C}_\text{P}$.
For the check to work we need to consider a special case of rank deformations where the spectral operator for the matrix model is available.
We present our studies by two examples.

The first example is the same rank deformation ${\bm M}=(M_1,M_2,0)$ as in \cite{MNN} with the second rank being the reference.
As studied in \eqref{Nchange} and \eqref{M123change} using the Hanany-Witten transition, this is equivalent to the rank deformation
\begin{align}
{\bm M}'=\biggl(k+M_1,-\frac{k}{2}+M_2,\frac{k}{2}\biggr),
\label{M222}
\end{align}
with the original reference frame on the matrix model side.
On the other hand, on the spectral theory side, by rearranging the second rank to be the reference cyclically for $\langle M_2\stackrel{2}{\bullet}M_1\stackrel{1}{\bullet}2M_1+M_2\stackrel{3}{\circ}M_1\stackrel{4}{\circ}\rangle$ and shifting the overall rank, the brane configuration of our interest is
\begin{align}
\langle M_1\stackrel{1}{\bullet}2M_1+M_2\stackrel{3}{\circ}M_1\stackrel{4}{\circ}M_2\stackrel{2}{\bullet}\rangle
\simeq\langle 0\stackrel{1}{\bullet}M_2+M_1\stackrel{3}{\circ}0\stackrel{4}{\circ}M_2-M_1\stackrel{2}{\bullet}\rangle,
\label{m2pmm1}
\end{align}
with $0$ being the lowest rank, $M_2\pm M_1\ge 0$.
As sketched in \cite{KMZ,MNN} and further clarified in appendices \ref{sec:M2DefApp} and \ref{sec:FGF}, the spectral operator for the current case is given by
\begin{align}
\bigl(\widehat H_{(2,2)}^{(2)}\bigr)^{-1}
=\bigl(\widehat H_{\stackrel{3}{\circ}\stackrel{1}{\bullet}}(M_2+M_1)\bigr)^{-1}
\bigl(\widehat H_{\stackrel{2}{\bullet}\stackrel{4}{\circ}}(M_2-M_1)\bigr)^{-1},
\label{H222}
\end{align}
with
\begin{align}
\widehat H_{\circ\bullet}(M)
&=e^{-\frac{\pi iM}{2}}\widehat Q^{\frac{1}{2}}\widehat P^{\frac{1}{2}}
+e^{\frac{\pi iM}{2}}\widehat Q^{-\frac{1}{2}}\widehat P^{\frac{1}{2}}
+e^{\frac{\pi iM}{2}}\widehat Q^{\frac{1}{2}}\widehat P^{-\frac{1}{2}}
+e^{-\frac{\pi iM}{2}}\widehat Q^{-\frac{1}{2}}\widehat P^{-\frac{1}{2}},\nonumber\\
\widehat H_{\bullet\circ}(M)
&=e^{\frac{\pi iM}{2}}\widehat P^{\frac{1}{2}}\widehat Q^{\frac{1}{2}}
+e^{-\frac{\pi iM}{2}}\widehat P^{\frac{1}{2}}\widehat Q^{-\frac{1}{2}}
+e^{-\frac{\pi iM}{2}}\widehat P^{-\frac{1}{2}}\widehat Q^{\frac{1}{2}}
+e^{\frac{\pi iM}{2}}\widehat P^{-\frac{1}{2}}\widehat Q^{-\frac{1}{2}}.
\label{Habjmst}
\end{align}
Here $\widehat H_{(2,2)}^{(2)}$ stands for the spectral operator for the $(2,2)$ model with the second rank being the reference.
In obtaining the expression \eqref{H222} by applying the computation in appendix \ref{fgf}, we have split the brane configuration \eqref{m2pmm1} into the former $\stackrel{1}{\bullet}\stackrel{3}{\circ}$ part and the latter $\stackrel{4}{\circ}\stackrel{2}{\bullet}$ part.
Then, we find
\begin{align}
\widehat H_{(2,2)}^{(2)}
=\widehat H_{\stackrel{2}{\bullet}\stackrel{4}{\circ}}(M_2-M_1)
\widehat H_{\stackrel{3}{\circ}\stackrel{1}{\bullet}}(M_2+M_1).
\end{align}
Schematically as in \eqref{schematic} we express the operator multiplication as
\begin{align}
&{\scriptstyle e^{\pi i(M_2-M_1)}}\hspace{-4mm}
\overset{e^{-\pi i(M_2-M_1)}}{\underset{e^{\pi i(M_2-M_1)}}{\sqbox{24}}}\hspace{-4mm}
{\scriptstyle e^{-\pi i(M_2-M_1)}}
\hspace{2mm}\times\hspace{2mm}
{\scriptstyle e^{-\pi i(M_2+M_1)}}\hspace{-4mm}
\overset{e^{\pi i(M_2+M_1)}}{\underset{e^{-\pi i(M_2+M_1)}}{\sqbox{31}}}\hspace{-4mm}
{\scriptstyle e^{\pi i(M_2+M_1)}}
\nonumber\\
&\quad
=\begin{subarray}{c}
{e^{-\pi i(M_2+M_1)}}\\[6mm]
{q^{-1}e^{\pi i(M_2-M_1)}}
\end{subarray}
\hspace{-14mm}
\overset{q^{\frac{1}{2}}e^{-\pi i(M_2-M_1)}\quad q^{\frac{1}{2}}e^{\pi i(M_2+M_1)}}
{\underset{q^{-\frac{1}{2}}e^{-\pi i(M_2+M_1)}\quad q^{-\frac{1}{2}}e^{\pi i(M_2-M_1)}}{\sqbox{\phantom{xx}2431\phantom{xx}}}}
\hspace{-14mm}
\begin{subarray}{c}
{q e^{-\pi i(M_2-M_1)}}\\[6mm]
{e^{\pi i(M_2+M_1)}}
\end{subarray}
=\begin{subarray}{c}
{q^{-1}e^{-2\pi iM_1}}\\[6mm]
{q^{-2}e^{2\pi i(M_2-M_1)}}
\end{subarray}
\hspace{-10mm}
\overset{1\quad e^{2\pi iM_2}}
{\underset{q^{-1}e^{-2\pi iM_1}\quad q^{-1}e^{2\pi i(M_2-M_1)}}{\sqbox{\phantom{xx}2431\phantom{xx}}}}
\hspace{-10mm}
\begin{subarray}{c}
{1}\\[6mm]
{q^{-1}e^{2\pi iM_2}}
\end{subarray},
\end{align}
where we have also fixed the gauge $e_2=e_4=e_6=e_8=1$ by using small similarity transformations.
By comparing with figure \ref{asymptoticvaluesfixed} we can identify the parameters as
\begin{align}
(h,e,f)=(e^{2\pi iM_1},q^{-\frac{1}{2}}e^{2\pi iM_2},q^{\frac{1}{2}}),
\end{align}
which is consistent with our expectation if we apply our identification of ${\cal C}_\text{B}$ in ${\cal C}_\text{P}$ \eqref{hefq} to \eqref{M222} as in
\begin{align}
(h,e,f)=(q^{-1}e^{2\pi iM'_1},e^{2\pi iM'_2},e^{2\pi iM'_3})=(e^{2\pi iM_1},q^{-\frac{1}{2}}e^{2\pi iM_2},q^{\frac{1}{2}}).
\end{align}

In our second example we consider the rank deformation ${\bm M}''=(\frac{k}{2}+M_1,\frac{k}{2},M_3)$ in the original reference and labels.
Then we find after applying the Hanany-Witten transition and shifting the overall rank
\begin{align}
&\Bigl\langle\frac{k}{2}+M_3\stackrel{2}{\bullet}
\frac{k}{2}+M_1+2M_3\stackrel{1}{\bullet}
\frac{3k}{2}+2M_1+M_3\stackrel{3}{\circ}
\frac{k}{2}+M_1\stackrel{4}{\circ}\Bigr\rangle\nonumber\\
&\simeq\Bigl\langle\frac{k}{2}+M_3\stackrel{2}{\bullet}
\frac{k}{2}+M_1+2M_3\stackrel{3}{\circ}
\frac{k}{2}+M_3\stackrel{1}{\bullet}
\frac{k}{2}+M_1\stackrel{4}{\circ}\Bigr\rangle\nonumber\\
&\simeq\bigl\langle 0\stackrel{2}{\bullet}
M_1+M_3\stackrel{3}{\circ}
0\stackrel{1}{\bullet}
M_1-M_3\stackrel{4}{\circ}\bigr\rangle.
\end{align}
Hence we find that in this case the spectral operator is
\begin{align}
\widehat H_{(2,2)}
=\widehat H_{\stackrel{4}{\circ}\stackrel{1}{\bullet}}(M_1-M_3)\widehat H_{\stackrel{3}{\circ}\stackrel{2}{\bullet}}(M_1+M_3).
\end{align}
As the previous example, after multiplying two spectral operators and fixing the gauge,
\begin{align}
&{\scriptstyle e^{-\pi i(M_1-M_3)}}\hspace{-4mm}
\overset{e^{\pi i(M_1-M_3)}}{\underset{e^{-\pi i(M_1-M_3)}}{\sqbox{41}}}\hspace{-4mm}
{\scriptstyle e^{\pi i(M_1-M_3)}}
\hspace{2mm}\times\hspace{2mm}
{\scriptstyle e^{-\pi i(M_1+M_3)}}\hspace{-4mm}
\overset{e^{\pi i(M_1+M_3)}}{\underset{e^{-\pi i(M_1+M_3)}}{\sqbox{32}}}\hspace{-4mm}
{\scriptstyle e^{\pi i(M_1+M_3)}}
\nonumber\\
&\quad
=\begin{subarray}{c}
{q^{-\frac{1}{2}}e^{-\pi i(M_1-M_3)}}\\[6mm]
{e^{-\pi i(M_1+M_3)}}
\end{subarray}
\hspace{-12mm}
\overset{e^{\pi i(M_1-M_3)}\quad q^{\frac{1}{2}}e^{\pi i(M_1+M_3)}}
{\underset{q^{-\frac{1}{2}}e^{-\pi i(M_1+M_3)}\quad e^{-\pi i(M_1-M_3)}}{\sqbox{\phantom{xx}4132\phantom{xx}}}}
\hspace{-12mm}
\begin{subarray}{c}
{e^{\pi i(M_1+M_3)}}\\[6mm]
{q^{\frac{1}{2}}e^{\pi i(M_1-M_3)}}
\end{subarray}
=\begin{subarray}{c}
{q^{-\frac{1}{2}}e^{-2\pi iM_1}}\\[6mm]
{e^{-2\pi i(M_1+M_3)}}
\end{subarray}
\hspace{-10mm}
\overset{1\quad q^{\frac{1}{2}}e^{2\pi iM_3}}
{\underset{q^{-\frac{1}{2}}e^{-2\pi iM_1}\quad e^{-2\pi i(M_1-M_3)}}{\sqbox{\phantom{xx}4132\phantom{xx}}}}
\hspace{-10mm}
\begin{subarray}{c}
{1}\\[6mm]
{q^{\frac{1}{2}}e^{-2\pi iM_3}}
\end{subarray},
\end{align}
we find
\begin{align}
(h,e,f)=(q^{-\frac{1}{2}}e^{2\pi iM_1},q^{\frac{1}{2}},e^{2\pi iM_3}),
\end{align}
by comparing with figure \ref{asymptoticvaluesfixed}, which again is exactly our expectation,
\begin{align}
(h,e,f)=(q^{-1}e^{2\pi iM''_1},e^{2\pi iM''_2},e^{2\pi iM''_3})=(q^{-\frac{1}{2}}e^{2\pi iM_1},q^{\frac{1}{2}},e^{2\pi iM_3}).
\end{align}

The computations in this subsection serve as non-trivial consistency checks for all of our proposals.
In section \ref{sec:MM_QC} we have identified the three-dimensional space of brane configurations ${\cal C}_\text{B}$ in the five-dimensional space of point configurations ${\cal C}_\text{P}$ from the configurations without rank deformations.
Here in this section we further relate the matrix models with general rank deformations to spectral theories in \eqref{hefq}.
Our computations show that all of these proposals are consistent with each other.

\section{Matrix models and topological strings\label{sec:QCandTS}}

In section \ref{secMatrixModel} we have stressed the importance of fixing a reference frame and in section \ref{sec:MM_QC} using the idea of fixing a reference frame we are able to identify the three-dimensional space of brane configurations ${\cal C}_\text{B}$ in the five-dimensional space of point configurations ${\cal C}_\text{P}$.
In the previous section we have established the relation between matrix models and spectral theories using the identification of ${\cal C}_\text{B}$ in ${\cal C}_\text{P}$ and provided non-trivial checks for it.
In this section we turn to the relation to the topological string theory.
Here the symmetry structure of ${\cal C}_\text{P}$ and the identification of ${\cal C}_\text{B}$ in ${\cal C}_\text{P}$ are critical to establish the explicit relation of parameters between matrix models and topological strings.

For the symmetry structure, following the proposal of the relation between matrix models and topological strings \cite{MPtop,MP,HMMO}, in \cite{MN3,MNN} the large $z$ expansion of the grand canonical partition functions of the (2,2) model and the (1,1,1,1) model was studied and it was found that the grand potential is given by the free energy of topological strings on local del Pezzo $D_5$ where the BPS indices are split.
In \cite{MNY} the BPS indices were further identified as representations of $D_5$ and the split was explained by assuming an unbroken subgroup of $D_5$ and decomposing the $D_5$ representations into this subgroup, which indicates that the free energy can be expressed by the characters of $D_5$.
To explain the unbroken subgroup in \cite{KMN} the spectral operators of the $(2,2)$ model and the $(1,1,1,1)$ model without rank deformations were studied.
After realizing the Weyl symmetries in the five-dimensional space of point configurations ${\cal C}_\text{P}$ and identifying the models in it, we can study the unbroken subgroup for each point.
From these studies we clearly observe that the symmetry structure of spectral operators and that of topological strings match with each other.

On the other hand, after we have identified rank deformations ${\cal C}_\text{B}$ in ${\cal C}_\text{P}$ in section \ref{sec:MM_QC}, in this section we can turn to the relation between matrix models and topological strings with rank deformations.
As we have noted in sections \ref{change} and \ref{unbrokensym}, the unbroken symmetry depends on the reference.
Due to this reason, it is important to fix a reference also in the description by topological strings.

\subsection{Topological strings from characters}

Although the relation between matrix models and topological strings was originally presented using K\"ahler parameters in \cite{HMMO}, it was found in \cite{MNY} that the expression using characters is more efficient after understanding that the BPS indices are split as representations are decomposed in the unbroken subgroup of $D_5$.
It was simply claimed in \cite{MNY} that the charges under the u$(1)$ actions in the characters are chosen suitably so that they respect the unbroken subgroup.
However, after identifying the three-dimensional space of brane configurations ${\cal C}_\text{B}$ in the five-dimensional space of point configurations ${\cal C}_\text{P}$ in sections \ref{subsec:BCPC} and \ref{change} and understanding the Weyl action in section \ref{unbrokensym}, in this section we can present a universal recipe for determining the u$(1)$ charges in the characters.
Thus we are able to present a completely group-theoretical description for the matrix models on the topological string theory side.
For readers unfamiliar with the progress of the relation between matrix models and topological strings in terms of K\"ahler parameters, we first summarize the results.

To present the relation, we define the reduced grand potential for the chemical potential $\mu=\log z$ from the grand canonical partition function as
\begin{align}
\Xi^{(n)}_{k,\bm{M}}(e^\mu)=\sum_{n=-\infty}^{\infty}e^{J^{(n)}_{k,\bm{M}}(\mu+2\pi in)},
\label{eq:GPdef}
\end{align}
by removing a trivial periodicity in the shift $\mu\to\mu+2\pi i$.
Here as we have stressed in \eqref{eq:GPF_SD22} the grand canonical partition function $\Xi^{(n)}_{k,\bm{M}}(z)$ depends on the reference rank $(n)$.
If we further redefine the chemical potential into an effective one, we can simplify the expression for $J^{(n)}_{k,\bm{M}}(\mu)$.
Namely, the reduced grand potential is decomposed into three parts for large $\mu$
\begin{align}
J^{(n)}_{k,\bm{M}}(\mu)
=J^{(n),\text{pert}}_{k,\bm{M}}(\mu_\text{eff})+J^{(n),\text{WS}}_{k,\bm{M}}(\mu_\text{eff})+J^{(n),\text{MB}}_{k,\bm{M}}(\mu_\text{eff}),
\label{eq:GP_FE}
\end{align}
where each part is called the perturbative part, the worldsheet instanton part and the membrane instanton part.
Then the perturbative part is given by
\begin{align}
J^{(n),\text{pert}}_{k,\bm{M}}(\mu_\text{eff})&=\frac{C_{k}}{3}\mu_\text{eff}^3+B_{k,\bm{M}}\mu_\text{eff}+A_{k,\bm{M}},
\end{align}
while the non-perturbative instanton parts are given in terms of the free energy of topological strings by
\begin{align}
J_{k,\bm{M}}^{(n),\text{WS}}(\mu_\text{eff})
&=\sum_{j_\text{L},j_\text{R}}\sum_{\bm{d}}N_{j_\text{L},j_\text{R}}^{\bm{d}}
\sum_n\frac{(-1)^{(s_\text{L}+s_\text{R}-1)n}s_\text{R}\sin(2\pi g_\text{s}ns_\text{L})}
{4n\sin^2(\pi g_\text{s}n)\sin(2\pi g_\text{s}n)}e^{-n\bm{d}\cdot\bm{T}},\nonumber\\
J_{k,\bm{M}}^{(n),\text{MB}}(\mu_\text{eff})
&=-\sum_{j_\text{L},j_\text{R}}\sum_{\bm{d}}N_{j_\text{L},j_\text{R}}^{\bm{d}}
\sum_n\frac{\partial}{\partial g_\text{s}}
\biggl[\frac{g_\text{s}\sin\bigl(\frac{\pi n}{g_\text{s}}s_\text{L}\bigr)\sin\bigl(\frac{\pi n}{g_\text{s}}s_\text{R}\bigr)}
{4\pi n^2\sin^3\bigl(\frac{\pi n}{g_\text{s}}\bigr)}e^{-n\frac{\bm{d}\cdot\bm{T}}{g_\text{s}}}\biggr],
\end{align}
with the quantities of topological strings being the coupling constant $g_\text{s}$, K\"ahler parameters $\bm{T}$, the corresponding degrees $\bm{d}$ and the BPS indices $N_{j_\text{L},j_\text{R}}^{\bm{d}}$.

For the example of the $(2,2)$ model with the rank deformation ${\bm M}=(M_1,M_2,0)$ and the first reference frame $n=1$, various quantities are given explicitly in \cite{MNN} including the effective chemical potential
\begin{align}
\mu_\text{eff}=\left\{\begin{array}{l}
\mu+4(-1)^{M_1}e^{-\mu}{}_4F_3\bigl(1,1,\frac{3}{2},\frac{3}{2};2,2,2;-16(-1)^{M_1}e^{-\mu}\bigr),\\
\hspace{4cm}\text{for }k:\text{even or }(M_1=0\text{ or }M_2=0),\\
\mu+2e^{-2\mu}{}_4F_3\bigl(1,1,\frac{3}{2},\frac{3}{2};2,2,2;-16e^{-2\mu}\bigr),\\
\hspace{4cm}\text{for }k:\text{odd and }(M_1=\frac{k}{2}\text{ or }M_2=\frac{k}{2}),
\end{array}\right.
\label{mueff}
\end{align}
the perturbative coefficients
\begin{align}
C_k=\frac{1}{2\pi^2k},\quad
B_{k,\bm{M}}=-\frac{1}{6k}-\frac{k}{3}+\frac{1}{2k}\bigl((M_1-k)^{2}+2M_2^{2}\bigr),
\label{eq:FEPert}
\end{align}
with $A_{k,{\bm{M}}}$ partially identified and the non-perturbative coefficients for the free energy of topological strings
\begin{align}
g_\text{s}=\frac{1}{k},\quad
\bm{T}=(T_1^+,T_1^-,T_2^+,T_2^-,T_3^+,T_3^-),\quad
\bm{d}=(d_1^+,d_1^-,d_2^+,d_2^-,d_3^+,d_3^-),
\end{align}
with
\begin{align}
T_1^{\pm}=\frac{\mu_{\mathrm{eff}}}{k}\pm\pi i(b_1+2b_2),\quad
T_2^{\pm}=\frac{\mu_{\mathrm{eff}}}{k}\pm\pi ib_1,\quad
T_3^{\pm}=\frac{\mu_{\mathrm{eff}}}{k}\pm\pi i(b_1-2b_2),
\end{align}
and
\begin{align}
(b_1,b_2)=\biggl(\frac{M_1}{k}-1,\frac{M_2}{k}\biggr).
\label{b1b2}
\end{align}
Here the definition of the K\"ahler parameters is slightly changed from \cite{MNN} for later convenience.
The total BPS indices $N_{j_\text{L},j_\text{R}}^{|\bm{d}|}$ are given by the tables of del Pezzo $D_5$ in \cite{HKP} and split by various combinations of degrees ${\bm d}$.

In \cite{MNY} it was pointed out that the split of the BPS indices can be regarded as the decomposition of the $D_5$ representations into the unbroken subgroup, which directly indicates that the non-perturbative part of the reduced grand potential is given by the characters.
Namely, the non-perturbative part can be expressed in terms of the $D_5$ characters as
\begin{align}
J_{k,\bm{M}}^\text{WS}(\mu_\text{eff})=\sum_{m=1}^\infty d_m(k,{\bm b})e^{-m\frac{\mu_\text{eff}}{k}},\quad
J_{k,\bm{M}}^\text{MB}(\mu_\text{eff})=\sum_{\ell=1}^\infty\Bigl(\widetilde b_\ell(k,{\bm b})\mu_\text{eff}+\widetilde c_\ell(k,{\bm b})\Bigr)
e^{-\ell\mu_\text{eff}}.
\label{eq:GPwithChara}
\end{align}
Here the instanton coefficients are given in their multi-covering components by
\begin{align}
&d_m(k,{\bm b})=(-1)^m\sum_{n|m}\frac{1}{n}\delta_{\frac{m}{n}}\biggl(\frac{k}{n},n{\bm b}\biggr),\nonumber\\
&\widetilde b_\ell(k,{\bm b})=\sum_{n|\ell}\frac{1}{n}\beta_{\frac{\ell}{n}}(nk,{\bm b}),\quad
\widetilde c_\ell(k,{\bm b})=-k^2\frac{\partial}{\partial k}\biggl[\frac{\widetilde b_\ell(k,{\bm b})}{\ell k}\biggr],
\end{align}
and the multi-covering components are given by
\begin{align}
\delta_d(k,{\bm b})&=\frac{(-1)^{d-1}}{(2\sin\frac{\pi}{k})^2}\sum_{j_\text{L},j_\text{R}}\sum_{\bf R}n^{d,{\bf R}}_{j_\text{L},j_\text{R}}
\chi_{\bf R}({\bm q})\chi_{j_\text{L}}(e^{\frac{2\pi i}{k}})\chi_{j_\text{R}}(1),\nonumber\\
\beta_d(k,{\bm b})&=\frac{(-1)^dd}{4\pi\sin\pi k}\sum_{j_\text{L},j_\text{R}}\sum_{\bf R}n^{d,{\bf R}}_{j_\text{L},j_\text{R}}
\chi_{\bf R}({\bm q}^k)\chi_{j_\text{L}}(e^{\pi ik})\chi_{j_\text{R}}(e^{\pi ik}),
\end{align}
with the su$(2)$ character
\begin{align}
\chi_j(q)=\frac{q^{2j+1}-q^{-(2j+1)}}{q-q^{-1}},
\end{align}
and
\begin{align}
{\bm q}=(1,e^{2\pi ib_2},e^{-2\pi ib_1},e^{2\pi ib_2},1).
\end{align}
The coefficient $n^{d,{\bf R}}_{j_\text{L},j_\text{R}}$ is the multiplicity of a representation ${\bf R}$ in degree $d$ and spins $(j_\text{L},j_\text{R})$ whose explicit values can be found in \cite{MNY}.

Now let us turn to our universal recipe for determining the u$(1)$ charges in the characters.
In \cite{MNY} it was claimed that the charges under the u$(1)$ actions ${\bm q}$ in the characters $\chi_{\bf R}({\bm q})$ can be identified from the unbroken subgroup.
After identifying the space of rank deformations ${\cal C}_\text{B}$ and the Weyl actions on it in section \ref{sec:MM_QC}, we can present the direct recipe.
First note that we do not denote the reference rank $(n)$ for $J_{k,{\bm M}}(\mu)$ in \eqref{eq:GPwithChara} deliberately since the reference rank $(n)$ and the rank difference ${\bm M}$ are translated to the parameter $(h,e,f)$ of quantum curves, as we have explained in the context of spectral theories in section \ref{sec:MM_ST}.
Here we further relate the parameter $(h,e,f)$ to the parameter of the characters.

For the identification of the u$(1)$ charges ${\bm q}$ in the characters $\chi_{\bf R}({\bm q})$, we first rewrite the parameters of ${\cal C}_\text{P}$ in \eqref{hef} as
\begin{align}
(\overline h_1,\overline h_2,e_1,e_3,e_5)=\biggl(\frac{ef}{h},\frac{hf}{e},\frac{f}{e},ef,\frac{f}{e}\biggr)
=h^{(-1,1,0,0,0)}e^{(1,-1,-1,1,-1)}f^{(1,1,1,1,1)},
\end{align}
where we have picked up the powers for $h$, $e$ and $f$ respectively.
If we relate the fundamental weights of $D_5$, $\omega_i$, identified from the Weyl actions on the five-dimensional space of point configurations ${\cal C}_\text{P}$ in \cite{KMN} with the canonical fundamental weights $\overline\omega_i$ (respecting the orthonormality of the Cartan matrix) used to construct characters in \cite{MNY} as
\begin{align}
\omega_1=(1,-1,0,0,-1)\quad&\leftrightarrow\quad
\overline\omega_5=\textstyle{(\frac{1}{2},\frac{1}{2},\frac{1}{2},\frac{1}{2},\frac{1}{2})},\nonumber\\
\omega_2=(1,-1,0,1,-1)\quad&\leftrightarrow\quad
\overline\omega_4=\textstyle{(\frac{1}{2},\frac{1}{2},\frac{1}{2},\frac{1}{2},-\frac{1}{2})},\nonumber\\
\omega_3=(1,-2,0,0,-2)\quad&\leftrightarrow\quad\overline\omega_3=(1,1,1,0,0),\nonumber\\
\omega_4=(0,-1,0,0,-2)\quad&\leftrightarrow\quad\overline\omega_2=(1,1,0,0,0),\nonumber\\
\omega_5=(0,0,1,0,-1)\quad&\leftrightarrow\quad\overline\omega_1=(1,0,0,0,0),
\end{align}
we can identify 
\begin{align}
&h^{(-1,1,0,0,0)}e^{(1,-1,-1,1,-1)}f^{(1,1,1,1,1)}
=h^{-\omega_3+\omega_4}
e^{\omega_1+\omega_2-\omega_3+\omega_4-\omega_5}
f^{\omega_1+\omega_2-\omega_3-\omega_4+\omega_5}\nonumber\\
&\quad\leftrightarrow
h^{-\overline\omega_3+\overline\omega_2}
e^{\overline\omega_5+\overline\omega_4-\overline\omega_3+\overline\omega_2-\overline\omega_1}
f^{\overline\omega_5+\overline\omega_4-\overline\omega_3-\overline\omega_2+\overline\omega_1}
=h^{(0,0,-1,0,0)}e^{(0,1,0,1,0)}f^{(0,-1,0,1,0)}.
\end{align}
Since we have identified the rank deformations of brane configurations $(M_1,M_2,M_3)$ (in the standard order 2134 of 5-branes) in the five-dimensional space of point configurations ${\cal C}_\text{P}$ in section \ref{sec:MM_QC} as in \eqref{hef2134}, we have
\begin{align}
(1,ef^{-1},h^{-1},ef,1)=(1,e^{2\pi i(M_2-M_3)},e^{-2\pi i(M_1-k)},e^{2\pi i(M_2+M_3)},1),
\end{align}
and obtain the arguments of the characters by rescaling correctly
\begin{align}
{\bm q}=(1,q_1,q_2,q_3,1)=\bigl(1,e^{2\pi i\frac{M_2-M_3}{k}},e^{-2\pi i(\frac{M_1}{k}-1)},e^{2\pi i\frac{M_2+M_3}{k}},1\bigr).
\label{qtopstr}
\end{align}
This is our main result for specifying the u$(1)$ charges ${\bm q}$ in the characters $\chi_{\bf R}({\bm q})$ to describe matrix models with topological strings.

Finally let us make a short conjecture on the perturbative part.
After rewriting the non-perturbative part in the group-theoretical language of $D_5$, it is also natural to rewrite the perturbative part.
The coefficient $B_{k,{\bm M}}$ \eqref{eq:FEPert} has a nice dependence on ${\bm M}$ and it is tantalizing to rewrite it as
\begin{align}
B_{k,{\bm M}}=-\frac{1}{6k}-\frac{k}{3}+\frac{1}{2k}\biggl\Vert\frac{k\log{\bm q}}{2\pi i}\biggr\Vert^2,
\label{Bconj}
\end{align}
where the norm $\Vert\cdot\Vert$ is defined by the Cartan matrix.

\subsection{Second frame}

After observing that changing references amounts to shifting the origin of the matrix models in section \ref{change}, as a non-trivial check we can consider the grand potential of the same rank deformations $(M_1,M_2)$ with the reference frame being the second rank and see whether the result can still be given by the characters with the u$(1)$ charges identified in \eqref{qtopstr}.
The rank deformation was restricted to the case $M_3=0$ in \cite{MNN} since the introduction of non-vanishing $M_3$ caused a severe divergence where the regularization was unclear.

In \cite{MNN} it was pointed out that it is important to fix a reference frame for the correspondence between matrix models and topological strings.
The reference was fixed to the first rank with the grand canonical partition function defined as
\begin{align}
\Xi_{k,{\bm M}}^{(1)}(z)=\sum_{N}^\infty z^{N+M_2}Z_k(N+M_2,N+M_1,N+2M_1+M_2,N+M_1),
\label{Xi1}
\end{align}
(with the overall normalization, the lower bound of the summation and the absolute values omitted) and the BPS indices used in describing the free energy of topological strings were split accordingly.
In \cite{MNY} the split of the BPS indices was understood from the decomposition of the $D_5$ representations into various subgroups and it was proposed to describe the reduced grand potential by characters, where the identification of the parameters is consistent with our general proposal \eqref{qtopstr} with $M_3=0$.

To fix the second rank as the reference, on the matrix model side we need to define the grand canonical partition functions as
\begin{align}
\Xi_{k,{\bm M}}^{(2)}(z)=\sum_{N}^\infty z^{N+M_1}Z_k(N+M_2,N+M_1,N+2M_1+M_2,N+M_1),
\label{Xi2}
\end{align}
instead of \eqref{Xi1}.
Due to this change, only $B_{k,{\bm M}}$ in the perturbative part and terms with powers of $e^{-\mu_\text{eff}}$ in the non-perturbative part of the reduced grand potential change.
On the topological string side, the identification of the parameters \eqref{M123change}
\begin{align}
\biggl(\frac{M'_1}{k}-1,\frac{M'_2}{k},\frac{M'_3}{k}\biggr)=\biggl(\frac{M_1}{k},\frac{M_2}{k}-\frac{1}{2},\frac{1}{2}\biggr),
\label{eq:Chara2}
\end{align}
is translated to that of the u$(1)$ charges in the characters as
\begin{align}
{\bm q}=\bigl(1,e^{2\pi i\frac{M'_2-M'_3}{k}},e^{-2\pi i(\frac{M'_1}{k}-1)},e^{2\pi i\frac{M'_2+M'_3}{k}},1\bigr)
=(1,e^{2\pi i(\frac{M_2}{k}-1)},e^{-2\pi i\frac{M_1}{k}},e^{2\pi i\frac{M_2}{k}},1),
\label{q123p}
\end{align}
by using our general proposal \eqref{qtopstr}.
For the perturbative part, the change of ${\bm q}$ from \eqref{qtopstr} to \eqref{q123p} is consistent with the change of the power of $z$ from \eqref{Xi1} to \eqref{Xi2} through \eqref{Bconj}.
For the non-perturbative part, this change of ${\bm q}$ does not affect the worldsheet instantons which is consistent with the above observation that only terms with powers of $e^{-\mu_\text{eff}}$ change.
Especially for the membrane instantons we can perform a very non-trivial check.
We have listed the corresponding numerical expansions of the grand potential in appendix \ref{secGPnumerical} and the characters in appendix \ref{characterlist}.
By substituting the characters into the expression of the free energy of topological strings \eqref{eq:GPwithChara}, we find an exact match.
Our computation in this subsection serves as another non-trivial check for our proposal on the idea of fixing a reference frame and the identification of the three-dimensional space of brane configurations ${\cal C}_\text{B}$ in the five-dimensional space of point configurations ${\cal C}_\text{P}$.

\section{Conclusion}

In this paper we have pointed out the importance in fixing a reference frame for the study of the super Chern-Simons matrix model.
After fixing references on all aspects of our analysis including brane configurations, matrix models, spectral theories and topological strings, we are able to construct a consistent correspondence among all of these aspects.
As several non-trivial checks, we find that the introduction of the idea of fixing a reference frame successfully specifies the three-dimensional subspace of rank deformations in brane configurations in the five-dimensional parameter space of point configurations of asymptotic values of quantum curves.
Also in section \ref{sec:MM_ST}, we find that, by fixing the lowest rank to be the reference, for a special class of rank deformations, the closed string formalism has been established and the spectral operators have been identified, whose parameters match exactly with those identified from the brane configurations.
Finally in section \ref{sec:QCandTS}, following previous computations, we present a universal expression for the free energy of topological strings corresponding to matrix models.
We can change frames and find that the unbroken subgroup also changes which is perfectly consistent with our identification of characters.
We shall list some further directions in the following.

First, by fixing a reference in brane configurations, we have identified the three-dimensional subspace of rank deformations in brane configurations ${\cal C}_\text{B}$ in the five-dimensional space of point configurations ${\cal C}_\text{P}$.
At the same time, since the five-dimensional space enjoys the full $D_5$ Weyl symmetry, it is perplexing what the role of the remaining two dimensions is.
We believe that this strongly suggests that our understanding of the fractional M2-branes is insufficient.
In this paper we only consider the situation where we have a clear picture of the brane configurations in type IIB string theory.
The existence of the extra two dimensions suggests that, in general, the fractional branes can be more subtle objects which change the geometrical backgrounds drastically so that the numbers of D3-branes or the order of the NS5-branes and the $(1,k)$5-branes does not make sense any more.

Secondly, even in the three-dimensional space of brane configurations ${\cal C}_\text{B}$, we have identified a new symmetry $s_3$ or $s_4$ \eqref{s3s4N} unknown from the Hanany-Witten transition or a few discrete symmetries.
We would like to give an interpretation to it from the study of brane physics.

Thirdly, it is interesting to see the implication of our work to the relation to the $q$-Painlev\'e system proposed in \cite{BGT}.
Especially, we would like to find out the relation between our shift symmetry \eqref{M123change} with the shift generator in the affine Weyl group for the Painlev\'e system \cite{KNY}.

Fourthly, our analysis is directly applicable to other genus one matrix models \cite{MNY,KMN}, higher genus matrix models \cite{HM,HHO,CGM} or even matrix models of $\widehat D$ type quiver \cite{ADF,MN4}.
Especially, in \cite{MNY,KMN} the $(2,1,2,1)$ matrix model was studied and it was found to correspond to the $E_7$ spectral theory.
By repeating our analysis for the $E_7$ theory we may find more examples of the correspondence.

Fifthly, the construction of the spectral operator by connecting canonical operators of 5-branes subsequently is reminiscent of the construction of the partition function in \cite{A}.
We believe that a larger framework of quantum curves will appear by clarifying the relation between these two constructions.

\appendix

\section{Fermi gas formalism\label{fgf}}

In this appendix we review the Fermi gas formalism for the super Chern-Simons matrix models \eqref{eq:PFdef} without rank deformations, propose a generalization with some rank deformations and relate the result to spectral operators.
The techniques are mostly taken from many previous works including \cite{MP,MM,MN1,MS2,KM,MNN,KMZ}, though we stress that our piecewise derivation is now much clearer.

\subsection{No rank deformations}\label{derivation}

In this subsection we derive the Fermi gas formalism for the super Chern-Simons matrix models \eqref{eq:PFdef} without rank deformations,
\begin{align}
Z^{\{s_a\}}_k(\{N\})=\int\frac{D^N\lambda_1D^N\lambda_2\cdots D^N\lambda_R}{(N!)^R(2\pi)^{NR}}
Z_k(N;\lambda_1,\lambda_2)Z_k(N;\lambda_2,\lambda_3)\cdots Z_k(N;\lambda_R,\lambda_1).
\label{Znodeform}
\end{align}
Here we have rescaled the integration variables by $k^{-1}(>0)$ and, for our application to the case of equal ranks, redefine $Z(N,N;\mu,\nu)$ and $D\lambda_a$ from \eqref{component} as
\begin{align}
Z_k(N;\mu,\nu)=\frac{1}{k^N}Z_k\Bigl(N,N;\frac{\mu}{k},\frac{\nu}{k}\Bigr)
&=\frac{\prod_{m<m'}^N2\sinh\frac{\mu_m-\mu_{m'}}{2k}
\prod_{n<n'}^N2\sinh\frac{\nu_n-\nu_{n'}}{2k}}
{k^N\prod_{m=1}^N\prod_{n=1}^N2\cosh\frac{\mu_m-\nu_n}{2k}},\nonumber\\
D\lambda_a&=d\lambda_ae^{\frac{i}{2\hbar}\sign(k_a)\lambda_a^2},
\label{Zequalrank}
\end{align}
with $\hbar=2\pi k$.
The derivation can be simplified as in \cite{MP,MN1} though we present in the current manner as a preparation for the next subsection with rank deformations.
It is convenient to proceed to the computation with the brane configuration in mind.
As explained in section \ref{subsec:MM}, each integration variable $\lambda_a$ corresponds to a stack of $N_a(=N)$ D3-branes and each factor of the integrand $Z_k(N;\lambda_a,\lambda_{a+1})$ corresponds to a 5-brane connecting two stacks of D3-branes with ranks $N_a$ and $N_{a+1}$ where the 5-brane can be the NS5-brane ($s_a=+1$) or the $(1,k)$5-brane ($s_a=-1$).

Let us first focus on the integrand $Z_k(N;\mu,\nu)$ in \eqref{Zequalrank}.
For later convenience we introduce eigenstates for the coordinate operator $\widehat q$ normalized as
\begin{align}
\langle q_1|q_2\rangle=2\pi\delta(q_1-q_2),\quad
\int\frac{dq}{2\pi}|q\rangle\langle q|=1.
\label{normalq}
\end{align}
Using the Cauchy determinant
\begin{align}
\frac{\prod_{m<m'}^N(x_m-x_{m'})
\prod_{n<n'}^N(y_n-y_{n'})}
{\prod_{m=1}^N\prod_{n=1}^N(x_m+y_n)}
=\det\begin{pmatrix}(x_m+y_n)^{-1}\end{pmatrix}_{1\le m,n\le N},
\end{align}
and the Fourier transformation
\begin{align}
\langle\mu|\frac{1}{2\cosh\frac{\widehat p}{2}}|\nu\rangle=\frac{1}{2k\cosh\frac{\mu-\nu}{2k}},
\end{align}
we find that $Z_k(N;\mu,\nu)$ defined in \eqref{Zequalrank} is given by
\begin{align}
Z_k(N;\mu,\nu)
=\det\biggl(\langle\mu|\frac{1}{2\cosh\frac{\widehat p}{2}}|\nu\rangle\biggr),
\label{Zeqra}
\end{align}
where tacitly the determinant is for the $N\times N$ matrix labeled by the subscripts $m,n$ of $\mu_m,\nu_n$.
Then using \eqref{Zeqra} we can rewrite the integrand of the partition function \eqref{Znodeform} into a product of determinants.

Let us next turn to the integration $D\lambda_{a}$ \eqref{Zequalrank} where $d\lambda_{a}$ is combined by a Fresnel factor $e^{\pm\frac{i}{2\hbar}\lambda_{a}^2}$ or an operator $e^{\pm\frac{i}{2\hbar}\widehat q^2}$ when acting on the ket states.
Since each bra state and each ket state always combine into an identity operator as in \eqref{normalq}, we are free to perform a similarity transformation.
We perform different similarity transformations depending on the types of 5-branes on the two sides of the integration variable.
With the operator $e^{\pm\frac{i}{2\hbar}\widehat q^2}$ coming from the integration $D\lambda_a$ taken into account, each integration now become
\begin{align}
\int\frac{d\lambda}{2\pi}|\lambda\rangle\langle\lambda|
&=\int\frac{d\lambda}{2\pi}e^{\frac{i}{2\hbar}\widehat p^2}|\lambda\rangle
\langle\lambda|e^{-\frac{i}{2\hbar}\widehat p^2},&
\text{for}\quad
(s_{a-1},s_a)&=(+1,+1),\nonumber\\
\int\frac{d\lambda}{2\pi}e^{\frac{i}{2\hbar}\widehat q^2}|\lambda\rangle\langle\lambda|
&=\int\frac{d\lambda}{2\pi}e^{\frac{i}{2\hbar}\widehat q^2}e^{\frac{i}{2\hbar}\widehat p^2}|\lambda\rangle
\langle\lambda|e^{-\frac{i}{2\hbar}\widehat p^2},&
\text{for}\quad
(s_{a-1},s_a)&=(-1,+1),\nonumber\\
\int\frac{d\lambda}{2\pi}e^{-\frac{i}{2\hbar}\widehat q^2}|\lambda\rangle\langle\lambda|
&=\int\frac{d\lambda}{2\pi}e^{\frac{i}{2\hbar}\widehat p^2}|\lambda\rangle
\langle\lambda|e^{-\frac{i}{2\hbar}\widehat p^2}e^{-\frac{i}{2\hbar}\widehat q^2},&
\text{for}\quad
(s_{a-1},s_a)&=(+1,-1),\nonumber\\
\int\frac{d\lambda}{2\pi}|\lambda\rangle\langle\lambda|
&=\int\frac{d\lambda}{2\pi}e^{\frac{i}{2\hbar}\widehat q^2}e^{\frac{i}{2\hbar}\widehat p^2}|\lambda\rangle
\langle\lambda|e^{-\frac{i}{2\hbar}\widehat p^2}e^{-\frac{i}{2\hbar}\widehat q^2},&
\text{for}\quad
(s_{a-1},s_a)&=(-1,-1).
\label{simitran}
\end{align}
Then we find that, after the similarity transformation, each operator $\bigl(2\cosh\frac{\widehat p}{2}\bigr)^{-1}$ in \eqref{Zeqra} becomes
\begin{align}
e^{-\frac{i}{2\hbar}\widehat p^2}\frac{1}{2\cosh\frac{\widehat p}{2}}e^{\frac{i}{2\hbar}\widehat p^2}
&=\frac{1}{2\cosh\frac{\widehat p}{2}},
&\text{for}\quad s_a=+1,\nonumber\\
e^{-\frac{i}{2\hbar}\widehat p^2}e^{-\frac{i}{2\hbar}\widehat q^2}\frac{1}{2\cosh\frac{\widehat p}{2}}
e^{\frac{i}{2\hbar}\widehat q^2}e^{\frac{i}{2\hbar}\widehat p^2}
&=\frac{1}{2\cosh\frac{\widehat q}{2}},
&\text{for}\quad s_a=-1.
\label{trivialize}
\end{align}
Note that for the $(1,k)$5-brane $(s_a=-1)$ the component in the determinant simplifies to the delta function
\begin{align}
\langle\lambda|\frac{1}{2\cosh\frac{\widehat q}{2}}|\lambda'\rangle=\frac{1}{2\cosh\frac{\lambda}{2}}\times 2\pi\delta(\lambda-\lambda').
\end{align}
In this sense we refer to our computation of the similarity transformation as trivializing the $(1,k)$5-branes.
It is of course a matter of convention whether to trivialize the $(1,k)$5-branes or the NS5-branes.

Then, we can combine all of the determinants into one by iterative uses of the continuous version of the Cauchy-Binet determinant in the operator formalism,
\begin{align}
\int\frac{d^N\lambda}{N!(2\pi)^N}\det\bigl(\langle\mu_m|\widehat M|\lambda_l\rangle\bigr)_{N\times N}
\det\bigl(\langle\lambda_l|\widehat N|\nu_n\rangle\bigr)_{N\times N}
=\det\bigl(\langle\mu_m|\widehat M\widehat N|\nu_n\rangle\bigr)_{N\times N}.
\end{align}
For the case of the $(p_1,q_1,p_2,q_2,\cdots)$ model, we obtain
\begin{align}
Z_k^{\{s_a\}}(\{N\})=\int\frac{d^N\lambda_1}{2\pi}\det(\langle\lambda_1|\widehat H^{-1}|\lambda_1\rangle),
\end{align}
with
\begin{align}
\widehat H^{-1}=\frac{1}{(2\cosh\frac{\widehat p}{2})^{p_1}}\frac{1}{(2\cosh\frac{\widehat q}{2})^{q_1}}
\frac{1}{(2\cosh\frac{\widehat p}{2})^{p_2}}\frac{1}{(2\cosh\frac{\widehat q}{2})^{q_2}}\cdots,
\end{align}
which directly implies \eqref{eq:NDQCGen}.
The above computation does not only present a derivation for \eqref{eq:NDQCGen}, but also explains clearly how the sequence of two types of 5-branes (NS5-branes $\bullet$ and $(1,k)$5-branes $\circ$) is translated to the sequence of two hyperbolic canonical operators ($\widehat{\cal P}$ and $\widehat{\cal Q}$) and how the computation for each 5-brane can be performed separately.

\subsection{Pairwise closed string formalism\label{sec:M2DefApp}}

In this subsection we generalize the computation in the previous subsection and present the closed string formalism for the super Chern-Simons matrix models \eqref{eq:PFdef} with rank deformations.
The closed string formation leads us to identifying the spectral operator in the next subsection.
As we have seen in the previous subsection without rank deformations, the computation can be performed locally for each 5-brane without referring to other 5-branes.
Here we shall present the closed string formalism with rank deformations locally as well.

For this purpose we consider the situation where 
\begin{itemize}
\item
(without loss of generality) the rank $N_1$ is not deformed, i.e., $N_1=N$ is the lowest rank,
\item
only ranks with non-vanishing Chern-Simons levels $k_a$ are deformed, and
\item
no simultaneous rank deformations happen for neighboring D3-branes.
\end{itemize}
Then the partition function we are considering is
\begin{align}
Z^{\{s_a\}}_k(\{N_a\})=\int\prod_{a'=1}^{R'}\frac{D^N\lambda_{a'}}{N!(2\pi)^N}\prod_{a'=1}^{R'}
Z_{k,M}^{\{s'_{a'}\}}(N;\lambda_{a'},\lambda_{a'+1}),
\end{align}
where $Z_{k,M}^{\{s'_{a'}\}}(N,\lambda_{a'},\lambda_{a'+1})$ can be either the previous case without rank deformations or the case with a rank deformation between a pair of 5-branes
\begin{align}
Z_{k,M}^{\{s'_{a'}\}}(N;\mu,\nu)=\begin{cases}Z_k(N;\mu,\nu),&\text{for }N\bullet N\text{ or }N\circ N,\\
Z_{k,M}^{(\circ,\bullet)}(N;\mu,\nu),&\text{for }N\circ N+M\bullet N,\\
Z_{k,M}^{(\bullet,\circ)}(N;\mu,\nu),&\text{for }N\bullet N+M\circ N.\end{cases}
\end{align}
Note that the primes in $\{s'_{a'}\}$ stand for a subset of original $\{s_a\}$ skipping those with rank deformations.
We introduce the notations
$Z_{k,M}^{(\circ,\bullet)}(N;\mu,\nu)=Z_{k,M}^{(-1,+1)}(N;\mu,\nu)$
and
$Z_{k,M}^{(\bullet,\circ)}(N;\mu,\nu)=Z_{k,M}^{(+1,-1)}(N;\mu,\nu)$
since we believe that the visualization is helpful in the computation.
For each case we define
\begin{align}
Z_k(N;\mu,\nu)&
=\frac{1}{k^N}Z_k\Bigl(N,N;\frac{\mu}{k},\frac{\nu}{k}\Bigr),
\nonumber\\
Z_{k,M}^{(\mp 1,\pm 1)}(N;\mu,\nu)
&=\frac{i^{\mp\frac{1}{2}((N+M)^2-N^2)}}{(N+M)!}\int\frac{d^{N+M}\lambda}{(2\pi)^{N+M}}e^{\pm\frac{i}{2\hbar}\sum_l\lambda_l^2}
\nonumber\\
&\quad\times\frac{1}{k^{N+\frac{M}{2}}}Z_k\Bigl(N,N+M;\frac{\mu}{k},\frac{\lambda}{k}\Bigr)
\frac{1}{k^{N+\frac{M}{2}}}Z_k\Bigl(N+M,N;\frac{\lambda}{k},\frac{\nu}{k}\Bigr).
\end{align}
Note that the phase factor is determined by the sign of the Chern-Simons level $\pm k$ with a rank deformation $M$, which is normalized so that it is absent by removing the rank deformation.

As in \eqref{Zeqra}, we introduce a determinant formula 
\begin{align}
\frac{1}{k^{N+\frac{M}{2}}}Z_k\Bigl(N,N+M;\frac{\mu}{k},\frac{\lambda}{k}\Bigr)
&=(-1)^{NM}\det\begin{pmatrix}
\displaystyle\biggl[\langle\mu_m|e^{\frac{M}{2k}\widehat q}\frac{1}{2\cosh\frac{\widehat p}{2}}e^{-\frac{M}{2k}\widehat q}|\lambda_l\rangle\biggr]_{m,l}\\
\bigl[\llangle 2\pi i\sigma_a|\lambda_l\rangle\bigr]_{a,l}
\end{pmatrix},\nonumber\\
\frac{1}{k^{N+\frac{M}{2}}}Z_k\Bigl(N+M,N;\frac{\lambda}{k},\frac{\nu}{k}\Bigr)
&=(-1)^{\frac{1}{2}M(M-1)}\nonumber\\
&\hspace{-5mm}\times\det\begin{pmatrix}
\displaystyle\biggl[\langle\lambda_l|e^{\frac{M}{2k}\widehat q}\frac{1}{2\cosh\frac{\widehat p}{2}}e^{-\frac{M}{2k}\widehat q}|\nu_n\rangle\biggr]_{l,n}&
\bigl[\langle\lambda_l|2\pi i\sigma_b\rrangle\bigr]_{l,b}
\end{pmatrix},
\end{align}
(with $\sigma_a=\frac{M+1}{2}-a$) where we have introduced eigenstates for the momentum operator $\widehat p$ normalized as
\begin{align}
\llangle p_1|p_2\rrangle=2\pi\delta(p_1-p_2),\quad
\langle q|p\rrangle=\frac{e^{\frac{i}{\hbar}qp}}{\sqrt{k}},
\quad\llangle p|q\rangle=\frac{e^{-\frac{i}{\hbar}qp}}{\sqrt{k}},
\label{normalp}
\end{align}
and considered imaginary momenta by analytical continuations, which needs justifications as explained in \cite{MS2}.
Then, by applying the similarity transformation \eqref{simitran} to all of the states including $\langle\mu_m|$ and $|\nu_n\rangle$, we find
\begin{align}
&Z_{k,M}^{(\circ,\bullet)}(N;\mu,\nu)
=\frac{i^{MN+\frac{1}{2}M(M-1)}e^{\frac{2\pi i}{k}\sum_\sigma\sigma^2}}{(N+M)!}\int\frac{d^{N+M}\lambda}{(2\pi)^{N+M}}
\nonumber\\
&\quad\times\det\begin{pmatrix}\displaystyle\biggl[\langle\mu_m|\frac{1}{2\cosh\frac{\widehat q+\pi iM}{2}}|\lambda_l\rangle\biggr]\\
\bigl[\langle 2\pi i\sigma_a|\lambda_l\rangle\bigr]\end{pmatrix}
\det\begin{pmatrix}
\biggl[\langle\lambda_l|e^{\frac{M}{2k}\widehat q}\frac{1}{2\cosh\frac{\widehat p}{2}}e^{-\frac{M}{2k}\widehat q}|\nu_n\rangle\biggr]&
\bigl[\langle\lambda_l|2\pi i\sigma_b\rrangle\bigr]\end{pmatrix},\nonumber\\
&Z_{k,M}^{(\bullet,\circ)}(N;\mu,\nu)
=\frac{i^{-MN-\frac{1}{2}M(M-1)}e^{-\frac{2\pi i}{k}\sum_\sigma\sigma^2}}{(N+M)!}\int\frac{d^{N+M}\lambda}{(2\pi)^{N+M}}
\nonumber\\
&\quad\times\det\begin{pmatrix}
\biggl[\langle\mu_m|e^{\frac{M}{2k}\widehat q}\frac{1}{2\cosh\frac{\widehat p}{2}}e^{-\frac{M}{2k}\widehat q}|\lambda_l\rangle\biggr]\\
\bigl[\llangle 2\pi i\sigma_a|\lambda_l\rangle\bigr]\end{pmatrix}
\det\begin{pmatrix}\displaystyle\biggl[\langle\lambda_l|\frac{1}{2\cosh\frac{\widehat q+\pi iM}{2}}|\nu_n\rangle\biggr]&
\bigl[\langle\lambda_l|2\pi i\sigma_b\rangle\bigr]\end{pmatrix},
\end{align}
where we have used the formulas
\begin{align}
\llangle p|e^{\frac{i}{2\hbar}\widehat q^2}e^{\frac{i}{2\hbar}\widehat p^2}=\sqrt{i}e^{-\frac{i}{2\hbar}p^2}\langle p|,\quad
e^{-\frac{i}{2\hbar}\widehat p^2}e^{-\frac{i}{2\hbar}\widehat q^2}|p\rrangle=\frac{1}{\sqrt{i}}e^{\frac{i}{2\hbar}p^2}|p\rangle,
\end{align}
to transform momentum eigenstates into coordinate eigenstates.

Note that the operators in the first determinant in $Z_{k,M}^{(\circ,\bullet)}(N;\mu,\nu)$ and the second determinant in $Z_{k,M}^{(\bullet,\circ)}(N;\mu,\nu)$ consist simply of the coordinate operator.
Hence the components in these determinants all fall into the delta functions and as in \eqref{trivialize} we have trivialized the $(1,k)$5-branes.
Using the expansion
\begin{align}
\int\frac{d^{N+M}\lambda}{(N+M)!}\det\bigl(f_m(\lambda_l)\bigr)\det\bigl(g_n(\lambda_l)\bigr)
=\int d^{N+M}\lambda\biggl(\prod_mf_m(\lambda_m)\biggr)\det\bigl(g_n(\lambda_l)\bigr),
\end{align}
and integrating out all of the delta functions we finally obtain
\begin{align}
&Z_{k,M}^{(\circ,\bullet)}(N;\mu,\nu)
=e^{-i\theta_{k,M}}Z_{k,M}^\text{(CS)}\det\bigl(\langle\mu|\bigl(\widehat H_{\bullet\circ}(M)\bigr)^{-1}|\nu\rangle\bigr),\nonumber\\
&Z_{k,M}^{(\bullet,\circ)}(N;\mu,\nu)
=e^{i\theta_{k,M}}Z_{k,M}^\text{(CS)}\det\bigl(\langle\mu|\bigl(\widehat H_{\circ\bullet}(M)\bigr)^{-1}|\nu\rangle\bigr),
\label{ZZHH}
\end{align}
where the operators are given by
\begin{align}
\bigl(\widehat H_{\bullet\circ}(M)\bigr)^{-1}
&=i^{M}\frac{\prod_\sigma 2\sinh\frac{\widehat q-2\pi i\sigma}{2k}}{2\cosh\frac{\widehat q+\pi iM}{2}}\frac{1}{2\cosh\frac{\widehat p}{2}}
\frac{1}{\prod_\sigma 2\cosh\frac{\widehat q-2\pi i\sigma}{2k}},
\nonumber\\
\bigl(\widehat H_{\circ\bullet}(M)\bigr)^{-1}
&=i^M\frac{1}{\prod_\sigma 2\cosh\frac{\widehat q-2\pi i\sigma}{2k}}\frac{1}{2\cosh\frac{\widehat p}{2}}
\frac{\prod_\sigma 2\sinh\frac{\widehat q-2\pi i\sigma}{2k}}{2\cosh\frac{\widehat q+\pi iM}{2}},
\label{Hoperators}
\end{align}
with the normalizations
\begin{align}
\theta_{k,M}=-\frac{\pi}{6k}(M^3-M),\quad
Z_{k,M}^\text{(CS)}&=\frac{1}{k^{\frac{M}{2}}}\prod_{a<a'}^M2\sin\frac{\pi(a'-a)}{k}.
\end{align}
Here the two spectral operators in \eqref{Hoperators} are conjugate to each other, 
\begin{align}
\widehat H_{\bullet\circ}(M)=\bigl(\widehat H_{\circ\bullet}(M)\bigr)^\dagger,
\label{hermiteconj}
\end{align}
as can be seen from $\cosh\frac{\widehat q-\pi iM}{2}=(-1)^M\cosh\frac{\widehat q+\pi iM}{2}$.
Note that in \eqref{ZZHH} we have deliberately reversed the order of $\bullet$ and $\circ$ from the partition function $Z_{k,M}(N;\mu,\nu)$ to the spectral operator $\widehat H(M)$ since the spectral operator becomes in the reverse order from the partition function after taking the inverse as in the previous subsection without rank deformations.

\subsection{Spectral operators\label{sec:FGF}}

In the previous subsection we have obtained the closed string formalism for special rank deformations.
Although the final result of the operators \eqref{Hoperators} is clean, it does not take the form of the spectral operators similar to the quantum curves \eqref{eqClassicalCurve1}.
Fortunately, a remarkable relation was found in \cite{MZ,KMZ,KS} and using it we can rewrite our results.
In general the derivation of the relation requires complicated computations of the quantum dilogarithm functions \cite{KMZ}, though for our present case we can utilize the result directly instead of repeating by ourselves.
Let us explain the rewriting in this subsection.

It was found in \cite{KMZ} that, for integral $M$, the operator equation
\begin{align}
m^{-\frac{1}{4}}\frac{e^{\frac{\widehat q}{4}}}
{\prod_\sigma(1+e^{\frac{\widehat q+2\pi i\sigma}{k}})}
\frac{1}{2\cosh\frac{\widehat p}{2}}\frac{e^{\frac{\widehat q}{4}}\prod_\sigma(1-e^{\frac{\widehat q+2\pi i\sigma}{k}})}{1+(-1)^Me^{\widehat q}}
=\frac{1}{e^{\widehat u}+me^{-\widehat u}+e^{\widehat v}+e^{-\widehat v}},
\end{align}
holds with
\begin{align}
m=e^{\pi i(k-2M)},\quad
\widehat u=\frac{\widehat q}{2}+\frac{\widehat p}{2}+\frac{3}{4}\log m,\quad
\widehat v=-\frac{\widehat q}{2}+\frac{\widehat p}{2}+\frac{1}{4}\log m.
\end{align}
The relation is obtained by equating the inverse operator of (2.6) and (2.104) in \cite{KMZ} and expressing the result in the $({\tt u},{\tt v})$ variables in that paper.
Then, it is not difficult to find
\begin{align}
\frac{1}{\widehat H_{\circ\bullet}(M)}
=m^{\frac{i}{4\pi k}\widehat q}
\frac{1}{m^{-\frac{1}{4}}(e^{\widehat u}+me^{-\widehat u}+e^{\widehat v}+e^{-\widehat v})}
m^{-\frac{i}{4\pi k}\widehat q},
\end{align}
where we have used
\begin{align}
\frac{e^{\frac{\widehat q}{4}}}{\prod_\sigma e^{\frac{\widehat q+2\pi i\sigma}{2k}}}=e^{-(\frac{1}{4}-\frac{M}{2k})\widehat q}
=m^{-\frac{i}{4\pi k}\widehat q}.
\end{align}
After applying the similarity transformation we find
\begin{align}
\widehat H_{\circ\bullet}(M)
=m^{\frac{1}{4}}e^{\frac{\widehat q+\widehat p}{2}}
+m^{-\frac{1}{4}}e^{\frac{-\widehat q+\widehat p}{2}}
+m^{-\frac{1}{4}}e^{\frac{\widehat q-\widehat p}{2}}
+m^{\frac{1}{4}}e^{\frac{-\widehat q-\widehat p}{2}},
\end{align}
which directly gives \eqref{Habjmst}
\begin{align}
\widehat H_{\circ\bullet}(M)
&=e^{-\frac{\pi iM}{2}}\widehat Q^{\frac{1}{2}}\widehat P^{\frac{1}{2}}
+e^{\frac{\pi iM}{2}}\widehat Q^{-\frac{1}{2}}\widehat P^{\frac{1}{2}}
+e^{\frac{\pi iM}{2}}\widehat Q^{\frac{1}{2}}\widehat P^{-\frac{1}{2}}
+e^{-\frac{\pi iM}{2}}\widehat Q^{-\frac{1}{2}}\widehat P^{-\frac{1}{2}},\nonumber\\
\widehat H_{\bullet\circ}(M)
&=e^{\frac{\pi iM}{2}}\widehat P^{\frac{1}{2}}\widehat Q^{\frac{1}{2}}
+e^{-\frac{\pi iM}{2}}\widehat P^{\frac{1}{2}}\widehat Q^{-\frac{1}{2}}
+e^{-\frac{\pi iM}{2}}\widehat P^{-\frac{1}{2}}\widehat Q^{\frac{1}{2}}
+e^{\frac{\pi iM}{2}}\widehat P^{-\frac{1}{2}}\widehat Q^{-\frac{1}{2}}.
\end{align}
Note that $\widehat H_{\bullet\circ}(M)$ is obtained from the conjugation \eqref{hermiteconj} and the results are consistent with the computation without rank deformations after setting $M=0$.
Schematically we can characterize them by the asymptotic values of zero points when regarding them as the defining equations of algebraic curves
\begin{align}
\widehat H_{\circ\bullet}(M)
={\scriptstyle e^{-\pi iM}}
\overset{e^{\pi iM}}{\underset{e^{-\pi iM}}{\sqbox{$\circ\bullet$}}}
{\scriptstyle e^{\pi iM}},\quad
\widehat H_{\bullet\circ}(M)
={\scriptstyle e^{\pi iM}}
\overset{e^{-\pi iM}}{\underset{e^{\pi iM}}{\sqbox{$\circ\bullet$}}}
{\scriptstyle e^{-\pi iM}},
\label{schematic}
\end{align}
where the axes of $\widehat Q$ and $\widehat P$ is the same as figures \ref{fig:asymptoticvalues} and \ref{asymptoticvaluesfixed}.
The asymptotic values depend of course on the order of the operators.
We adopt the standard normal ordering as the $D_5$ curve for $\widehat H_{\circ\bullet}(M)$ while the inverse normal ordering for $\widehat H_{\bullet\circ}(M)$.
For a general sequence of 5-branes all we have to do is to multiply these operators reversely as in \eqref{eq:NDQCGen}, though we need to take care of the normal ordering again.

\section{Instanton effects and topological strings\label{tsappendix}}

\subsection{Grand potential from matrix models\label{secGPnumerical}}

In this section, we list the grand potential \eqref{eq:GPdef} defined from the grand canonical partition function with the second reference frame \eqref{Xi2} for various combinations of $k$ and $(M_1,M_2)$.
In the following expression we always omit displaying the reference frame $(2)$ in $J_{k,(M_1,M_2)}^{(2),\text{np}}(\mu_\text{eff})=J_{k,(M_1,M_2)}^{(2),\text{WS}}(\mu_\text{eff})+J_{k,(M_1,M_2)}^{(2),\text{MB}}(\mu_\text{eff})$.
In redefining the chemical potential $\mu$ to $\mu_\text{eff}$, we adopt \eqref{mueff}, the same relation as that for the first reference.
Although only terms related to membrane instantons $e^{-\mu_\text{eff}}$ are deformed compared with the expression with the first reference we record the whole expressions to avoid confusions.

\noindent
$\bullet\;k=1$
\begin{align}
J_{1,(0,0)}^{\text{np}}
&=\frac{2(\mu_{\text{eff}}^{2}+2\mu_{\text{eff}}+2)}{\pi^{2}}e^{-\mu_{\text{eff}}}
+\biggl[-\frac{9(2\mu_{\text{eff}}^{2}+2\mu_{\text{eff}}+1)}{2\pi^{2}}+2\biggr]e^{-2\mu_{\text{eff}}}\nonumber\\
&\quad+\biggl[\frac{164(9\mu_{\text{eff}}^{2}+6\mu_{\text{eff}}+2)}{27\pi^{2}}-16\biggr]e^{-3\mu_{\text{eff}}}
+\biggl[-\frac{777(8\mu_{\text{eff}}^{2}+4\mu_{\text{eff}}+1)}{16\pi^{2}}+138\biggr]e^{-4\mu_{\text{eff}}}\nonumber\\
&\quad+{\cal O}(e^{-5\mu_{\text{eff}}}),\nonumber\\
J_{1,(\frac{1}{2},\frac{1}{2})}^{\text{np}}
&=\biggl[\frac{2\mu_{\text{eff}}^{2}+2\mu_{\text{eff}}+1}{2\pi^{2}}-\frac{7}{4}\biggr]e^{-2\mu_{\text{eff}}}
+\biggl[-\frac{9(8\mu_{\text{eff}}^{2}+4\mu_{\text{eff}}+1)}{16\pi^{2}}+\frac{79}{8}\biggr]e^{-4\mu_{\text{eff}}}
+{\cal O}(e^{-5\mu_{\text{eff}}}).
\end{align}

\noindent
$\bullet\;k=2$
\begin{align}
J_{2,(0,0)}^{\text{np}}
&=4e^{-\frac{1}{2}\mu_{\text{eff}}}+\biggl[\frac{\mu_{\text{eff}}^{2}+2\mu_{\text{eff}}+2}{\pi^{2}}-7\biggr]e^{-\mu_{\text{eff}}}
+\frac{40}{3}e^{-\frac{3}{2}\mu_{\text{eff}}}\nonumber\\
&\quad+\biggl[-\frac{9(2\mu_{\text{eff}}^{2}+2\mu_{\text{eff}}+1)}{4\pi^{2}}-\frac{75}{2}\biggr]e^{-2\mu_{\text{eff}}}
+\frac{724}{5}e^{-\frac{5}{2}\mu_{\text{eff}}}\nonumber\\
&\quad+\biggl[\frac{82(9\mu_{\text{eff}}^{2}+6\mu_{\text{eff}}+2)}{27\pi^{2}}-\frac{1318}{3}\biggr]e^{-3\mu_{\text{eff}}}
+{\cal O}(e^{-\frac{7}{2}\mu_{\text{eff}}}),\nonumber\\
J_{2,(1,0)}^{\text{np}}
&=\biggl[-\frac{\mu_{\text{eff}}^{2}+2\mu_{\text{eff}}+2}{\pi^{2}}\biggr]e^{-\mu_{\text{eff}}}
+\biggl[-\frac{9(2\mu_{\text{eff}}^{2}+2\mu_{\text{eff}}+1)}{4\pi^{2}}-5\biggr]e^{-2\mu_{\text{eff}}}\nonumber\\
&\quad+\biggl[-\frac{82(9\mu_{\text{eff}}^{2}+6\mu_{\text{eff}}+2)}{27\pi^{2}}-32\biggr]e^{-3\mu_{\text{eff}}}
+{\cal O}(e^{-4\mu_{\text{eff}}}),\nonumber\\
J_{2,(1,1)}^{\text{np}}
&=\biggl[-\frac{\mu_{\text{eff}}^{2}+2\mu_{\text{eff}}+2}{\pi^{2}}+2\biggr]e^{-\mu_{\text{eff}}}
+\biggl[-\frac{9(2\mu_{\text{eff}}^{2}+2\mu_{\text{eff}}+1)}{4\pi^{2}}+10\biggr]e^{-2\mu_{\text{eff}}}\nonumber\\
&\quad+\biggl[-\frac{82(9\mu_{\text{eff}}^{2}+6\mu_{\text{eff}}+2)}{27\pi^{2}}+\frac{212}{3}\biggr]e^{-3\mu_{\text{eff}}}
+{\cal O}(e^{-4\mu_{\text{eff}}}),\nonumber\\
J_{2,(0,1)}^{\text{np}}
&=\biggl[\frac{\mu_{\text{eff}}^{2}+2\mu_{\text{eff}}+2}{\pi^{2}}-1\biggr]e^{-\mu_{\text{eff}}}
+\biggl[-\frac{9(2\mu_{\text{eff}}^{2}+2\mu_{\text{eff}}+1)}{4\pi^{2}}+\frac{19}{2}\biggr]e^{-2\mu_{\text{eff}}}\nonumber\\
&\quad+\biggl[\frac{82(9\mu_{\text{eff}}^{2}+6\mu_{\text{eff}}+2)}{27\pi^{2}}-\frac{202}{3}\biggr]e^{-3\mu_{\text{eff}}}
+{\cal O}(e^{-4\mu_{\text{eff}}}).
\end{align}

\noindent
$\bullet\;k=3$
\begin{align}
J_{3,(0,0)}^{\text{np}}
&=\frac{16}{3}e^{-\frac{1}{3}\mu_{\text{eff}}}-4e^{-\frac{2}{3}\mu_{\text{eff}}}
+\biggl[\frac{2(\mu_{\text{eff}}^{2}+2\mu_{\text{eff}}+2)}{3\pi^{2}}+\frac{112}{9}\biggr]e^{-\mu_{\text{eff}}}
-61e^{-\frac{4}{3}\mu_{\text{eff}}}+\frac{3376}{15}e^{-\frac{5}{3}\mu_{\text{eff}}}\nonumber\\
&\quad+\biggl[-\frac{3(2\mu_{\text{eff}}^{2}+2\mu_{\text{eff}}+1)}{2\pi^{2}}-\frac{2266}{3}\biggr]e^{-2\mu_{\text{eff}}}
+\frac{52880}{21}e^{-\frac{7}{3}\mu_{\text{eff}}}+{\cal O}(e^{-\frac{8}{3}\mu_{\text{eff}}}),\nonumber\\
J_{3,(1,0)}^{\text{np}}
&=\frac{8}{3}e^{-\frac{1}{3}\mu_{\text{eff}}}-6e^{-\frac{2}{3}\mu_{\text{eff}}}
+\biggl[-\frac{2(\mu_{\text{eff}}^{2}+2\mu_{\text{eff}}+2)}{3\pi^{2}}+\frac{110}{9}\biggr]e^{-\mu_{\text{eff}}}
-30e^{-\frac{4}{3}\mu_{\text{eff}}}+\frac{1088}{15}e^{-\frac{5}{3}\mu_{\text{eff}}}\nonumber\\
&\quad+\biggl[-\frac{3(2\mu_{\text{eff}}^{2}+2\mu_{\text{eff}}+1)}{2\pi^{2}}-209\biggr]e^{-2\mu_{\text{eff}}}
+\frac{12160}{21}e^{-\frac{7}{3}\mu_{\text{eff}}}+{\cal O}(e^{-\frac{8}{3}\mu_{\text{eff}}}),\nonumber\\
J_{3,(\frac{3}{2},\frac{1}{2})}^{\text{np}}
&=-\frac{10}{3}e^{-\frac{2}{3}\mu_{\text{eff}}}-8e^{-\frac{4}{3}\mu_{\text{eff}}}
+\biggl[\frac{2\mu_{\text{eff}}^{2}+2\mu_{\text{eff}}+1}{6\pi^{2}}-\frac{1045}{36}\biggr]e^{-2\mu_{\text{eff}}}
+{\cal O}(e^{-\frac{8}{3}\mu_{\text{eff}}}),\nonumber\\
J_{3,(\frac{3}{2},\frac{3}{2})}^{\text{np}}
&=-\frac{4}{3}e^{-\frac{2}{3}\mu_{\text{eff}}}+3e^{-\frac{4}{3}\mu_{\text{eff}}}
+\biggl[\frac{2\mu_{\text{eff}}^{2}+2\mu_{\text{eff}}+1}{6\pi^{2}}+\frac{275}{36}\biggr]e^{-2\mu_{\text{eff}}}
+{\cal O}(e^{-\frac{8}{3}\mu_{\text{eff}}}),\nonumber\\
J_{3,(\frac{1}{2},\frac{3}{2})}^{\text{np}}
&=\frac{2}{3}e^{-\frac{2}{3}\mu_{\text{eff}}}+2e^{-\frac{4}{3}\mu_{\text{eff}}}
+\biggl[\frac{2\mu_{\text{eff}}^{2}+2\mu_{\text{eff}}+1}{6\pi^{2}}-\frac{349}{36}\biggr]e^{-2\mu_{\text{eff}}}
+{\cal O}(e^{-\frac{8}{3}\mu_{\text{eff}}}),\nonumber\\
J_{3,(0,1)}^{\text{np}}
&=\frac{4}{3}e^{-\frac{1}{3}\mu_{\text{eff}}}-2e^{-\frac{2}{3}\mu_{\text{eff}}}
+\biggl[\frac{2(\mu_{\text{eff}}^{2}+2\mu_{\text{eff}}+2)}{3\pi^{2}}+\frac{28}{9}\biggr]e^{-\mu_{\text{eff}}}
-8e^{-\frac{4}{3}\mu_{\text{eff}}}+\frac{244}{15}e^{-\frac{5}{3}\mu_{\text{eff}}}\nonumber\\
&\quad+\biggl[-\frac{3(2\mu_{\text{eff}}^{2}+2\mu_{\text{eff}}+1)}{2\pi^{2}}-\frac{74}{3}\biggr]e^{-2\mu_{\text{eff}}}
+\frac{1712}{21}e^{-\frac{7}{3}\mu_{\text{eff}}}+{\cal O}(e^{-\frac{8}{3}\mu_{\text{eff}}}).
\end{align}

\noindent
$\bullet\;k=4$
\begin{align}
J_{4,(0,0)}^{\text{np}}
&=8e^{-\frac{1}{4}\mu_{\text{eff}}}-8e^{-\frac{1}{2}\mu_{\text{eff}}}+\frac{80}{3}e^{-\frac{3}{4}\mu_{\text{eff}}}
+\biggl[\frac{\mu_{\text{eff}}^{2}+2\mu_{\text{eff}}+2}{2\pi^{2}}-\frac{197}{2}\biggr]e^{-\mu_{\text{eff}}}
+\frac{1928}{5}e^{-\frac{5}{4}\mu_{\text{eff}}}\nonumber\\
&\quad-\frac{4784}{3}e^{-\frac{3}{2}\mu_{\text{eff}}}+\frac{44976}{7}e^{-\frac{7}{4}\mu_{\text{eff}}}
+{\mathcal{O}}(e^{-2\mu_{\text{eff}}}),\nonumber\\
J_{4,(1,0)}^{\text{np}}
&=4\sqrt{2}e^{-\frac{1}{4}\mu_{\text{eff}}}-8e^{-\frac{1}{2}\mu_{\text{eff}}}+\frac{32\sqrt{2}}{3}e^{-\frac{3}{4}\mu_{\text{eff}}}
+\biggl[-\frac{\mu_{\text{eff}}^{2}+2\mu_{\text{eff}}+2}{2\pi^{2}}-55\biggr]e^{-\mu_{\text{eff}}}\nonumber\\
&\quad+\frac{756\sqrt{2}}{5}e^{-\frac{5}{4}\mu_{\text{eff}}}-\frac{2384}{3}e^{-\frac{3}{2}\mu_{\text{eff}}}
+\frac{13920\sqrt{2}}{7}e^{-\frac{7}{4}\mu_{\text{eff}}}+{\cal O}(e^{-2\mu_{\text{eff}}}),\nonumber\\
J_{4,(2,0)}^{\text{np}}
&=-8e^{-\frac{1}{2}\mu_{\text{eff}}}
+\biggl[\frac{\mu_{\text{eff}}^{2}+2\mu_{\text{eff}}+2}{2\pi^{2}}-\frac{73}{2}\biggr]e^{-\mu_{\text{eff}}}
-\frac{560}{3}e^{-\frac{3}{2}\mu_{\text{eff}}}+{\cal O}(e^{-2\mu_{\text{eff}}}),\nonumber\\
J_{4,(2,1)}^{\text{np}}
&=-2e^{-\frac{1}{2}\mu_{\text{eff}}}
+\biggl[\frac{\mu_{\text{eff}}^{2}+2\mu_{\text{eff}}+2}{2\pi^{2}}-\frac{15}{2}\biggr]e^{-\mu_{\text{eff}}}
+\frac{4}{3}e^{-\frac{3}{2}\mu_{\text{eff}}}+{\cal O}(e^{-2\mu_{\text{eff}}}),\nonumber\\
J_{4,(2,2)}^{\text{np}}
&=\biggl[\frac{\mu_{\text{eff}}^{2}+2\mu_{\text{eff}}+2}{2\pi^{2}}-\frac{1}{2}\biggr]e^{-\mu_{\text{eff}}}
+{\cal O}(e^{-2\mu_{\text{eff}}}),\nonumber\\
J_{4,(1,2)}^{\text{np}}
&=\biggl[-\frac{\mu_{\text{eff}}^{2}+2\mu_{\text{eff}}+2}{2\pi^{2}}+5\biggr]e^{-\mu_{\text{eff}}}
+{\cal O}(e^{-2\mu_{\text{eff}}}),\nonumber\\
J_{4,(0,2)}^{\text{np}}
&=\biggl[\frac{\mu_{\text{eff}}^{2}+2\mu_{\text{eff}}+2}{2\pi^{2}}+\frac{3}{2}\biggr]e^{-\mu_{\text{eff}}}
+{\cal O}(e^{-2\mu_{\text{eff}}}),\nonumber\\
J_{4,(0,1)}^{\text{np}}
&=4e^{-\frac{1}{4}\mu_{\text{eff}}}-6e^{-\frac{1}{2}\mu_{\text{eff}}}+\frac{40}{3}e^{-\frac{3}{4}\mu_{\text{eff}}}
+\biggl[\frac{\mu_{\text{eff}}^{2}+2\mu_{\text{eff}}+2}{2\pi^{2}}-\frac{75}{2}\biggr]e^{-\mu_{\text{eff}}}\nonumber\\
&\quad+\frac{564}{5}e^{-\frac{5}{4}\mu_{\text{eff}}}-348e^{-\frac{3}{2}\mu_{\text{eff}}}+\frac{7480}{7}e^{-\frac{7}{4}\mu_{\text{eff}}}
+{\cal O}(e^{-2\mu_{\text{eff}}}).
\end{align}

\noindent
$\bullet\;k=6$
\begin{align}
J_{6,(0,0)}^{\text{np}}
&=16e^{-\frac{1}{6}\mu_{\text{eff}}}-\frac{52}{3}e^{-\frac{1}{3}\mu_{\text{eff}}}+\frac{148}{3}e^{-\frac{1}{2}\mu_{\text{eff}}}
-189e^{-\frac{2}{3}\mu_{\text{eff}}}+\frac{4336}{5}e^{-\frac{5}{6}\mu_{\text{eff}}}\nonumber\\
&\quad+\biggl[\frac{\mu_{\text{eff}}^{2}+2\mu_{\text{eff}}+2}{3\pi^{2}}-\frac{38137}{9}\biggr]e^{-\mu_{\text{eff}}}
+\frac{148752}{7}e^{-\frac{7}{6}\mu_{\text{eff}}}+{\mathcal{O}}(e^{-\frac{4}{3}\mu_{\text{eff}}}),\nonumber\\
J_{6,(1,0)}^{\text{np}}
&=8\sqrt{3}e^{-\frac{1}{6}\mu_{\text{eff}}}-\frac{50}{3}e^{-\frac{1}{3}\mu_{\text{eff}}}+24\sqrt{3}e^{-\frac{1}{2}\mu_{\text{eff}}}
-158e^{-\frac{2}{3}\mu_{\text{eff}}}+\frac{1952\sqrt{3}}{5}e^{-\frac{5}{6}\mu_{\text{eff}}}\nonumber\\
&\quad+\biggl[-\frac{\mu_{\text{eff}}^{2}+2\mu_{\text{eff}}+2}{3\pi^{2}}-\frac{28394}{9}\biggr]e^{-\mu_{\text{eff}}}
+\frac{60976\sqrt{3}}{7}e^{-\frac{7}{6}\mu_{\text{eff}}}+{\cal O}(e^{-\frac{4}{3}\mu_{\text{eff}}}),\nonumber\\
J_{6,(2,0)}^{\text{np}}
&=8e^{-\frac{1}{6}\mu_{\text{eff}}}-\frac{46}{3}e^{-\frac{1}{3}\mu_{\text{eff}}}+\frac{68}{3}e^{-\frac{1}{2}\mu_{\text{eff}}}
-94e^{-\frac{2}{3}\mu_{\text{eff}}}+\frac{1568}{5}e^{-\frac{5}{6}\mu_{\text{eff}}}\nonumber\\
&\quad+\biggl[\frac{\mu_{\text{eff}}^{2}+2\mu_{\text{eff}}+2}{3\pi^{2}}-\frac{11959}{9}\biggr]e^{-\mu_{\text{eff}}}
+\frac{36576}{7}e^{-\frac{7}{6}\mu_{\text{eff}}}+{\cal O}(e^{-\frac{4}{3}\mu_{\text{eff}}}),\nonumber\\
J_{6,(3,0)}^{\text{np}}
&=-\frac{44}{3}e^{-\frac{1}{3}\mu_{\text{eff}}}-61e^{-\frac{2}{3}\mu_{\text{eff}}}
+\biggl[-\frac{\mu_{\text{eff}}^{2}+2\mu_{\text{eff}}+2}{3\pi^{2}}-\frac{4772}{9}\biggr]e^{-\mu_{\text{eff}}}
+{\cal O}(e^{-\frac{4}{3}\mu_{\text{eff}}}),\nonumber\\
J_{6,(3,1)}^{\text{np}}
&=-\frac{26}{3}e^{-\frac{1}{3}\mu_{\text{eff}}}-32e^{-\frac{2}{3}\mu_{\text{eff}}}
+\biggl[-\frac{\mu_{\text{eff}}^{2}+2\mu_{\text{eff}}+2}{3\pi^{2}}-\frac{1658}{9}\biggr]e^{-\mu_{\text{eff}}}
+{\cal O}(e^{-\frac{4}{3}\mu_{\text{eff}}}),\nonumber\\
J_{6,(3,2)}^{\text{np}}
&=-\frac{2}{3}e^{-\frac{1}{3}\mu_{\text{eff}}}
+\biggl[-\frac{\mu_{\text{eff}}^{2}+2\mu_{\text{eff}}+2}{3\pi^{2}}+\frac{40}{9}\biggr]e^{-\mu_{\text{eff}}}
+{\cal O}(e^{-\frac{4}{3}\mu_{\text{eff}}}),\nonumber\\
J_{6,(3,3)}^{\text{np}}
&=\frac{4}{3}e^{-\frac{1}{3}\mu_{\text{eff}}}+3e^{-\frac{2}{3}\mu_{\text{eff}}}
+\biggl[-\frac{\mu_{\text{eff}}^{2}+2\mu_{\text{eff}}+2}{3\pi^{2}}-\frac{62}{9}\biggr]e^{-\mu_{\text{eff}}}
+{\cal O}(e^{-\frac{4}{3}\mu_{\text{eff}}}),\nonumber\\
J_{6,(2,3)}^{\text{np}}
&=\frac{2}{3}e^{-\frac{1}{3}\mu_{\text{eff}}}+2e^{-\frac{2}{3}\mu_{\text{eff}}}
+\biggl[\frac{\mu_{\text{eff}}^{2}+2\mu_{\text{eff}}+2}{3\pi^{2}}-\frac{85}{9}\biggr]e^{-\mu_{\text{eff}}}
+{\cal O}(e^{-\frac{4}{3}\mu_{\text{eff}}}),\nonumber\\
J_{6,(1,3)}^{\text{np}}
&=-\frac{2}{3}e^{-\frac{1}{3}\mu_{\text{eff}}}+2e^{-\frac{2}{3}\mu_{\text{eff}}}
+\biggl[-\frac{\mu_{\text{eff}}^{2}+2\mu_{\text{eff}}+2}{3\pi^{2}}+\frac{76}{9}\biggr]e^{-\mu_{\text{eff}}}
+{\cal O}(e^{-\frac{4}{3}\mu_{\text{eff}}}),\nonumber\\
J_{6,(0,3)}^{\text{np}}
&=-\frac{4}{3}e^{-\frac{1}{3}\mu_{\text{eff}}}+3e^{-\frac{2}{3}\mu_{\text{eff}}}
+\biggl[\frac{\mu_{\text{eff}}^{2}+2\mu_{\text{eff}}+2}{3\pi^{2}}+\frac{89}{9}\biggr]e^{-\mu_{\text{eff}}}
+{\cal O}(e^{-\frac{4}{3}\mu_{\text{eff}}}),\nonumber\\
J_{6,(0,2)}^{\text{np}}
&=4e^{-\frac{1}{6}\mu_{\text{eff}}}-\frac{22}{3}e^{-\frac{1}{3}\mu_{\text{eff}}}+\frac{40}{3}e^{-\frac{1}{2}\mu_{\text{eff}}}
-32e^{-\frac{2}{3}\mu_{\text{eff}}}+\frac{484}{5}e^{-\frac{5}{6}\mu_{\text{eff}}}\nonumber\\
&\quad+\biggl[\frac{\mu_{\text{eff}}^{2}+2\mu_{\text{eff}}+2}{3\pi^{2}}-\frac{2557}{9}\biggr]e^{-\mu_{\text{eff}}}
+\frac{6192}{7}e^{-\frac{7}{6}\mu_{\text{eff}}}+{\cal O}(e^{-\frac{4}{3}\mu_{\text{eff}}}),\nonumber\\
J_{6,(0,1)}^{\text{np}}
&=12e^{-\frac{1}{6}\mu_{\text{eff}}}-\frac{46}{3}e^{-\frac{1}{3}\mu_{\text{eff}}}+36e^{-\frac{1}{2}\mu_{\text{eff}}}
-128e^{-\frac{2}{3}\mu_{\text{eff}}}+\frac{2652}{5}e^{-\frac{5}{6}\mu_{\text{eff}}}\nonumber\\
&\quad+\biggl[\frac{\mu_{\text{eff}}^{2}+2\mu_{\text{eff}}+2}{3\pi^{2}}-\frac{20995}{9}\biggr]e^{-\mu_{\text{eff}}}
+\frac{73680}{7}e^{-\frac{7}{6}\mu_{\text{eff}}}+{\cal O}(e^{-\frac{4}{3}\mu_{\text{eff}}}).
\end{align}

\subsection{Characters\label{characterlist}}

In this subsection we list characters for various representations of $D_5$ in order to study the non-perturbative effects in the previous subsection.
In the following we abbreviate the arguments of the characters $\chi_{\bf R}(1,q_1,q_2,q_3,1)$ by $\chi_{\bf R}(q_1,q_2,q_3)$ for simplicity.

\noindent
$\bullet$\;{conjugacy class 0}
\begin{align}
\chi_{\bf 1}(q_1,q_2,q_3)&=1,\nonumber\\
\chi_{\bf 45}(q_1,q_2,q_3)&=9+4(q_2+q_2^{-1})(q_3+q_3^{-1})+q_2^2+q_2^{-2}+q_3^2+q_3^{-2}\nonumber\\
&\hspace{-10mm}+(q_1+q_1^{-1})\bigl(4+(q_2+q_2^{-1})(q_3+q_3^{-1})\bigr),\nonumber\\
\chi_{\bf 54}(q_1,q_2,q_3)&=12+4(q_2+q_2^{-1})(q_3+q_3^{-1})+q_2^2+q_2^{-2}+q_3^2+q_3^{-2}
+(q_2^2+q_2^{-2})(q_3^2+q_3^{-2})\nonumber\\
&\hspace{-10mm}+(q_1+q_1^{-1})\bigl(4+(q_2+q_2^{-1})(q_3+q_3^{-1})\bigr)+q_1^2+q_1^{-2}.
\end{align}

\noindent
$\bullet$\;{conjugacy class 2}
\begin{align}
\chi_{\bf 10}(q_1,q_2,q_3)&=4+(q_2+q_2^{-1})(q_3+q_3^{-1})+q_1+q_1^{-1},\nonumber\\
\chi_{\bf 120}(q_1,q_2,q_3)&=16+8(q_2+q_2^{-1})(q_3+q_3^{-1})+4(q_2^2+q_2^{-2}+q_3^2+q_3^{-2})\nonumber\\
&\hspace{-20mm}+(q_1+q_1^{-1})(8+4(q_2+q_2^{-1})(q_3+q_3^{-1})+q_2^2+q_2^{-2}+q_3^2+q_3^{-2}),\nonumber\\
\chi_{\bf 126}(q_1,q_2,q_3)&=12+7(q_2+q_2^{-1})(q_3+q_3^{-1})+4(q_2^2+q_2^{-2}+q_3^2+q_3^{-2})\nonumber\\
&\hspace{-20mm}+(q_1+q_1^{-1})(7+4(q_2+q_2^{-1})(q_3+q_3^{-1})+3(q_2^2+q_2^{-2}+q_3^2+q_3^{-2})),\nonumber\\
\chi_{\bf 320}(q_1,q_2,q_3)&=40+18(q_2+q_2^{-1})(q_3+q_3^{-1})+8(q_2^2+q_2^{-2}+q_3^2+q_3^{-2})
+4(q_2^2+q_2^{-2})(q_3^2+q_3^{-2})\nonumber\\
&\hspace{-20mm}
+(q_2+q_2^{-1})(q_3+q_3^{-1})(q_2q_3+q_2^{-1}q_3^{-1})(q_2q_3^{-1}+q_2^{-1}q_3)\nonumber\\
&\hspace{-20mm}+(q_1+q_1^{-1})\bigl(20+8(q_2+q_2^{-1})(q_3+q_3^{-1})+2(q_2^2+q_2^{-2}+q_3^2+q_3^{-2})
+(q_2^2+q_2^{-2})(q_3^2+q_3^{-2})\bigr)\nonumber\\
&\hspace{-20mm}+(q_1^2+q_1^{-2})\bigl(4+(q_2+q_2^{-1})(q_3+q_3^{-1})\bigr).
\end{align}

\noindent
$\bullet$\;{conjugacy class 1 or 3}
\begin{align}
\chi_{\bf 16}(q_1,q_2,q_3)&=2(q_1^{\frac{1}{2}}+q_1^{-\frac{1}{2}})
(q_2^{\frac{1}{2}}q_3^{\frac{1}{2}}+q_2^{-\frac{1}{2}}q_3^{-\frac{1}{2}})
(q_2^{\frac{1}{2}}q_3^{-\frac{1}{2}}+q_2^{-\frac{1}{2}}q_3^{-\frac{1}{2}}),\nonumber\\
\frac{\chi_{\bf 144}(q_1,q_2,q_3)}{\chi_{\bf 16}(q_1,q_2,q_3)}
&=3+(q_2+q_2^{-1})(q_3+q_3^{-1})+q_1+q_1^{-1},\nonumber\\
\frac{\chi_{\bf 560}(q_1,q_2,q_3)}{\chi_{\bf 16}(q_1,q_2,q_3)}
&=5+3(q_2+q_2^{-1})(q_3+q_3^{-1})+q_2^2+q_2^{-2}+q_3^2+q_3^{-2}\nonumber\\
&\quad+(q_1+q_1^{-1})\bigl(3+(q_2+q_2^{-1})(q_3+q_3^{-1})\bigr),\nonumber\\
\frac{\chi_{\bf 720}(q_1,q_2,q_3)}{\chi_{\bf 16}(q_1,q_2,q_3)}
&=9+3(q_2+q_2^{-1})(q_3+q_3^{-1})+q_2^2+q_2^{-2}+q_3^2+q_3^{-2}
+(q_2^2+q_2^{-2})(q_3^2+q_3^{-2})\nonumber\\
&\quad+(q_1+q_1^{-1})\bigl(3+(q_2+q_2^{-1})(q_3+q_3^{-1})\bigr)+q_1^2+q_1^{-2}.
\end{align}

\section*{Acknowledgements}
We are grateful to Tomohiro Furukawa, Tomoki Nosaka, Shigeki Sugimoto, Yuji Sugimoto and Yasuhiko Yamada for valuable discussions and comments.
The work of S.M. is supported by JSPS Grant-in-Aid for Scientific Research (C) \#26400245 and \#19K03829.
S.M. would like to thank Yukawa Institute for Theoretical Physics at Kyoto University for warm hospitality.

\end{document}